\documentclass[aps,pra,twocolumn,showpacs,floatfix,longbibliography]{revtex4-1}

\usepackage{epsfig,amsmath,amssymb}
\usepackage{graphics} 
\usepackage{subfigure}
\usepackage{color}
\usepackage{multirow}
\usepackage{mathrsfs}
\usepackage{natbib}
\usepackage{bbold}

\usepackage{wasysym}

\usepackage[
colorlinks=true,
linkcolor=blue,
urlcolor=blue,
citecolor=blue,
hyperindex=true
]{hyperref}

\begin{document}

\title
{
Frustrated spin-$\frac{1}{2}$ Heisenberg magnet on a square-lattice bilayer: High-order study of the quantum critical behavior of the $J_{1}$--$J_{2}$--$J_{1}^{\perp}$ model
}

\author
{R. F. Bishop$^{1,2}$}
\email{raymond.bishop@manchester.ac.uk}
\author
{P. H. Y. Li$^{1,2}$}
\email{peggyhyli@gmail.com}
\author
{O. G\"{o}tze$^{3}$}
\email{oliver.goetze@ovgu.de}
\author
{J. Richter$^{3,4}$}
\email{Johannes.Richter@physik.uni-magdeburg.de}
\affiliation
{$^{1}$School of Physics and Astronomy, Schuster Building, The University of Manchester, Manchester, M13 9PL, UK}
\affiliation
{$^{2}$School of Physics and Astronomy, University of Minnesota, 116 Church Street SE, Minneapolis, Minnesota 55455, USA}
\affiliation
{$^{3}$Institut f\"{u}r Theoretische Physik, Otto-von-Guericke Universit\"{a}t Magdeburg, P.O.Box 4120, 39016 Magdeburg, Germany}
\affiliation
{$^{4}$Max-Planck-Institut f\"{u}r Physik Komphexer Systeme, N\"{o}thnitzer Stra{\ss}e 38, 01187 Dresden, Germany}

\begin{abstract}
The zero-temperature phase diagram of the spin-$\frac{1}{2}$ $J_{1}$--$J_{2}$--$J_{1}^{\perp}$ model on an $AA$-stacked square-lattice bilayer is studied using the coupled cluster method implemented to very high orders.  Both nearest-neighbor (NN) and frustrating next-nearest-neighbor Heisenberg exchange interactions, of strengths $J_{1}>0$ and $J_{2} \equiv \kappa J_{1}>0$, respectively, are included in each layer.  The two layers are coupled via a NN interlayer Heisenberg exchange interaction with a strength $J_{1}^{\perp} \equiv \delta J_{1}$.  The magnetic order parameter $M$ (viz., the sublattice magnetization) is calculated directly in the thermodynamic (infinite-lattice) limit for the two cases when both layers have antiferromagnetic ordering of either the N\'{e}el or the striped kind, and with the layers coupled so that NN spins between them are either parallel (when $\delta < 0$) or antiparallel (when $\delta > 0$) to one another.  Calculations are performed at $n$th order in a well-defined sequence of approximations, which exactly preserve both the Goldstone linked cluster theorem and the Hellmann-Feynman theorem, with $n \leq 10$.  The sole approximation made is to extrapolate such sequences of $n$th-order results for $M$ to the exact limit, $n \to \infty$.  By thus locating the points where $M$ vanishes, we calculate the full phase boundaries of the two collinear AFM phases in the $\kappa$--$\delta$ half-plane with $\kappa > 0$.  In particular, we provide the accurate estimate, ($\kappa \approx 0.547,\delta \approx -0.45$), for the position of the quantum triple point (QTP) in the region $\delta < 0$.  We also show that there is no counterpart of such a QTP in the region $\delta > 0$, where the two quasiclassical phase boundaries show instead an ``avoided crossing'' behavior, such that the entire region that contains the nonclassical paramagnetic phases is singly connected.  
\end{abstract}

%\pacs{75.10.Jm, 75.10.Kt, 75.30.Kz, 75.40.Cx}

\maketitle

\section{INTRODUCTION}
\label{introd_sec}
The frustrated spin-$\frac{1}{2}$ $J_{1}$--$J_{2}$ Heisenberg antiferromagnet on the square lattice, which contains isotropic Heisenberg exchange interactions with strengths $J_{1}>0$ between all nearest-neighbor (NN) pairs of spins and $J_{2}>0$ between all next-nearest-neighbor (NNN) pairs, has become a paradigmatic model of quantum magnetism.  It has received enormous attention over the last thirty or so years \cite{Chandra:1988,Dagotto:1989,Gelfand:1989,Sachdev:1990,Chubukov:1991,Read:1991,Richter:1993,Richter:1994,Ivanov:1994_J1J2mod,Schulz:1996,Oitmaa:1996,Zhitomirsky:1996,Trumper:1997_J1J2mod,Bishop:1998_J1J2mod,Singh:1999,Kotov:1999,Capriotti:2000,Capriotti:2001,Takano:2003,Roscilde:2004,Lante:2006_J1J2mod_sqLatt,Sirker:2006,Schm:2006_stackSqLatt,Mambrini:2006,Bi:2008_JPCM_J1J1primeJ2,Bi:2008_PRB_J1xxzJ2xxz,Darradi:2008_J1J2mod,Isaev:2009_J1J2mod,Murg:2009_peps,Ralko:2009,Richter:2010_ED40,Reuther:2010_J1J2mod,Reuther:2011_J1J2J3mod,Yu:2012,Gotze:2012,Jiang:2012,Mezzacapo:2012,LiT:2012,Wang:2013,Zhang:2013:J1J2SqLatt,Hu:2013,Gong:2014_J1J2mod_sqLatt,Doretto:2014_J1J2mod_sqLatt,Qi:2014_J1J2SqLatt,Metavitsiadis_Eggert:2014_J1J2mod_sqLatt,Ren:2014_J1J2SqLatt,Wang:2014_J1J2SqLatt,Chou:2014_J1J2SqLatt,Morita:2015_J1J2SqLatt,Richter:2015_ccm_J1J2sq_spinGap,Wang:2016_J1J2mod,Poilblanc:2017_J1J2mod,Haghshenas:2018_J1J2mod,Yu:2018_J1J2mod,Wang:2018_J1J2mod,Liu:2018_J1J2mod}, starting with its proposed relationship to the disappearance of antiferromagnetic (AFM) long-range order (LRO) in the high-$T_{c}$ cuprate superconductors.  The conjecture here was that frustrated AFM exchange couplings might lead to a quantum spin liquid (QSL) state in which preformed pairs, or resonating valence bonds, could become superconducting upon doping \cite{Anderson:1987_QSL,Lee:2006_QSL}.  More recently, as frustrated quantum magnets have emerged as an active research field in their own right, the model has become recognized as one of the most challenging quantum spin-lattice systems.  Accordingly, it has been widely studied \cite{Chandra:1988,Dagotto:1989,Gelfand:1989,Sachdev:1990,Chubukov:1991,Read:1991,Richter:1993,Richter:1994,Ivanov:1994_J1J2mod,Schulz:1996,Oitmaa:1996,Zhitomirsky:1996,Trumper:1997_J1J2mod,Bishop:1998_J1J2mod,Singh:1999,Kotov:1999,Capriotti:2000,Capriotti:2001,Takano:2003,Roscilde:2004,Lante:2006_J1J2mod_sqLatt,Sirker:2006,Schm:2006_stackSqLatt,Mambrini:2006,Bi:2008_JPCM_J1J1primeJ2,Bi:2008_PRB_J1xxzJ2xxz,Darradi:2008_J1J2mod,Isaev:2009_J1J2mod,Murg:2009_peps,Ralko:2009,Richter:2010_ED40,Reuther:2010_J1J2mod,Reuther:2011_J1J2J3mod,Yu:2012,Gotze:2012,Jiang:2012,Mezzacapo:2012,LiT:2012,Wang:2013,Zhang:2013:J1J2SqLatt,Hu:2013,Gong:2014_J1J2mod_sqLatt,Doretto:2014_J1J2mod_sqLatt,Qi:2014_J1J2SqLatt,Metavitsiadis_Eggert:2014_J1J2mod_sqLatt,Ren:2014_J1J2SqLatt,Wang:2014_J1J2SqLatt,Chou:2014_J1J2SqLatt,Morita:2015_J1J2SqLatt,Richter:2015_ccm_J1J2sq_spinGap,Wang:2016_J1J2mod,Poilblanc:2017_J1J2mod,Haghshenas:2018_J1J2mod,Yu:2018_J1J2mod,Wang:2018_J1J2mod,Liu:2018_J1J2mod}, by a large number of theoretical techniques, as a prototypical system in which to examine quantum phase transition (QPTs) between quasiclassical ground-state (GS) phases with magnetic LRO and magnetically disordered (paramagnetic) quantum phases that are driven by frustration.

In addition to this extensive theoretical interest, it is also worth noting that several good experimental realizations of spin-$\frac{1}{2}$ $J_{1}$--$J_{2}$ models on a quasi-two-dimensional square lattice exist with $J_{1}>0$ and $J_{2}>0$.  Examples include the vanadium-layered oxide materials Li$_{2}$VO(Si,Ge)O$_{4}$ \cite{Melzi:2000_sqLatt_J1J2mod_merge,*Melzi:2001_sqLatt_J1J2mod_merge,*Rosner:2003_sqLatt_J1J2mod_merge} and the $B$-site ordered double-perovskite oxides Ba$_{2}$CuWO$_{6}$ \cite{Todate:2007_sqLatt_J1J2mod}, Sr$_{2}$CuMoO$_{6}$ \cite{Vasala:2014_sqLatt_J1J2mod}, Sr$_{2}$CuWO$_{6}$ \cite{Vasala:2014_sqLatt_J1J2mod,Vasala:2014_sqLatt_J1J2mod_JPCM26}, and Sr$_{2}$CuTeO$_{6}$ \cite{Koga:2016_sqLatt_J1J2mod}.

Despite the intense interest in this model from both theorists and
experimentalists, as outlined above, the nature of its GS phase around
the value $\kappa=\frac{1}{2}$ of the frustration parameter,
$\kappa\equiv J_{2}/J_{1}$, which represents the point of maximum
frustration in the classical version of the model, still remains
largely unresolved.  Thus, if the spins on the square-lattice sites
carry spin quantum number $s$, the model becomes classical in the
limit $s \to \infty$.  In this classical limit the GS phase is simple
in the two limiting cases $\kappa=0$ and $\kappa \to \infty$.
Clearly, when $\kappa=0$, the model has N\'{e}el AFM order [i.e., with
a magnetic wave vector $\mathbf{Q}=(\pi,\pi)$].  For nonzero values of $\kappa$ the N\'{e}el-ordered state has a GS energy per
spin given by $E_{{\rm cl}}/N=2J_{1}(-1+\kappa)s^{2}$.  By contrast,
when $\kappa \to \infty$, the classical ordering is such that on each
of the two equivalent interpenetrating sublattices (i.e., which
comprise NNN sites on the original lattice connected by $J_{2}$ bonds)
the spins are separately N\'{e}el-ordered, and with a relative angle
$\theta$ between the ordering directions on the two sublattices.  The
GS energy per spin in this case is given by
$E_{{\rm cl}}/N=-2J_{1}\kappa s^{2}$, {\it independent} of $\theta$, for any value of $\kappa$.

Clearly, the classical $J_{1}$--$J_{2}$ model on the square lattice
thus has a first-order phase transition at
$\kappa=\kappa_{{\rm cl}}=\frac{1}{2}$ between two AFM states, viz.,
the N\'{e}el state for $\kappa<\frac{1}{2}$ and an infinitely
degenerate family of ground states specified by the relative angle
$\theta$ between the ordering directions on the two interpenetrating
sublattices, for $\kappa>\frac{1}{2}$.  Thus, for
$\kappa>\frac{1}{2}$, the classical GS manifold has
$\text{SU(2)} \times \text{SU(2)}$ symmetry, which is larger than the $\text{SU(2)}$
symmetry of the Hamiltonian.  In this latter case, although the
effects of the exchange fields ($J_{1}$) between the two sublattices
cancel out, the zero-point quantum fluctuations, as well as the thermal
fluctuations, {\it will} depend on the angle $\theta$ between the two
sublattice spin orientations.  This leads to a prototypical example
\cite{Chandra:1990_ordByDisord_experimental} of the phenomenon of {\it
  order by disorder}
\cite{Villain:1977_ordByDisord_merge,*Villain:1980_ordByDisord_merge,Shender:1982_ordByDisord},
whereby the GS degeneracy is lifted by quantum fluctuations with the
angle $\theta$ now selected to be 0 or $\pi$.  The AFM GS ordering is
now collinear, and the corresponding GS phase is a striped one
consisting of successive alternating columns (or rows) of parallel
spins [i.e., with a magnetic wave vector $\mathbf{Q}=(\pi,0)$ or
$\mathbf{Q}=(0,\pi)$, respectively].  The GS symmetry is thereby
reduced from $\text{SU(2)} \times \text{SU(2)}$ to $\text{SU(2)} \times \text{Z}_{2}$, and the collinear striped state breaks the invariance of the Heisenberg
Hamiltonian under both spin rotations [$\text{SU(2)}$] and rotations by
90$^{\circ}$ of the square lattice [$\text{Z}_{2}$].

In the classical, $s \to \infty$, limit of the model, lowest-order
spin-wave theory, wherein the effects of quantum fluctuations are
taken into account perturbatively at $O(s^{-1})$, thus shows
\cite{Chandra:1988} that the critical coupling
$\kappa_{{\rm cl}}=\frac{1}{2}$ marks a first-order transition between
the N\'{e}el and striped collinear AFM phases.  In the extreme quantum
case, $s=\frac{1}{2}$, in which we are interested here, where quantum
fluctuations now have to be taken fully into account beyond
perturbation theory, it may be anticipated that these two
quasiclassical AFM phases persist, but are now separated by one or
more intermediate paramagnetic phases with no classical counterparts
(i.e., without magnetic LRO).  While there is essentially complete
consensus that this scenario is realized in the spin-$\frac{1}{2}$
$J_{1}$--$J_{2}$ model on the square lattice, the nature of both the
phase (or phases) in the intermediate regime and their associated
QPTs, as well as the precise critical values of $\kappa$ at which the
latter occur, are still not completely resolved, despite many
calculations over the last thirty or so years.  These have included
investigations of the model using a wide diversity of modern
theoretical techniques and numerical tools of ever increasing
sophistication.  Examples include those based on mean-field theories
of various (e.g., cluster, hierarchical) types
\cite{Gelfand:1989,Isaev:2009_J1J2mod,Reuther:2010_J1J2mod,Ren:2014_J1J2SqLatt},
the exact diagonalization (ED) of finite-sized clusters
\cite{Dagotto:1989,Ivanov:1994_J1J2mod,Schulz:1996,Capriotti:2000,Roscilde:2004,Mambrini:2006,Richter:2010_ED40,Gotze:2012},
linked-cluster series expansions
\cite{Gelfand:1989,Oitmaa:1996,Singh:1999,Sirker:2006,Reuther:2011_J1J2J3mod},
the bond-operator formalism
\cite{Sachdev:1990,Zhitomirsky:1996,Doretto:2014_J1J2mod_sqLatt},
resonating valence bond (RVB) approaches
\cite{Capriotti:2001,LiT:2012,Wang:2013,Zhang:2013:J1J2SqLatt,Hu:2013,Wang:2014_J1J2SqLatt,Chou:2014_J1J2SqLatt},
variational Monte Carlo (VMC) approaches based on various families of trial
GS wave functions (e.g., RVB states, entangled plaquette states)
\cite{Capriotti:2001,Mezzacapo:2012,LiT:2012,Wang:2013,Zhang:2013:J1J2SqLatt,Hu:2013,Wang:2014_J1J2SqLatt,Chou:2014_J1J2SqLatt,Morita:2015_J1J2SqLatt,Yu:2018_J1J2mod},
various quantum field-theoretical approaches
\cite{Takano:2003,Lante:2006_J1J2mod_sqLatt,Ralko:2009,Chandra:1990_ordByDisord_experimental}
including the dynamic functional renormalization group
\cite{Reuther:2010_J1J2mod,Reuther:2011_J1J2J3mod}, the density-matrix
renormalization group (DMRG)
\cite{Jiang:2012,Gong:2014_J1J2mod_sqLatt,Wang:2018_J1J2mod},
matrix-product or tensor-network approaches
\cite{Murg:2009_peps,Yu:2012,Wang:2013,Wang:2016_J1J2mod,Poilblanc:2017_J1J2mod,Haghshenas:2018_J1J2mod,Liu:2018_J1J2mod},
and the coupled cluster method (CCM)
\cite{Schm:2006_stackSqLatt,Bi:2008_JPCM_J1J1primeJ2,Bi:2008_PRB_J1xxzJ2xxz,Darradi:2008_J1J2mod,Reuther:2011_J1J2J3mod,Gotze:2012,Richter:2015_ccm_J1J2sq_spinGap}.

While, in the intermediate region, where the GS phase or phases are
non-magnetic, the $\text{SU(2)}$ spin symmetry is not broken, various
symmetries of the lattice still may or may not be broken.  In the
former case one can have various valence-bond crystalline (VBC) phases
where the lattice symmetries are broken by the formation of some
pattern of spin singlets.  Examples include the columnar dimer VBC
phase, which breaks both translational and rotational lattice
symmetries, and the plaquette VBC phase, which breaks only the
translational symmetry.  Alternatively, one could have a QSL phase
that conserves {\it all} lattice symmetries.  Such a QSL phase could
be either gapped or gapless (e.g., of the $\text{Z}_{2}$ type).

Each of these phases has been proposed to form the stable GS in part
of all of the paramagnetic intermediate regime of the
spin-$\frac{1}{2}$ $J_{1}$--$J_{2}$ model on the square lattice by
various of the above-cited references, with no overall consensus
having yet emerged.  Part of the reason for this uncertainty
undoubtedly must lie in the fact that of the various methods discussed
above that have high potential accuracy and/or are capable of
systematic improvement via some well-defined hierarchical
approximation scheme, almost all are either intrinsically biased in
favor of some particular GS phase and/or are not directly performed in
the thermodynamic (infinite-lattice) limit of interest.  In the latter
regard, for example, the great majority of the techniques employed are
performed on lattices of a finite size ($N$ spins), and some form of
finite-size scaling is then used to extrapolate to the thermodynamic
($N \to \infty$) limit.

As has been very rigorously and authoritatively demonstrated in a
recent study \cite{Sandvik:2012_SqLatt_J-Q_model} of the
spin-$\frac{1}{2}$ $J$--$Q$ model on the square lattice, for which the
infamous quantum Monte Carlo (QMC) minus-sign problem is absent, and
hence where large-scale QMC calculations can be undertaken, by
contrast with the corresponding $J_{1}$--$J_{2}$ model of interest
here, such extrapolations to the thermodynamic limit can have great
uncertainties.  This is specially true in cases where is little or no
analytic guidance from theoretical considerations, as is often the
case, but can also even hold when such guidance is present.  In this
context it is particularly noteworthy that the CCM
\cite{Coester:1958_ccm,Coester:1960_ccm,Cizek:1966_ccm,Kummel:1978_ccm,Bishop:1978_ccm,Bishop:1982_ccm,Arponen:1983_ccm,Bishop:1987_ccm,Arponen:1987_ccm,Arponen:1987_ccm_2,Bartlett:1989_ccm,Arponen:1991_ccm,Bishop:1991_TheorChimActa_QMBT,Bishop:1998_QMBT_coll,Zeng:1998_SqLatt_TrianLatt,Fa:2004_QM-coll,Bartlett:2007_ccm,Bishop:2014_honey_XXZ_nmp14}
provides a rather singular example of a theoretical quantum many-body
technique that can and does study arbitrary spin-lattice models
directly in the thermodynamic limit.  It is precisely for that reason
that we employ it here.

Furthermore, in view of the still puzzling nature of the phase or phases present in the intermediate paramagnetic regime of the spin-$\frac{1}{2}$ $J_{1}$--$J_{2}$ model on the square lattice around the value $\kappa=\frac{1}{2}$ of the frustration parameter, it is also potentially useful to examine a larger class of systems for which this model reduces to a special case.  Thus, we are strongly motivated to consider the corresponding spin-$\frac{1}{2}$ $J_{1}$--$J_{2}$--$J_{1}^{\perp}$ model on a square-lattice bilayer.  Each of the two monolayers is just a frustrated $J_{1}$--$J_{2}$ system, but the two layers are now connected by Heisenberg exchange bonds of strength $J_{1}^{\perp}\equiv \delta J_{1}$ between NN interlayer pairs of spins, with the two layer arranged in $AA$ stacking [i.e., with each site of one (horizontal) monolayer placed immediately above its counterpart on the other monolayer].  The original $J_{1}$--$J_{2}$ model is then just the special case $\delta=0$ of the larger $J_{1}$--$J_{2}$--$J_{1}^{\perp}$ model.  In this paper we use the CCM to study the spin-$\frac{1}{2}$ $J_{1}$--$J_{2}$--$J_{1}^{\perp}$ model on a square-lattice bilayer, for both signs of the interlayer coupling parameter $\delta$.  In particular, we will concentrate our efforts on examining the complete phase boundaries of the two quasiclassical collinear (AFM) phases (viz., the phases with N\'{e}el and striped AFM order on each of the coupled monolayers) in the $\kappa$--$\delta$ half-plane with $\kappa > 0$ (and $J_{1}>0$), and specifically in the window $0 \leq \kappa \leq 1$ that contains the intermediate paramagnetic regime in the case $\delta=0$.

In this context it is interesting to note too that the spin-$\frac{1}{2}$  $J_{1}$--$J_{2}$--$J_{1}^{\perp}$ model has also been previously studied on a stacked square lattice (i.e., where the number of layers $n \to \infty$, rather than the case $n=2$ studied here) \cite{Schm:2006_stackSqLatt}, where use was also made of the CCM.  Thus, the bilayer model we study here lies, in some sense, between the strictly two-dimensional square-lattice $J_{1}$--$J_{2}$ model (i.e., where $\delta=0$) and the strictly three-dimensional $J_{1}$--$J_{2}$--$J_{1}^{\perp}$ model on the stacked square lattice with an infinite number of layers.  Furthermore, unlike the latter case, the bilayer case also exhibits the additional physical phenomenon of dimerization between NN interlayer pairs, as discussed more fully in Sec.\ \ref{model_sec}.  These features thus provide considerable additional motivation to study the bilayer model.

The plan for the remainder of this paper is as follows.  The $J_{1}$--$J_{2}$--$J_{1}^{\perp}$ model is itself first described in Sec.\ \ref{model_sec}, where we also discuss more fully the main features of the limiting case, $J_{1}^{\perp}=0$, of the monolayer model.  We also give there some discussion of what we might expect to be some of the main features of the phase boundaries of the two quasiclassical AFM phases as the interlayer coupling parameter, $\delta\equiv J_{1}^{\perp}/J_{1}$, is introduced.  The main features of the CCM as applied to quantum spin-lattice problems are then reviewed in Sec.\ \ref{ccm_sec} before our numerical results are presented in Sec.\ \ref{results_sec}.  Finally, our findings are summarized and discussed in Sec.\ \ref{discuss_summary_sec}, where we also make comparisons with the results of others. 

\section{THE MODEL}
\label{model_sec}
The Hamiltonian of the $J_{1}$--$J_{2}$--$J_{1}^{\perp}$ model on a square-lattice
bilayer is specified as
\begin{equation}
\begin{aligned}
H & =  J_{1}\sum_{{\langle i,j \rangle},\alpha} \mathbf{s}_{i,\alpha}\cdot\mathbf{s}_{j,\alpha} + 
J_{2}\sum_{{\langle\langle i,k \rangle\rangle},\alpha} \mathbf{s}_{i,\alpha}\cdot\mathbf{s}_{k,\alpha}  \\
& \quad + J_{1}^{\perp}\sum_{i} \mathbf{s}_{i,1}\cdot\mathbf{s}_{i,2} \\
& \equiv  J_{1}h(\kappa,\delta)\,; \quad  \kappa \equiv J_{2}/J_{1}\,, \quad \delta \equiv J_{1}^{\perp}/J_{1}\,,
\label{H_eq}
\end{aligned}
\end{equation}
such that the sites on each (horizontal) monolayer are labelled by the index $i$ (i.e., with the two layers in $AA$ stacking such that sites $i$ on the top layer lie vertically above those on the bottom layer), and the two layers are labelled by the index $\alpha = 1,2$.  Every site $(i,\alpha)$ is occupied by a spin-$s$ particle described in terms of the usual $\text{SU(2)}$ operators ${\bf
  s}_{i,\alpha}\equiv(s^{x}_{i,\alpha},s^{y}_{i,\alpha},s^{z}_{i,\alpha})$,
with ${\bf s}^{2}_{i,\alpha} = s(s+1)\mathbb{1}$, and where we restrict discussion here to the case $s=\frac{1}{2}$.  The first two sums over $\langle i,j \rangle$ and
$\langle \langle i,k \rangle \rangle$ in Eq.\ (\ref{H_eq}) run over all NN and
NNN intralayer pairs of spins, respectively, with each Heisenberg bond (with respective strengths $J_{1}$ and $J_{2}$) counted once and once only.  The third sum in Eq.\ (\ref{H_eq}) over the index $i$ counts all corresponding 
interlayer NN Heisenberg bonds of strength $J_{1}^{\perp}$.  We shall be interested here in the case when both intralayer bonds are AFM in nature (i.e., $J_{1}>0$ and $J_{2}\equiv \kappa J_{1}>0$), such that frustration is present in each monolayer, but where the interlayer coupling parameter, $J_{1}^{\perp} \equiv \delta J_{1}$, may be either AFM ($\delta>0$) or ferromagnetic (FM) ($\delta<0$) in nature.  Since the parameter $J_{1}$ merely sets the overall energy scale, the Hamiltonian may be expressed as in the last line of Eq.\ (\ref{H_eq}), such that the relevant parameters of the model are $\kappa$ and $\delta$.  

\begin{figure*}[t]
\begin{center}
\mbox{
\subfigure[]{\includegraphics[width=5.2cm]{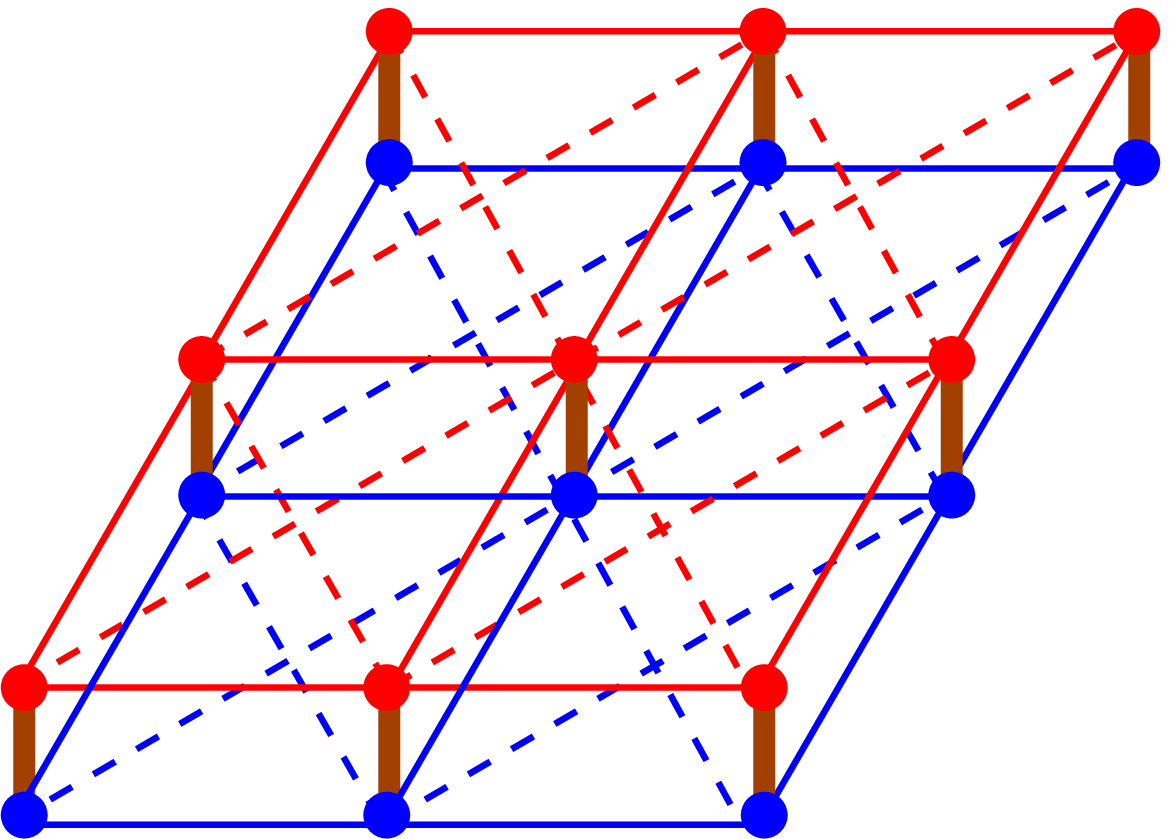}}
\hspace{0.9cm} \subfigure[]{\includegraphics[width=3.5cm]{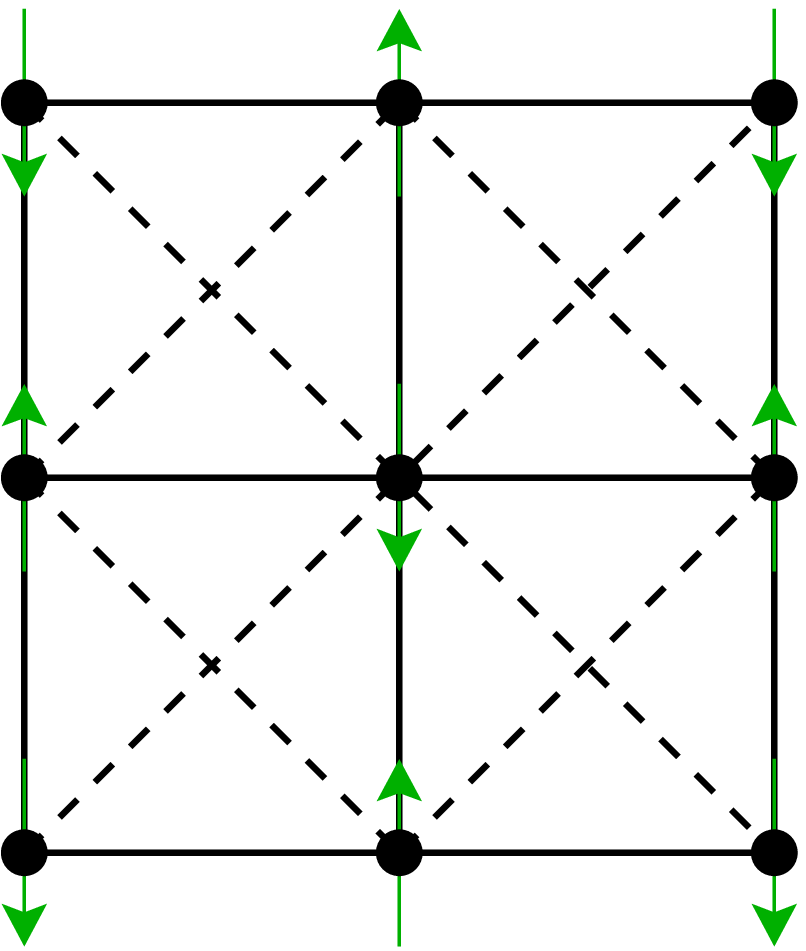}}
\hspace{0.9cm} \subfigure[]{\includegraphics[width=3.5cm]{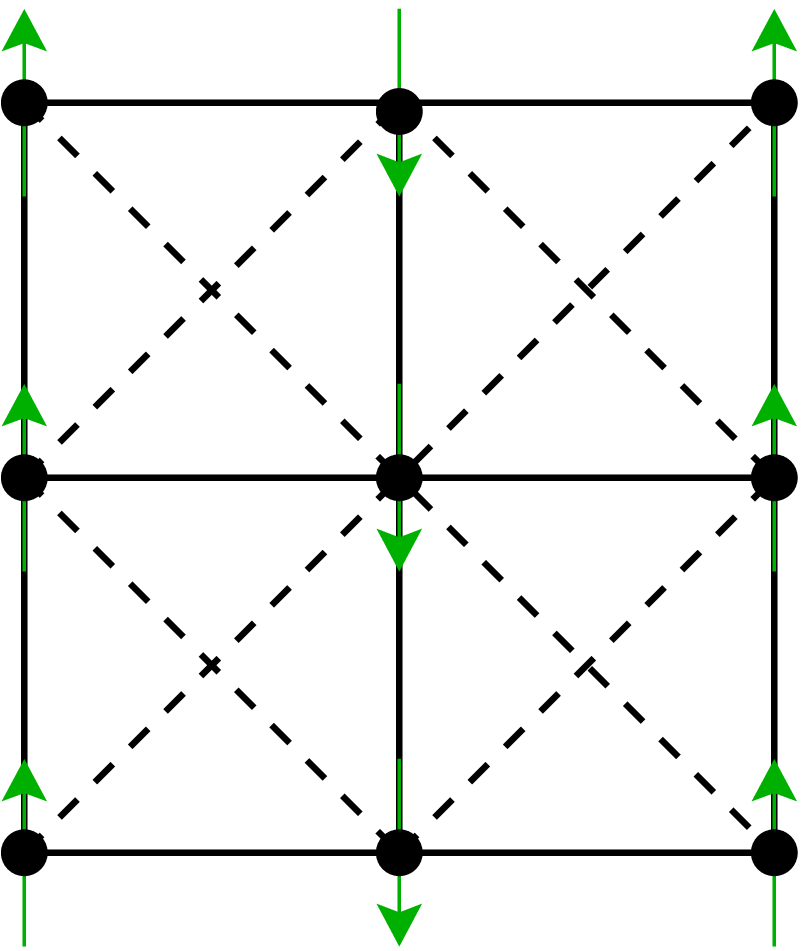}}
}
  \caption{The $J_{1}$--$J_{2}$--$J_{1}^{\perp}$ model on the square-lattice bilayer, showing (a) the two layers $1$ (red) and $2$ (blue), the nearest-neighbor intralayer $J_{1}$ bonds as thin (red or blue) solid lines; the nearest-neighbor interlayer $J_{1}^{\perp}$ bonds as thick (brown) solid lines, and the next-nearest-neighbor intralayer $J_{2}$ bonds as dashed (red or blue) lines; (b) the monolayer N\'{e}el state; and (c) the monolayer columnar striped state.  Lattice sites are shown by filled circles (\CIRCLE) and spins by the (green) arrows.}
\label{model_pattern}
\end{center}
\end{figure*}

Our main interest here will thus be to investigate the regions of stability of the two collinear AFM phases in each monolayer (i.e., the quasiclassical N\'{e}el and striped phases) in the $\kappa$--$\delta$ half-plane with $\kappa>0$, as the interlayer coupling, $\delta$, is turned on.  The square-lattice bilayer is illustrated in Fig.\ \ref{model_pattern}(a), while the patterns of spins of the two quasiclassical AFM phases on each monolayer are shown in Fig.\ \ref{model_pattern}(b) and \ref{model_pattern}(c), respectively.  
%%%%%%%%%%%%%%%%
%%%%%%%%%%%%%%%%%%
For both AFM phases the original square lattice (with spacing $d$) is decomposed into two equivalent sublattices.  For the N\'{e}el state each sublattice is itself square (i.e., with spacing $\sqrt{2}d\times \sqrt{2}d$), such that each site on one sublattice has its 4 NN sites on the original lattice on the other sublattice.  By contrast, for the striped states, the original lattice is decomposed into equivalent sublattices chosen either as alternating columns (each with spacing $2d \times d$) or as alternating rows (each with spacing $d \times 2d$).  Each of the corresponding classical AFM phases then has its spins on one sublattice pointing in a given, arbitrary (say, down) direction, and those on the other sublattice pointing in the opposite (say, up) direction.  The classical N\'{e}el state is thus as shown in Fig.\ \ref{model_pattern}(b), while the classical columnar striped state is as shown in Fig.\ \ref{model_pattern}(c).

For the case of the spin-$\frac{1}{2}$ square-lattice monolayer ($\delta=0$) there is essentially complete agreement that N\'{e}el order persists for $\kappa < \kappa_{c_{1}}$, striped order persists for $\kappa > \kappa_{c_{2}}$, and some paramagnetic phase (or phases) form the stable GS phase in the intermediate regime $\kappa_{c_{1}} < \kappa < \kappa_{c_{2}}$.  While the nature of the phase(s) in the intermediate regime remains unresolved, as noted in Sec.\ \ref{introd_sec}, modern high-quality calculations do seem to be converging on values for the two critical points of $\kappa_{c_{1}} \approx 0.43(3)$ and $\kappa_{c_{2}} \approx 0.605(15).$

Thus, for example, three recent independent DMRG calculations yielded the values $\kappa_{c_{1}} \approx 0.41$, $\kappa_{c_{2}} \approx 0.62$ \cite{Jiang:2012}, $\kappa_{c_{1}} \approx 0.44$, $\kappa_{c_{2}} \approx 0.61$ \cite{Gong:2014_J1J2mod_sqLatt}, and $\kappa_{c_{1}} \approx 0.46$, $\kappa_{c_{2}} \approx 0.62$ \cite{Wang:2018_J1J2mod}, while two high-order CCM calculations yielded the values $\kappa_{c_{1}} \approx 0.447$, $\kappa_{c_{2}} \approx 0.586$ \cite{Gotze:2012} and $\kappa_{c_{1}} \approx 0.454$, $\kappa_{c_{2}} \approx 0.588$ \cite{Richter:2015_ccm_J1J2sq_spinGap}.  Similar results have also been found, for example, from a plaquette-renormalized tensor-network study \cite{Yu:2012} that gave values $\kappa_{c_{1}} \approx 0.40$, $\kappa_{c_{2}} \approx 0.62$; a renormalization group (RG) approach \cite{Metavitsiadis_Eggert:2014_J1J2mod_sqLatt}, in which the RG flows were numerically integrated, that gave values $\kappa_{c_{1}} \approx 0.416$, $\kappa_{c_{2}} \approx 0.606$; a cluster mean-field theory approach \cite{Ren:2014_J1J2SqLatt} that gave values $\kappa_{c_{1}} \approx 0.42$, $\kappa_{c_{2}} \approx 0.59$; a VMC calculation using an AFM fermionic RVB class of trial wave functions \cite{Chou:2014_J1J2SqLatt} that gave values $\kappa_{c_{1}} \approx 0.45$, $\kappa_{c_{2}} \approx 0.6$; and a separate many-variable VMC calculation combined with quantum-number projections \cite{Morita:2015_J1J2SqLatt} that gave values $\kappa_{c_{1}} \approx 0.4$, $\kappa_{c_{2}} \approx 0.6$.  

We should note, however, that while there is broad agreement on the value for $\kappa_{c_{2}}$, there are still outlier calculations for $\kappa_{c_{1}}$.  For example, a bond-operator formalism approach that included cubic and quartic interactions beyond the harmonic approximation \cite{Doretto:2014_J1J2mod_sqLatt} yielded a lower value of $\kappa_{c_{1}} \approx 0.34$ (and  $\kappa_{c_{2}} \approx 0.59$), while a recent approach using the cluster update algorithm for tensor product states \cite{Wang:2016_J1J2mod} yielded the much higher value of $\kappa_{c_{1}} \approx 0.572$.  In this context it is interesting to note too that a large-scale ED calculation using finite-size scaling to the thermodynamic limit ($N \to \infty$) on finite square lattices of up to $N=40$ sites \cite{Richter:2010_ED40} yielded values $\kappa_{c_{1}} \approx 0.35$, $\kappa_{c_{2}} \approx 0.66$ based on the points where the N\'{e}el and striped order parameters vanish, respectively, but also gave values $\kappa_{c_{1}} \approx 0.46$, $\kappa_{c_{2}} \approx 0.60$ based on points where the respective zero-field transverse (uniform) magnetic susceptibility vanishes.  The latter estimates are clearly in much better agreement with the modern consensual values.  On the other hand we should note that the vanishing of the magnetic susceptibility only denotes the opening up of a new {\it gapped} phase.  Any non-magnetic  gapless state (e.g., of the QSL variety) would not, of course, be seen by calculations of the susceptibility alone.

Turning our attention now to the bilayer, it is clear that the interlayer $J_{1}^{\perp}$ bonds have no additional frustrating effect on the intralayer magnetic LRO.  Indeed, in the classical limit ($s \to \infty$) they have zero effect.  However, for finite spin quantum numbers $s$, if we consider first the case of zero frustration ($\kappa = 0$), the $J_{1}$ and $J_{1}^{\perp}$ bonds do still compete with one another since the $J_{1}^{\perp}$ bonds by themselves promote the formation of NN interlayer dimers.  For the present spin-$\frac{1}{2}$ case when $J_{1}^{\perp}>0$ these are spin-singlet pairs, while for $J_{1}^{\perp}<0$ they are spin-triplet pairs.  Thus, even with zero frustration ($J_{2}=0$), the introduction of  AFM $J_{1}^{\perp}$ bonds induces a competition between a GS magnetic phase with N\'{e}el LRO and a nonclassical paramagnetic phase of the VBC kind, which is formed of interlayer dimers.  The resulting spin-$\frac{1}{2}$ $J_{1}$--$J_{1}^{\perp}$ model on a square-lattice bilayer has been studied previously \cite{Kazuo:1990_SqLatt_bilayer,Kazuo:1992_SqLatt_bilayer,Millis:1993_SqLatt_bilayer,Millis:1994_SqLatt_bilayer,Sandvik:1994_SqLatt_bilayer,Sandvik:1995_SqLatt_bilayer,Chubukov:1995_SqLatt_bilayer,Zheng:1997_SqLatt_bilayer,Shevchenko:1999_SqLatt_bilayer,Shevchenko:2000_SqLatt_bilayer,Wang:2006_SqLatt_bilayer,Collins:2008_SqLatt_bilayer,Fritz:2011_dimerized_AFM,Ganesh:2011_honey_bilayer_PRB84,Helmes:2014_honey_bilayer,Devakul:2014_honey_bilayer,Lohofer:2015_honey_bilayer}.  Since QMC calculations can be performed in this case (i.e., when $\kappa=0$), the position $\delta_{c_{1}}^{>}(\kappa=0)$ of the QPT between the N\'{e}el-ordered state and the quantum disordered interlayer-dimer VBC (IDVBC) state can be ascertained with high accuracy.  For example, a finite-size scaling of QMC results on lattices with $2L^{2}$ spins with $L \leq 10$ \cite{Sandvik:1994_SqLatt_bilayer} gave a value $\delta_{c_{1}}^{>}(\kappa=0)=2.51(2)$, while a more recent QMC calculation of Wang {\it et al}. \cite{Wang:2006_SqLatt_bilayer} using the improved stochastic series-expansion algorithm with operator-loop updates and finite-size scaling on $L\times L\times 2$ lattices with $L\leq 42$ gave the very precise value $\delta_{c_{1}}^{>}(\kappa=0)=2.5220(1)$.  An exponent-biased SE analysis of Zheng \cite{Zheng:1997_SqLatt_bilayer} gave the comparable result $\delta_{c_{1}}^{>}(\kappa=0)=2.537(5)$.

We turn now finally to the case of interest here where we also introduce intralayer frustration via the NNN AFM $J_{2}$ bonds.  The resulting spin-$\frac{1}{2}$ $J_{1}$--$J_{2}$--$J_{1}^{\perp}$ model on a square-lattice bilayer has received much less attention \cite{Kazuo:1996_SqLatt_bilayer_J1J2J1perp,Kazuo:1998_SqLatt_bilayer_J1J2J1perp} than either of the limiting cases $\delta=0$ \cite{Chandra:1988,Dagotto:1989,Gelfand:1989,Sachdev:1990,Chubukov:1991,Read:1991,Richter:1993,Richter:1994,Ivanov:1994_J1J2mod,Schulz:1996,Oitmaa:1996,Zhitomirsky:1996,Trumper:1997_J1J2mod,Bishop:1998_J1J2mod,Singh:1999,Kotov:1999,Capriotti:2000,Capriotti:2001,Takano:2003,Roscilde:2004,Lante:2006_J1J2mod_sqLatt,Sirker:2006,Schm:2006_stackSqLatt,Mambrini:2006,Bi:2008_JPCM_J1J1primeJ2,Bi:2008_PRB_J1xxzJ2xxz,Darradi:2008_J1J2mod,Isaev:2009_J1J2mod,Murg:2009_peps,Ralko:2009,Richter:2010_ED40,Reuther:2010_J1J2mod,Reuther:2011_J1J2J3mod,Yu:2012,Gotze:2012,Jiang:2012,Mezzacapo:2012,LiT:2012,Wang:2013,Zhang:2013:J1J2SqLatt,Hu:2013,Gong:2014_J1J2mod_sqLatt,Doretto:2014_J1J2mod_sqLatt,Qi:2014_J1J2SqLatt,Metavitsiadis_Eggert:2014_J1J2mod_sqLatt,Ren:2014_J1J2SqLatt,Wang:2014_J1J2SqLatt,Chou:2014_J1J2SqLatt,Morita:2015_J1J2SqLatt,Richter:2015_ccm_J1J2sq_spinGap,Wang:2016_J1J2mod,Poilblanc:2017_J1J2mod,Haghshenas:2018_J1J2mod,Yu:2018_J1J2mod,Wang:2018_J1J2mod,Liu:2018_J1J2mod} or $\kappa=0$ \cite{Kazuo:1990_SqLatt_bilayer,Kazuo:1992_SqLatt_bilayer,Millis:1993_SqLatt_bilayer,Millis:1994_SqLatt_bilayer,Sandvik:1994_SqLatt_bilayer,Sandvik:1995_SqLatt_bilayer,Chubukov:1995_SqLatt_bilayer,Zheng:1997_SqLatt_bilayer,Shevchenko:1999_SqLatt_bilayer,Shevchenko:2000_SqLatt_bilayer,Wang:2006_SqLatt_bilayer,Collins:2008_SqLatt_bilayer,Fritz:2011_dimerized_AFM,Ganesh:2011_honey_bilayer_PRB84,Helmes:2014_honey_bilayer,Devakul:2014_honey_bilayer,Lohofer:2015_honey_bilayer} discussed above.  We note, however, that a very recent paper \cite{Stapmanns:2018_SqLatt_bilayer_J1J1perpJ2perp} studied the case where frustration is introduced instead via an interlayer NNN AFM $J_{2}^{\perp}$ bond, resulting in a $J_{1}$--$J_{1}^{\perp}$--$J_{2}^{\perp}$ model, with very different properties and behavior (and see also Ref.\ \cite{Alet:2016_SqLatt_bilayer}).

Before presenting our results in Sec.\ \ref{results_sec} for the $J_{1}$--$J_{2}$--$J_{1}^{\perp}$ model it may be worthwhile to outline first our aims and expectations.  As the interlayer coupling parameter $\delta$ is turned on we anticipate that its initial effect at a fixed value of $\kappa$ at which magnetic order of either the N\'{e}el or striped sort exists, will be to enhance the stability of the corresponding quasiclassical state, since the effect of the $J_{1}^{\perp}$ bonds is to increase the number of NN bonds and hence to take a step towards three-dimensionality.  {\it A priori}, one can expect that this effect is roughly symmetric with respect to small positive and negative values of $\delta$.  Hence, we anticipate that on a $\kappa\delta$ plot the N\'{e}el and striped phase boundaries will both show a cusp at $\delta=0$.  Thus, we expect that $\kappa_{c_{1}}(\delta)$ will initially {\it increase} and $\kappa_{c_{2}}(\delta)$ will initially {\it decrease} as $\delta$ is either increased to small positive values or decreased to small negative values.  Accordingly, we are particularly interested in what then happens as $|\delta|$ is increased further in both the FM ($\delta<0$) and AFM ($\delta>0$) regimes of interlayer coupling.  The situation is expected to be very different in the two cases.

Thus, firstly, in the region where $\delta<0$ it is evident that in the limit $\delta \to -\infty$ the system will simply behave as a spin-1 $J_{1}$--$J_{2}$ model on a square-lattice monolayer.  Unlike the corresponding spin-$\frac{1}{2}$ case the spin-1 $J_{1}$--$J_{2}$ model on the square-lattice seems to show \cite{Bi:2008_EPL_J1J1primeJ2_s1,Bi:2008_JPCM_J1xxzJ2xxz_s1,Haghshenas:2018_SqLatt_J1J2mod_s1} a direct transition between the N\'{e}el and striped phases at a critical value $\kappa\approx 0.55$, although an early DMRG calculation\cite{Jiang:2009_SqLatt_J1J2mod_s1} indicated a disordered paramagnetic phase in the narrow region $0.525 \lesssim \kappa \lesssim 0.555$.  Interestingly, a later and very recent DMRG calculation \cite{Haghshenas:2018_SqLatt_J1J2mod_s1} using larger finite lattices showed that if such an intermediate region did exist it could do so only in the much smaller regime  $0.545 \lesssim \kappa \lesssim 0.550$.  While the system sizes in the DMRG calculations \cite{Haghshenas:2018_SqLatt_J1J2mod_s1,Jiang:2009_SqLatt_J1J2mod_s1} were too small for a critical analysis, both the CCM analysis \cite{Bi:2008_EPL_J1J1primeJ2_s1,Bi:2008_JPCM_J1xxzJ2xxz_s1} and an infinite projected entangled-pair state analysis \cite{Haghshenas:2018_SqLatt_J1J2mod_s1}, have shown that the direct transition between the N\'{e}el and striped phases for the spin-1 case is a first-order transition.  The best estimate for the critical coupling of the transition is $\kappa \approx 0.549$ \cite{Haghshenas:2018_SqLatt_J1J2mod_s1}.

Returning to our bilayer model, let us denote by $\delta_{c_{1}}^{{\mathrm F}}(\kappa)$ and $\delta_{c_{2}}^{{\mathrm F}}(\kappa)$ the critical values of $\delta$ (for a given value of $\kappa$) at which N\'{e}el order and striped order, respectively, melt in the regime of FM interlayer coupling ($\delta<0$).  Equivalently, these phase boundaries, $\delta=\delta_{c_{1}}^{{\mathrm F}}(\kappa)$ and $\delta=\delta_{c_{2}}^{{\mathrm F}}(\kappa)$, are also denoted, respectively as $\kappa=\kappa_{c_{1}}^{{\mathrm F}}(\delta)$ and $\kappa=\kappa_{c_{2}}^{{\mathrm F}}(\delta)$.  In the light of the above discussion it seems clear that in the half-plane $\delta<0$ there must exist a quantum triple point (QTP) that occurs at a value $\delta=\delta_{{\mathrm T}}^{{\mathrm F}}$ such that $\kappa_{c_{1}}^{{\mathrm F}}(\delta_{{\mathrm T}}^{{\mathrm F}})=\kappa_{c_{2}}^{{\mathrm F}}(\delta_{{\mathrm T}}^{{\mathrm F}})$ or, equivalently, when $\delta_{c_{1}}^{{\mathrm F}}(\kappa_{{\mathrm T}}^{{\mathrm F}})=\delta_{c_{2}}^{{\mathrm F}}(\kappa_{{\mathrm T}}^{{\mathrm F}})$.  Thus, if the position of this QTP is ($\kappa_{{\mathrm T}}^{{\mathrm F}},\delta_{{\mathrm T}}^{{\mathrm F}}$), then for all values $\delta<\delta_{{\mathrm T}}^{{\mathrm F}}$ there will be a direct transition between the N\'{e}el and striped phases at a value $\kappa^{{\mathrm F}}(\delta)$, where we expect $\lim_{\delta\to -\infty} \kappa^{{\mathrm F}}(\delta)\approx 0.549$.  One of our aims will be to evaluate accurately the position ($\kappa_{{\mathrm T}}^{{\mathrm F}},\delta_{{\mathrm T}}^{{\mathrm F}}$) of the QTP where the N\'{e}el, striped, and disordered paramagnetic phases meet in the half-plane $\delta<0$.  It seems almost certain that $\kappa_{{\mathrm T}}^{{\mathrm F}}$ will lie between the monolayer values $\kappa_{c_{1}}(\delta=0)\approx 0.43(3)$ and $\kappa_{c_{2}}(\delta=0)\approx 0.605(15)$.

The possible scenarios in the half-plane $\delta>0$ are even more interesting.  One possibility is the obvious analog to that discussed above, with another QTP between the N\'{e}el, striped, and intermediate paramagnetic phases.  However, in this scenario, such a QTP would presumably have to be accompanied by another QTP, at a larger value of $\delta$, now between the N\'{e}el and striped phases together with the gapped IDVBC state that we know must physically occur for large enough values of $\delta$ at any fixed value of $\kappa$.  Such a scenario (at least as far as the first QTP is concerned) was obtained in an earlier CCM calculation \cite{Schm:2006_stackSqLatt} of the spin-$\frac{1}{2}$ $J_{1}$--$J_{2}$--$J_{1}^{\perp}$ model on the stacked square lattice (i.e., the same model as considered here but with an infinite number of layers in which all NN interlayer pairs are connected via $J_{1}^{\perp}$ bonds).  In this case, of course, no IDVBC state occurs, and the N\'{e}el and striped AFM phases simply undergo a first-order transition for all values of the coupling parameter $\delta$ beyond the first (and only) QTP in this case. 

An alternative, perhaps more intriguing, scenario in the half-plane $\delta>0$ is one in which the boundaries of the two quasiclassical AFM phases turn back on themselves, in a reentrant fashion, sufficiently rapidly as $\delta$ is increased so that they avoid crossing.  One of our major aims here is to perform sufficiently accurate calculations as to be able to distinguish with confidence between such different scenarios.  Before we present our findings in Sec.\ \ref{results_sec}, however, we first briefly discuss in Sec.\ \ref{ccm_sec} the most important features of the CCM that we use to obtain them.

\section{THE COUPLED CLUSTER METHOD}
\label{ccm_sec}
The CCM \cite{Coester:1958_ccm,Coester:1960_ccm,Cizek:1966_ccm,Kummel:1978_ccm,Bishop:1978_ccm,Bishop:1982_ccm,Arponen:1983_ccm,Bishop:1987_ccm,Arponen:1987_ccm,Arponen:1987_ccm_2,Bartlett:1989_ccm,Arponen:1991_ccm,Bishop:1991_TheorChimActa_QMBT,Bishop:1998_QMBT_coll,Zeng:1998_SqLatt_TrianLatt,Fa:2004_QM-coll,Bartlett:2007_ccm,Bishop:2014_honey_XXZ_nmp14} is widely recognized as providing one of the most flexible, most widely utilized, and most accurate of all available {\it ab initio} techniques in modern microscopic quantum many-body theory.  One of the keys to its success is the fact that it preserves size-extensivity and size-consistency at every level of approximation, thereby enabling it to be implemented from the very outset in the thermodynamic ($N \to \infty$) limit.  Hence any errors associated with finite-size scaling, as needs to be performed in almost all competing methods, are obviated.  A second key to the success of the CCM lies in the fact that it also exactly preserves at all levels of approximation the very important Hellmann-Feynman theorem as well as the Goldstone linked-cluster theorem.  A third key to its success is that there exist well-defined, systematic, and very widely tested hierarchies of truncations within which the method can be computationally implemented to very high orders of approximation, as will be done here.  Since the CCM becomes exact within such a truncation hierarchy as the order $n$ of the approximation tends to infinity ($n \to \infty$), the only approximation ever made is in the extrapolation of such a sequence of approximants for any physical parameter calculated for the system under study.  The combination of these features ensures that the CCM yields accurate and self-consistent sets of results for all GS and excited-state (ES) quantities calculated.

Amongst many applications to quantum many-body problems in fields as diverse as nuclear physics, subnuclear physics, quantum chemistry, atomic and molecular physics, quantum optics, and condensed matter physics, the CCM has, in particular, by now been applied to a wide variety of spin-lattice systems of interest in quantum magnetism (see, e.g., Refs.\ \cite{Bishop:1998_J1J2mod,Schm:2006_stackSqLatt,Bi:2008_JPCM_J1J1primeJ2,Bi:2008_PRB_J1xxzJ2xxz,Darradi:2008_J1J2mod,Reuther:2011_J1J2J3mod,Gotze:2012,Richter:2015_ccm_J1J2sq_spinGap,Zeng:1998_SqLatt_TrianLatt,Fa:2004_QM-coll,Bishop:2014_honey_XXZ_nmp14,Bi:2008_EPL_J1J1primeJ2_s1,Bi:2008_JPCM_J1xxzJ2xxz_s1,Li:2019_honeycomb_bilayer_J1J2J1perp_neel-II} and references therein).  Since its application to such systems has already been widely described in the literature, therefore we content ourselves here with presenting a brief overview of only those features that are most relevant to us now.

The first step in any implementation of the CCM is to choose a suitable model (or reference) state $|\Phi\rangle$ for the $N$-body system with ($N \to \infty$) under consideration, together with a complete set of mutually commuting, multiconfigurational creation operators, $C^{+}_{I} \equiv (C_{I}^{-})^{\dagger}$.  The main requirement on $|\Phi\rangle$ is that it should be a cyclic vector (or, equivalently, a generalized vacuum state) with respect to the set of operators $\{C_{I}^{+}\}$.  The set-index $I$ here is used to indicate a complete labelling of the many-particle configuration created in the state $C_{I}^{+}|\Phi\rangle$.  We thus require the set $\{|\Phi\rangle; C_{I}^{+}\}$ to obey the conditions,
\begin{equation}
\sum_{I}C_{I}^{+}|\Phi\rangle\langle\Phi|C_{I}^{-}=\mathbb{1}\,,
\end{equation}

\begin{equation}
\langle\Phi|C_{I}^{+} = 0 = C_{I}^{-}|\Phi\rangle\,, \quad \forall I \neq 0\,; \quad C_{0}^{+}\equiv\mathbb{1}\,,
\end{equation}
%%%%%%%
\begin{equation}
[C_{I}^{+},C_{J}^{+}]=0=[C_{I}^{-},C_{J}^{-}]\,,  \label{creation-destruction_operators_commutation_relation}
\end{equation}
where $\mathbb{1}$ is the unit vector in the $N$-particle Hilbert space.  It is also convenient to choose the states $\{C_{I}^{+}|\Phi\rangle\}$ that so span the $N$-body Hilbert space to be an orthonormal set,
\begin{equation}
\langle\Phi|C_{I}^{-}C_{J}^{+}|\Phi\rangle=\delta_{I,J}\,,   \label{create_destruct_operators_orthonornmal_Eq}
\end{equation}
with $\delta_{I,J}$ a suitably generalized Kronecker symbol.

The exact many-body GS ket and bra states, $|\Psi\rangle$ and $\langle\tilde{\Psi}|$ ($=\langle\Psi|/\langle\Psi|\Psi\rangle$), respectively, which satisfy the respective GS Schr\"{o}dinger equations,
\begin{equation}
H|\Psi\rangle = E|\Psi\rangle\,; \quad \langle\tilde{\Psi}|H = E\langle\tilde{\Psi}|\,,   \label{schrondinger_eq}
\end{equation}
with $|\Psi\rangle$ now satisfying the intermediate normalization condition,
$\langle\Phi|\Psi\rangle = 1 = \langle\Phi|\Phi\rangle$, together with $\langle\tilde{\Psi}|\Psi\rangle = 1$, are now parametrized within the CCM with respect to the model state $|\Phi\rangle$ via the distinctive
exponentiated
forms of correlation operators,
\begin{equation}
|\Psi\rangle = {\mathrm e}^{S}|\Phi\rangle\,; \quad S=\sum_{I\neq 0}{\cal{S}}_{I}C_{I}^{+}\,,  \label{ket_parametrization_eq}
\end{equation}
\begin{equation}
\langle\tilde{\Psi}|=\langle\Phi|\tilde{S}{\mathrm e}^{-S}\,;  \quad \tilde{S}=1 + \sum_{I \neq 0} {\cal{\tilde{S}}}_{I}C_{I}^{-}\,,  \label{bra_parametrization_eq}
\end{equation}
that are one of the distinguishing features of the method.  Although 
Hermiticity implies that the destruction correlation operator $\tilde{S}$ is formally related to its creation counterpart $S$ via the relation
\begin{equation}
\langle\Phi|\tilde{S} = \frac{\langle\Phi|{\mathrm e}^{S^{\dagger}}{\mathrm e}^{S}}{\langle\Phi|{\mathrm e}^{S^{\dagger}}{\mathrm e}^{S}|\Phi\rangle}\,,  \label{Destruct_operator_hermiticity_Eq}
\end{equation}
the CCM treats $S$ and $\tilde{S}$ as independent operators.  They will clearly satisfy Eq.\ (\ref{Destruct_operator_hermiticity_Eq}) when no approximations are made, but may violate it when truncations are made in the sums over the index $I$ in Eqs.\ (\ref{ket_parametrization_eq}) and (\ref{bra_parametrization_eq}), as described below in practical implementations.  The compensation paid for this loss of explicit Hermiticity is the huge advantage that in all such truncations the Hellmann-Feynman theorem {\it is} now manifestly maintained.

All GS properties of the system may thus be calculated in terms of the set of real $c$-number correlation coefficients $\{{\cal S}_{I},\tilde{{\cal S}}_{I}\}$.  In turn, these may be found by insertion of the parametrizations of Eqs.\ (\ref{ket_parametrization_eq}) and (\ref{bra_parametrization_eq}) into the respective Schr\"{o}dinger equations (\ref{schrondinger_eq}), followed by projection onto the complete sets of states $\langle\Phi|C_{I}^{-}$ and $C_{I}^{+}|\Phi\rangle$, respectively.  As a completely equivalent alternative procedure we may derive $\{{\cal S}_{I}, \tilde{{\cal S}}_{I}\}$ by demanding that the GS energy expectation value functional, $\bar{H}=\bar{H}({\cal S}_{I},\tilde{{\cal S}}_{I})$, defined as
\begin{equation}
\bar{H}\equiv\langle\tilde{\Psi}|H|\Psi\rangle=
\langle\Phi|\tilde{S}{\mathrm e}^{-S}H{\mathrm e}^{S}|\Phi\rangle\,,  
\end{equation}
be an extremum with respect to the entire set of parameters $\{{\cal S}_{I},{\cal \tilde{S}}_{I}\}$.  By either method we may readily derive the sets of equations,
\begin{equation}
\langle\Phi|C^{-}_{I}{\rm e}^{-S}H{\rm e}^{S}|\Phi\rangle = 0\,, \quad \forall I \neq 0\,,  \label{non_linear_ket_Eq}
\end{equation}
%%%%%%%%%
\begin{equation}
\langle\Phi|\tilde{S}({\mathrm e}^{-S}[H,C^{+}_{I}]{\mathrm e}^{S}|\Phi\rangle=0\,, \quad \forall I \neq 0\,,   \label{linear_bra_Eq}
\end{equation}
%%%%%%%%%%
By using Eq.\ (\ref{non_linear_ket_Eq}) we may also show that the GS energy at the stationary point may be expressed purely in terms of the set of creation coefficients $\{{\cal S}_{I}\}$ as 
\begin{equation}
E=E({\cal S}_{I})=\langle\Phi|{\mathrm e}^{-S}H{\mathrm e}^{S}|\Phi\rangle\,.  \label{E_GS_Eq}
\end{equation}
Correspondingly, Eq.\ (\ref{linear_bra_Eq}) may be written in the equivalent form,
\begin{equation}
\langle\Phi|\tilde{S}({\mathrm e}^{-S}H{\mathrm e}^{S}-E)C^{+}_{I}|\Phi\rangle=0\,, \quad \forall I \neq 0\,.   \label{linear_bra_Eq_equivalentForm}
\end{equation}
By contrast, the GS expectation value, $\bar{A}\equiv\langle\tilde{\Psi}|A|\Psi\rangle$, of any other operator $A$ requires both sets of GS CCM coefficients for its evaluation,
\begin{equation}
\bar{A}=\bar{A}({\cal S}_{I},\tilde{{\cal S}}_{I})=
\langle\Phi|\tilde{S}{\mathrm e}^{-S}A{\mathrm e}^{S}|\Phi\rangle\,.
\end{equation}

We note that the characteristic CCM exponentiated operators ${\mathrm e}^{\pm S}$ only enter into Eqs.\ (\ref{non_linear_ket_Eq}) and (\ref{linear_bra_Eq}), which need to be solved for the GS coefficients $\{{\cal S}_{I}, \tilde{{\cal S}}_{I}\}$, in the form of the associated similarity transform of the Hamiltonian, ${\mathrm e}^{-{S}}H{\mathrm e}^{S}$.  In order to solve Eqs.\ (\ref{non_linear_ket_Eq}) and (\ref{linear_bra_Eq}) in practice we utilize the nested commutator expansion, 
\begin{equation}
{\mathrm e}^{-S}H{\mathrm e}^{S}=\sum_{n=0}^{\infty}\frac{1}{n!}[H,S]_{n}\,,  \label{H_similarity_transform_expansion_Eq}
\end{equation}
where the $n$-fold nested commutators $[H,S]_{n}$ are defined
iteratively  as
\begin{equation}
[H,S]_{n} \equiv [[H,S]_{n-1},S]\,; \quad [H,S]_{0}=H\,.
\end{equation}

Another key feature of the CCM parametrizations of Eqs.\ (\ref{ket_parametrization_eq}) and (\ref{bra_parametrization_eq}) is now that the otherwise infinite sum in Eq.\ (\ref{H_similarity_transform_expansion_Eq}) terminates in practice for all Hamiltonians that contain only finite-order multinomials in the appropriate single-particle operators, as in the present case.  The reason for this is simple, namely that all components in the expansion of Eq.\ (\ref{ket_parametrization_eq}) mutually commute by construction, as in Eq.\ (\ref{creation-destruction_operators_commutation_relation}).  For the present Hamiltonian of Eq.\ (\ref{H_eq}), which is bilinear in the basic one-body operators $(s^{x}_{i,\alpha},s^{y}_{i,\alpha},s^{z}_{i,\alpha})$, the sum in Eq.\ (\ref{H_similarity_transform_expansion_Eq}) will terminate with the term $n=2$ for the choices for $\{|\Phi\rangle;C_{I}^{+}\}$ that we describe below, due to the basic $\text{SU(2)}$ commutation relations.  Thus, all nested commutators with $n>2$ simply vanish identically.

For the same reason, all terms in the expansion of $\bar{H}$ are linked, and it is this fact that leads to the CCM satisfying the Goldstone linked-cluster theorem (and consequently being size-extensive) at all levels of truncation in the expansions of Eqs.\ (\ref{ket_parametrization_eq}) and (\ref{bra_parametrization_eq}).  Thus, in the solutions of Eqs.\ (\ref{non_linear_ket_Eq}) and (\ref{linear_bra_Eq}) the sole approximation made is in what set of configurations $\{I\}$ will be retained in the expansions of Eqs.\ (\ref{ket_parametrization_eq}) and (\ref{bra_parametrization_eq}), as described below.

We turn now to the choice of model state $|\Phi\rangle$ and the associated set of multiconfigurational creation operators $\{C_{I}^{+}\}$ for the present case.  In any spin-lattice application of the CCM a convenient (but not the only) choice for $|\Phi\rangle$ is always any quasiclassical state with perfect magnetic LRO, i.e., one for which the spin on every lattice site is specified independently via its given spin projection onto some specified spin quantization axis.  Here we will thus use both the N\'{e}el and striped AFM states as our independent CCM model states.  It is highly convenient to treat all such states in the same way, so that all sites may be treated equivalently.  A simple means of doing so is to choose a local spin quantization axis independently on each site (i.e., equivalently, by making a suitable passive rotation of each spin separately) so that in these local axes the reference state is a tensor product of spin-down states, $|\Phi\rangle =
|\downarrow\downarrow\downarrow\cdots\downarrow\rangle$, such that all spins point along the negative $z_{s}$ direction in these local sets of axes.  A beneficial effect of choosing such rotations is that all cases may henceforward be treated on an equal footing and by a universal computational code.  The cases are distinguished only by that the spin Hamiltonian needs to be rewritten in terms of the particular local axes needed for each specific model state.

Such passive rotations are unitary transformations that leave the basic $\text{SU}(2)$ algebra unchanged, but also have the other beneficial effect of allowing us to choose the operators $C_{I}^{+}$ as products of single-spin raising operators $s_{k,\alpha}^{+}\equiv s_{k,\alpha}^{x}+is_{k,\alpha}^{y}$.  Thus, we have that the set-index $I$ becomes a set of lattice indices, $\{I\} \to \{l_{1},l_{2},\cdots,l_{n}\,; n=1,2,\cdots,2sN\}$, where $l_{i}\equiv (k_{i},\alpha)$, and in which any given lattice site index $l_{i}$ may be repeated so that it appears no more than $2s$ times, where $s$ is the spin quantum number of the spins in the general case.  Thus, we have $C_{I}^{+} \to s_{l_{1}}^{+}s_{l_{2}}^{+}\cdots s_{l_{n}}^{+}$, with $n=1,2,\cdots,2sN$.  Here, we take $s=\frac{1}{2}$, so that in each configuration $I$ no lattice site may appear more than once.  

In these local rotated spin axes the order parameter (i.e., the sublattice magnetization) takes the universal form,
\begin{equation}
M = -\frac{1}{N}\sum_{i=1}^{N}\langle\Phi|\tilde{S}{\mathrm e}^{-S}s_{l_{i}}^{z}{\mathrm e}^{S}|\Phi\rangle\,.  \label{M_definition_eq}
\end{equation}
Once again, the expression ${\mathrm e}^{-S}s_{l_{i}}^{z}{\mathrm e}^{S}$ may be evaluated exactly via a nested commutator expansion akin to Eq.\ (\ref{H_similarity_transform_expansion_Eq}) that now terminates at the term with $n=1$.

As we have indicated above, the only approximation that we make in implementing the CCM is the truncation of the expansion of Eqs.\ (\ref{ket_parametrization_eq}) and (\ref{bra_parametrization_eq}) for the correlation operators $S$ and $\tilde{S}$.  We shall use here the well-established localised (lattice-animal-based subsystem) LSUB$n$ scheme wherein we retain at $n$th order all multispin-flip correlations on the lattice over no more than $n$ contiguous sites.  A set of sites is contiguous in this sense if every site in the set is NN to at least one other in the set, in some specified geometry.  As the truncation index $n$ tends to infinity $(n \to \infty)$, the corresponding LSUB$n$ approximation becomes exact.

\begin{figure*}[t]
\begin{center}
\mbox{
\subfigure[]{\includegraphics[width=8.5cm]{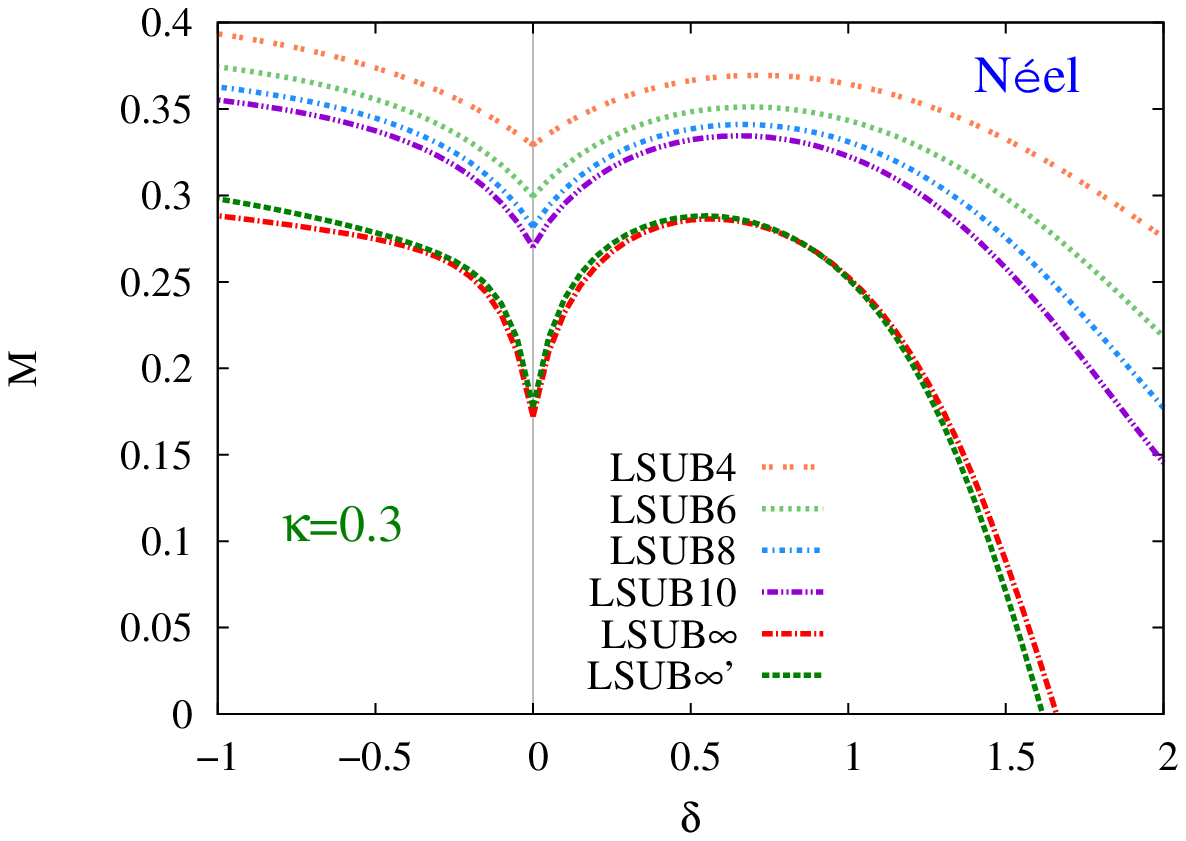}}
\hspace{0.1cm} \subfigure[]{\includegraphics[width=8.5cm]{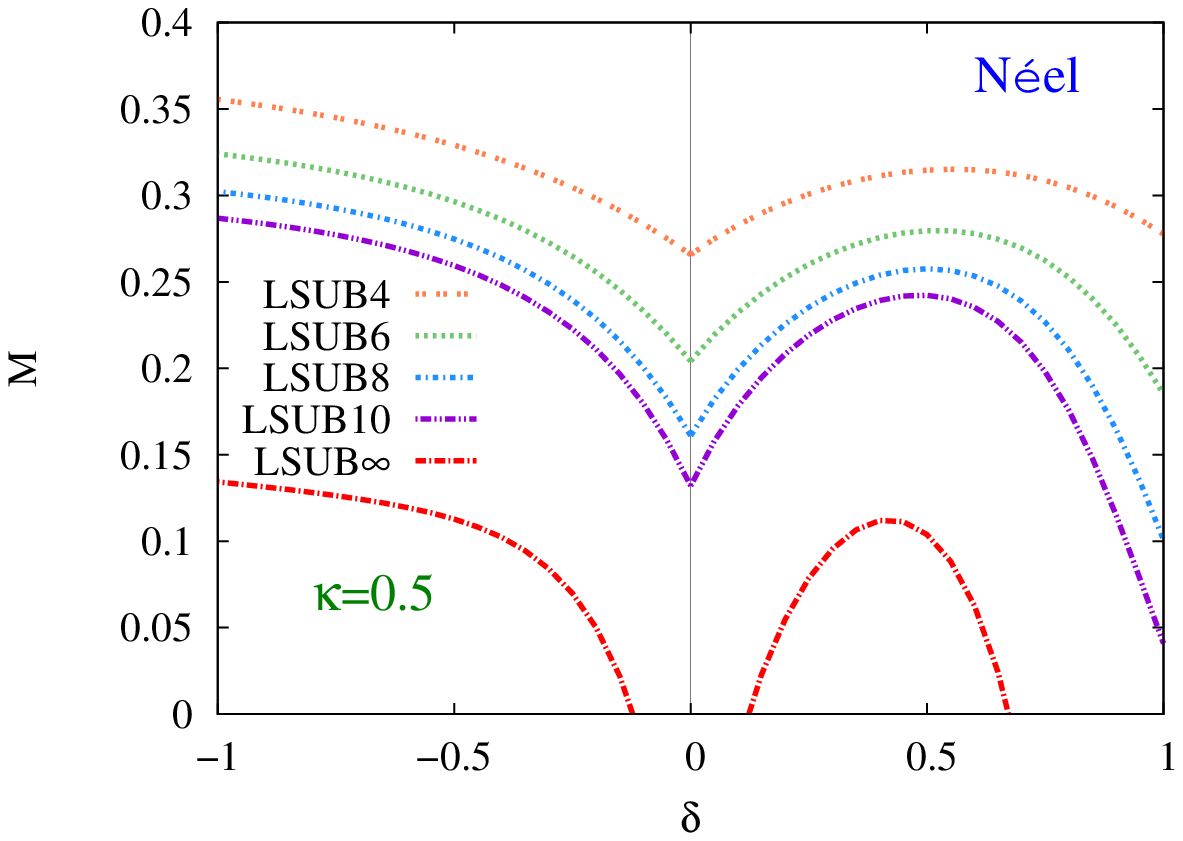}}
}
  \caption{CCM results for the GS magnetic order parameter $M$ vs the scaled interlayer exchange coupling constant, $\delta \equiv J_{1}^{\perp}/J_{1}$, for the spin-$\frac{1}{2}$ $J_{1}$--$J_{2}$--$J_{1}^{\perp}$ model on the square-lattice bilayer (with $J_{1}>0$), for two selected values of the intralayer frustration parameter, $\kappa \equiv J_{2}/J_{1}$;  (a) $\kappa=0.3$, and (b) $\kappa=0.5$.  Results based on the N\'{e}el state on each monolayer, and the two layers coupled so that NN spins between them are antiparallel (parallel) to one another for $\delta > 0$ ($\delta < 0$), as CCM model state are shown in LSUB$n$ approximations with $n=4,6,8,10$, together with the corresponding LSUB$\infty$ extrapolated result based on Eq.\ (\ref{M_extrapo_frustrated}) and the LSUB$n$ data sets $n=\{4,6,8,10\}$.  For case (a) we also show the respective LSUB$\infty$' extrapolated curve using the restricted LSUB$n$ data set $n=\{6,8,10\}$.}
\label{M_neel_fix-J2}
\end{center}
\end{figure*}

One may use the space- and point-group symmetries of the lattice and the particular CCM model state $|\Phi\rangle$ being used, as well as any pertinent conservation laws, to reduce the number of independent configurations retained at any order $n$ of approximation.  For example, for each of the model states considered in Sec.\ \ref{results_sec} (i.e., the N\'{e}el and striped states on each monolayer), the Hamiltonian of Eq.\ (\ref{H_eq}) conserves the total $z$ component of spin, $s_{T}^{z}\equiv\sum_{i=1}^{N}s_{l_{i}}^{z}$ (where {\it global} spin axes are now assumed), to the sector $s_{T}^{z}=0$.  We denote by $N_{f}(n)$ the minimal number of distinct (and nonzero) fundamental multispin-slip configurations that are retained at a given LSUB$n$ level of approximation after all such symmetries and conservation laws are take into account.  The number $N_{f}(n)$ typically still grows rapidly with $n$, and available computational resources then determine the maximum order $n$ that can be computed.

In the present case we are able to perform LSUB$n$ calculations up to the very high order $n=10$.  Thus, for the spin-$\frac{1}{2}$ square-lattice bilayer model under consideration we have $N_{f}(10)=239\,021$ $(443\,813)$ when the CCM model state is chosen so that each monolayer has N\'{e}el (striped) AFM order.  To derive and then to solve such larger sets of coupled nonlinear multinomial equations (\ref{non_linear_ket_Eq}) for $S$ and linear equations (\ref{linear_bra_Eq}) for $\tilde{S}$ we use both massive parallelization and large-scale supercomputer resources.  In order to derive the equations (and see Ref.\ \cite{Zeng:1998_SqLatt_TrianLatt}) we also use a purpose-built and customized computer algebra package \cite{ccm_code}, without which it would not be possible to go to such large orders $n$ of LSUB$n$ truncation.

Since no approximations have been made in the evaluation of any finite-order LSUB$n$ truncation of our basic CCM equations, nor in the subsequent evaluation of any GS parameter of the system, our {\it only} approximation is now made at the last step where we extrapolate an LSUB$n$ sequence of approximants to the (exact) $n \to \infty$ limit.  By now there is a great deal of empirical evidence on how to do so.  For example, for the LSUB$n$ approximants $M(n)$ to the magnetic order parameter $M$ of Eq.\ (\ref{M_definition_eq}), a well-tested scheme for systems with strong frustration, and/or for which the system has a QPT between states with and without magnetic LRO, has been found (and see, e.g., Refs.\ \cite{Bi:2008_JPCM_J1J1primeJ2,Bi:2008_PRB_J1xxzJ2xxz,Darradi:2008_J1J2mod,Reuther:2011_J1J2J3mod,Gotze:2012,Bishop:2014_honey_XXZ_nmp14,Bi:2008_EPL_J1J1primeJ2_s1,Bi:2008_JPCM_J1xxzJ2xxz_s1,Li:2019_honeycomb_bilayer_J1J2J1perp_neel-II} and references cited therein) to be given by
\begin{equation}
M(n) = \mu_{0}+\mu_{1}n^{-1/2}+\mu_{2}n^{-3/2}\,.   \label{M_extrapo_frustrated}
\end{equation}
By fitting a sequence of LSUB$n$ approximants $M(n)$ to Eq.\ (\ref{M_extrapo_frustrated}) we thus extract the corresponding extrapolated (LSUB$\infty$) value $\mu_{0}$ for $M$.

\section{RESULTS}
\label{results_sec}
We show first in Fig.\ \ref{M_neel_fix-J2} our CCM results for the magnetic order parameter $M$ based on a model state in which each monolayer has N\'{e}el order.
For values of the interlayer coupling parameter $\delta>0$ ($\delta<0$) the two layers are coupled so that NN spins are antiparallel (parallel).  Results are shown as functions of $\delta$ in Figs.\ \ref{M_neel_fix-J2}(a) and \ref{M_neel_fix-J2}(b) respectively for two fixed values of the intralayer frustration parameter, $\kappa=0.3$ and $\kappa=0.5$.  These values are chosen to lie on either side of the critical value $\kappa_{c_{1}}(\delta=0)\approx 0.45$, at which N\'{e}el order melts in the monolayer.  In both cases results are shown for even-order LSUB$n$ approximations with $4 \leq n \leq 10$, as well as for the (LSUB$\infty$) estimates $\mu_{0}$ obtained from fitting these data to the extrapolation scheme of Eq.\ (\ref{M_extrapo_frustrated}).  The LSUB2 data are omitted since, in principle, they are expected to be of too low order to fit well to such a three-term extrapolation scheme.  Nevertheless, when separate fits are made that include them (i.e., to the LSUB$n$ data sets $n=\{2,4,6,8,10\}$), the extrapolated values for $\mu_{0}$ hardly change at all, thereby demonstrating the robustness of the fits.  To enable readers to judge by eye for themselves how robust and accurate are our extrapolations, we also show in Fig.\ \ref{M_neel_fix-J2}(a) the corresponding extrapolation (labelled LSUB$\infty$') based on the restricted LSUB$n$ data set $n=\{6,8,10\}$.  Although a fit based on only three data points to a three-term extrapolation scheme such as that of Eq.\ (\ref{M_extrapo_frustrated}) is not, {\it a priori}, expected to be as robust as one based on more data points, the agreement between the LSUB$\infty$ and LSUB$\infty$' curves can be clearly seen.  Similar agreement is observed for all other values of the intralayer frustration parameter $\kappa$.

\begin{figure*}[t]
\begin{center}
\mbox{
\subfigure[]{\includegraphics[width=8.5cm]{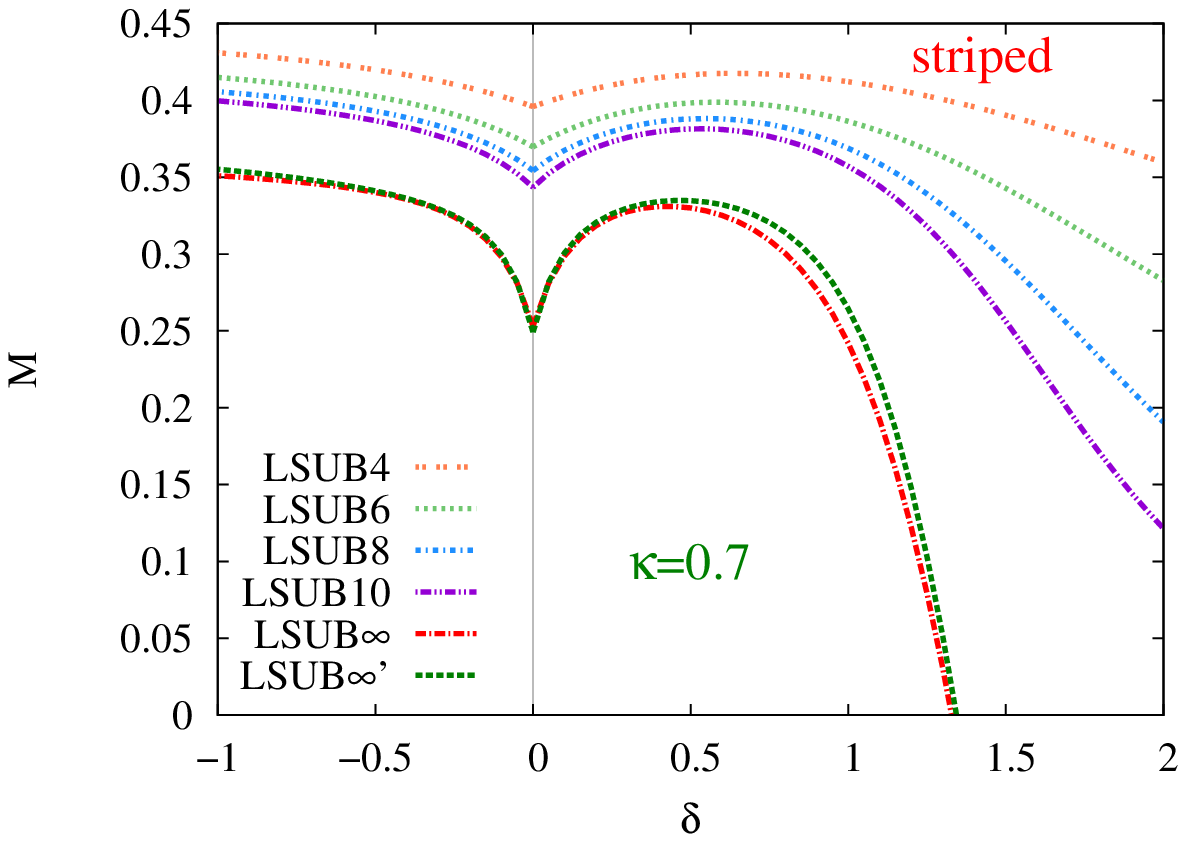}}
\hspace{0.1cm} \subfigure[]{\includegraphics[width=8.5cm]{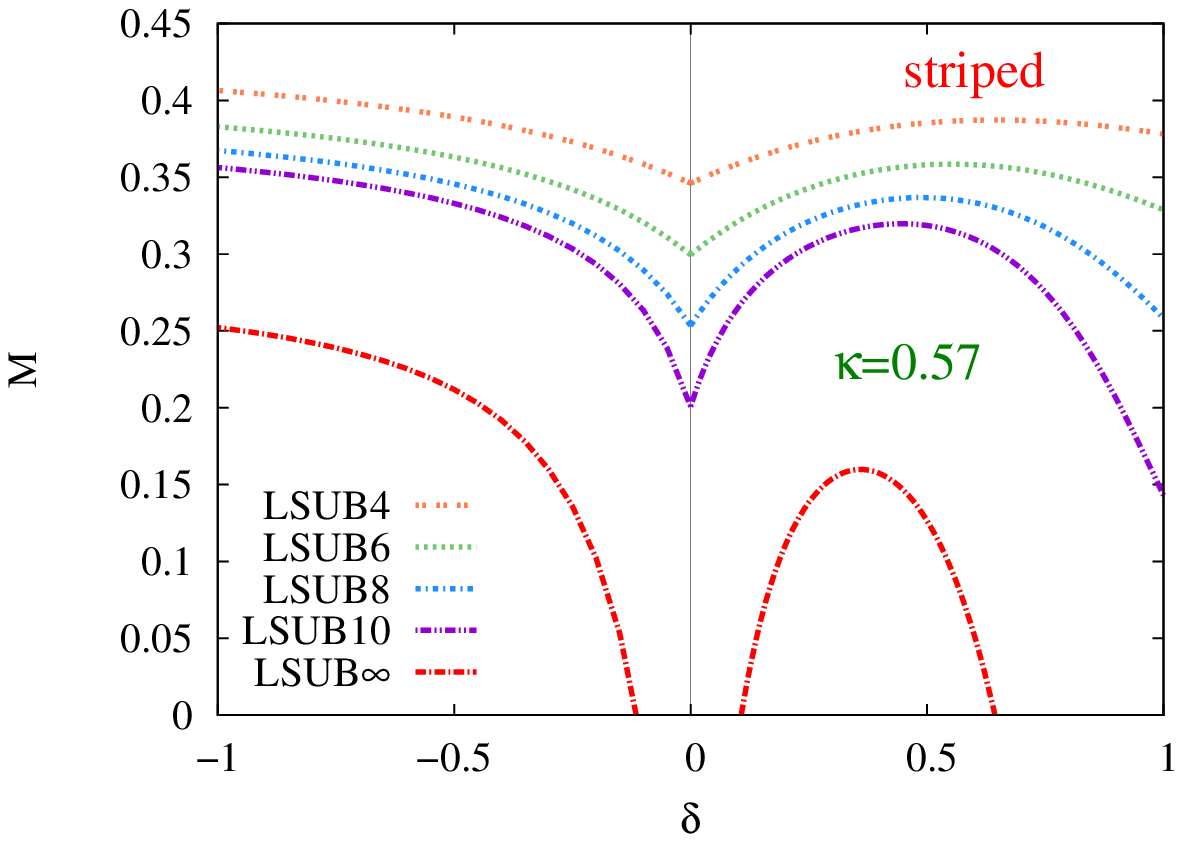}}
}
  \caption{CCM results for the GS magnetic order parameter $M$ vs the scaled interlayer exchange coupling constant, $\delta \equiv J_{1}^{\perp}/J_{1}$, for the spin-$\frac{1}{2}$ $J_{1}$--$J_{2}$--$J_{1}^{\perp}$ model on the square-lattice bilayer (with $J_{1}>0$), for two selected values of the intralayer frustration parameter, $\kappa \equiv J_{2}/J_{1}$;  (a) $\kappa=0.7$, and (b) $\kappa=0.57$.  Results based on the striped state on each monolayer, and the two layers coupled so that NN spins between them are antiparallel (parallel) to one another for $\delta > 0$ ($\delta < 0$), as CCM model state are shown in LSUB$n$ approximations with $n=4,6,8,10$, together with the corresponding LSUB$\infty$ extrapolated result based on Eq.\ (\ref{M_extrapo_frustrated}) and the LSUB$n$ data sets $n=\{4,6,8,10\}$.  For case (a) we also show the respective LSUB$\infty$' extrapolated curve using the restricted LSUB$n$ data set $n=\{6,8,10\}$.}
\label{M_striped_fix-J2}
\end{center}
\end{figure*}

Turning first to Fig.\ \ref{M_neel_fix-J2}(a), the LSUB$\infty$ extrapolated curve exhibits all of the features we expect from our discussion in Sec.\ \ref{model_sec}.  Thus, firstly, the cusp at $\delta=0$ (where $M \approx 0.17$ for the value $\kappa=0.3$ shown) is exactly as expected from the observation that as $|\delta|$ is increased from zero in either direction, the order is first enhanced due to the increase in the number of NN bonds and the consequent step towards increasing the dimensionality of the system.  Secondly, we observe that as $\delta$ is increased further, in the regime $\delta > 0$ of AFM coupling between the two layers, $M$ attains a maximum value of about 0.29 at a value $\delta \approx 0.57$ before the effects of interlayer dimerization become sufficiently strong to start to weaken the intralayer N\'{e}el order as $\delta$ is increased further beyond that point.  This continues up to an upper critical value, $\delta_{c_{1}}^{>}=\delta_{c_{1}}^{>}(\kappa)$, above which N\'{e}el order disappears entirely.  For the value $\kappa=0.3$ shown in Fig.\ \ref{M_neel_fix-J2}(a), this upper critical value is seen to be at $\delta_{c_{1}}^{>}(0.3)\approx 1.66$.

As the value of $\kappa$ is increased beyond 0.3 the cusp at $\delta=0$ in the LSUB$\infty$ curve of Fig.\ \ref{M_neel_fix-J2}(a) is lowered until, at the value $\kappa \approx 0.447$, it reaches the $\delta=0$ axis.  This is precisely the value we obtain for $\kappa_{c_{1}}(\delta=0)$ within this same extrapolation scheme, using the LSUB$n$ data set $n=\{4,6,8,10\}$.  We note in passing that for the monolayer case ($\delta=0$) it has also been possible to perform LSUB12 calculations \cite{Richter:2015_ccm_J1J2sq_spinGap}.  The corresponding value obtained from fitting to the LSUB$n$ data set $n=\{4,6,8,10,12\}$ is $\kappa_{c_{1}}(\delta=0)\approx 0.454$ \cite{Richter:2015_ccm_J1J2sq_spinGap}, which again demonstrates the accuracy and robustness of our extrapolations.

If we now slightly increase $\kappa$ beyond this value $\kappa_{c_{1}}(0)$, we obtain curves such as those shown in Fig.\ \ref{M_neel_fix-J2}(b).  Thus, for a certain range of values above $\kappa_{c_{1}}(0)$, for which N\'{e}el order is absent for the monolayer $(\delta=0)$, as $\delta$ is either decreased or increased from this value, N\'{e}el order becomes re-established at certain critical values, $\delta_{c_{1}}^{{\mathrm F}}(\kappa)$ for $\delta<0$ and $\delta_{c_{1}}^{<}(\kappa)$ for $\delta>0$.  For the value $\kappa=0.5$ shown in Fig.\ \ref{M_neel_fix-J2}(b) these values are seen to be $\delta_{c_{1}}^{{\mathrm F}}(0.5) \approx -0.12$ and $\delta_{c_{1}}^{<}(0.5) \approx 0.12$.  Once again, as $\delta$ is now increased beyond the value $\delta_{c_{1}}^{<}(\kappa)$, N\'{e}el order is at first enhanced until $M$ attains a maximal value.  For the value $\kappa=0.5$ shown in Fig.\ \ref{M_neel_fix-J2}(b) this maximum value for $M$, for the case of AFM interlayer coupling, is about 0.11 at a value $\delta\approx 0.42$.  Further increase in $\delta$ then reduce the magnetic order until it again melts entirely at a value $\delta_{c_{1}}^{>}(\kappa)$.  For $\kappa=0.5$ this upper critical value is seen from Fig.\ \ref{M_neel_fix-J2}(b) to be at $\delta_{c_{1}}^{>}(0.5) \approx 0.67$.

If we continue to increase $\kappa$ slowly beyond the value $\kappa=0.5$ shown in Fig.\ \ref{M_neel_fix-J2}(b), the lower and upper critical values,  $\delta_{c_{1}}^{<}(\kappa)$ and $\delta_{c_{1}}^{>}(\kappa)$, move towards one another until at some value $\kappa_{1}^{\mathrm{max}} \approx 0.535$ they merge, $\delta_{c_{1}}^{<}(\kappa_{1}^{\mathrm{max}})=\delta_{c_{1}}^{>}(\kappa_{1}^{\mathrm{max}})\approx 0.3$.  N\'{e}el order is then wholly absent for any value $\kappa > \kappa_{1}^{\mathrm{max}}$ of the intralayer frustration parameter and for any value $\delta>0$ of the AFM interlayer coupling.

\begin{figure*}[t]
\begin{center}
\mbox{
\subfigure[]{\includegraphics[width=8.0cm]{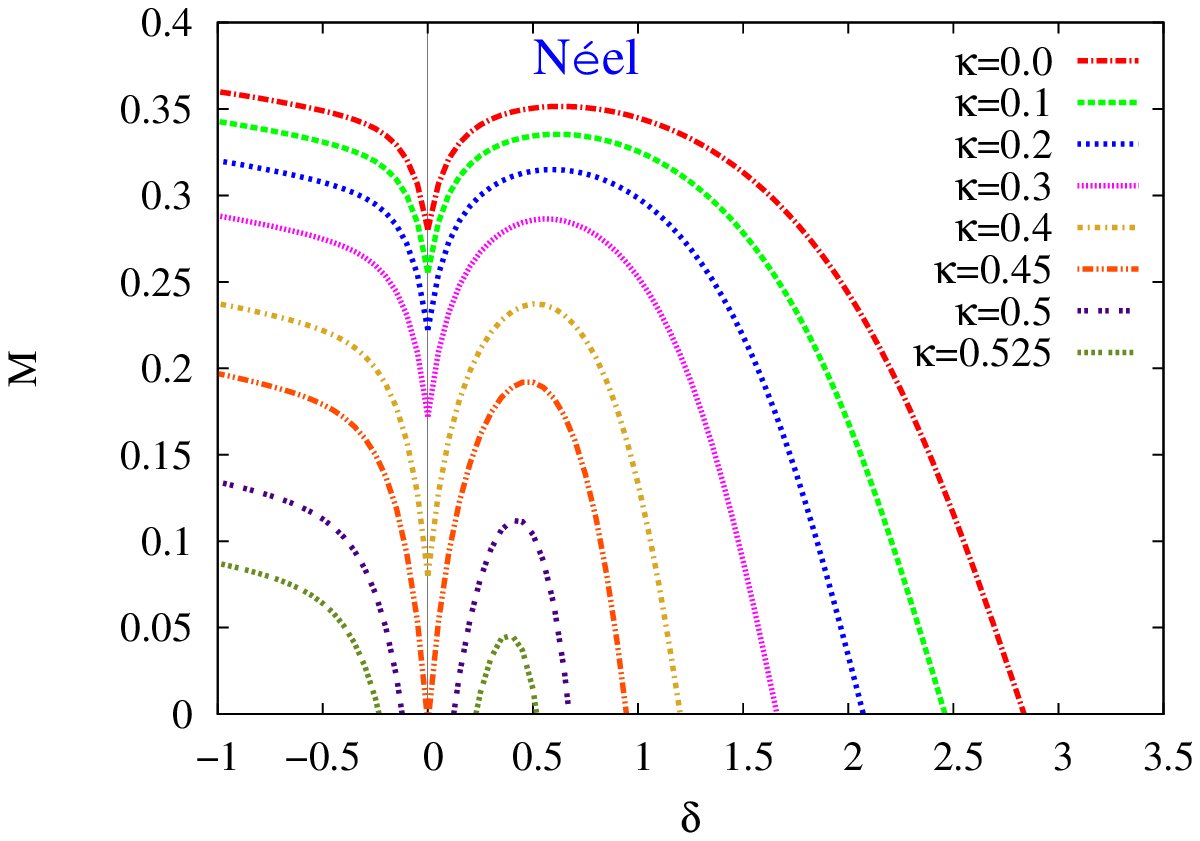}}
\hspace{0.1cm} \subfigure[]{\includegraphics[width=8.0cm]{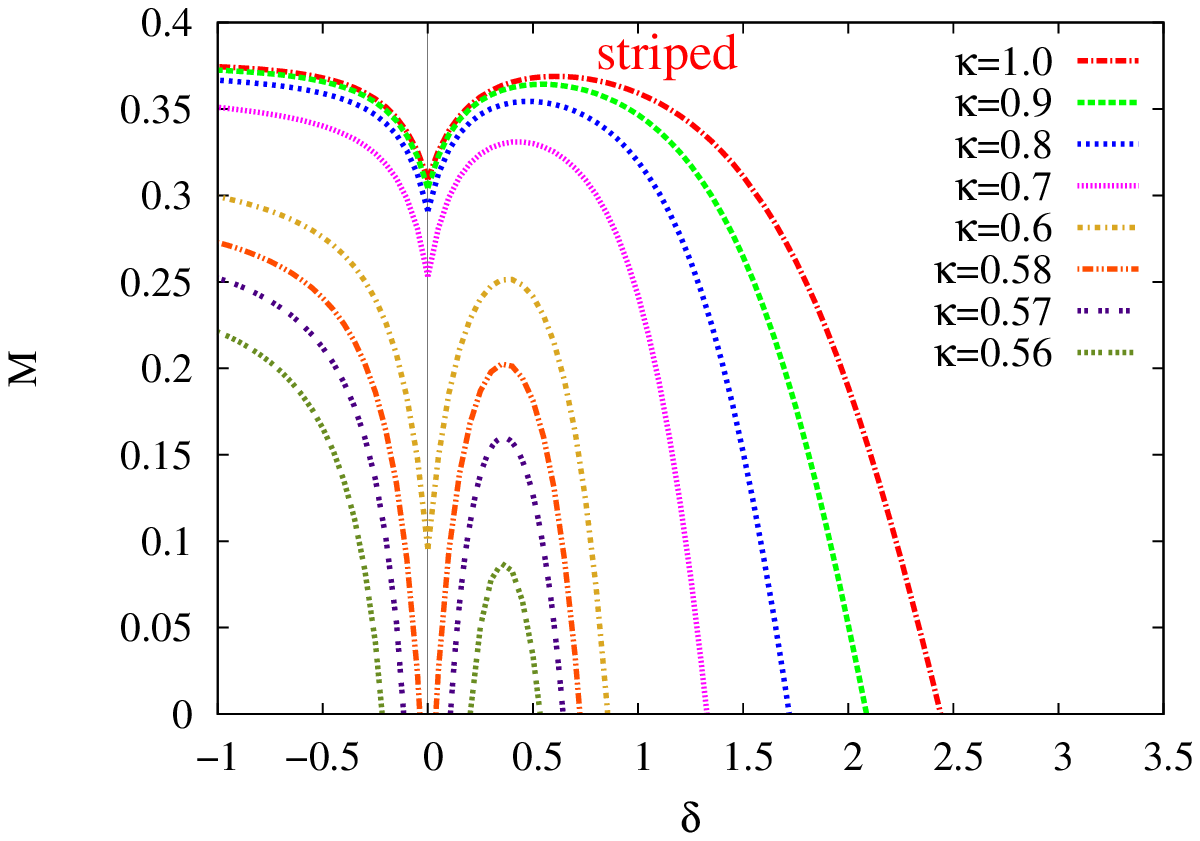}}
}
  \caption{CCM results for the GS magnetic order parameter $M$ vs the scaled interlayer exchange coupling constant, $\delta \equiv J_{1}^{\perp}/J_{1}$, for the spin-$\frac{1}{2}$ $J_{1}$--$J_{2}$--$J_{1}^{\perp}$ model on the square-lattice bilayer (with $J_{1}>0$), for a variety of values of the intralayer frustration parameter, $\kappa \equiv J_{2}/J_{1}$, using (a) the N\'{e}el state and (b) the striped state as the CCM model state on each monolayer, and the two layers coupled so that NN spins between them are antiparallel (parallel) to one another for $\delta > 0$ ($\delta < 0$).  In each case we show extrapolated results, obtained from using Eq.\ (\ref{M_extrapo_frustrated}) with the respective LSUB$n$ data sets $n=\{4,6,8,10\}$.}
\label{M_contours_fix-J2}
\end{center}
\end{figure*}

Corresponding results to those shown in Fig.\ \ref{M_neel_fix-J2}, which are based on a CCM model state with N\'{e}el order on each monolayer, as in Fig.\ \ref{model_pattern}(b), are now shown in Fig.\ \ref{M_striped_fix-J2} based on a corresponding model state with striped AFM order on each monolayer, as in Fig.\ \ref{model_pattern}(c).
The two layers are again coupled so that NN interlayer spins are antiparallel for $\delta>0$ and parallel for $\delta<0$.  The two values of $\kappa$ shown, viz., $\kappa=0.7$ in Fig.\ \ref{M_striped_fix-J2}(a) and $\kappa=0.57$ in Fig.\ \ref{M_striped_fix-J2}(b) are now chosen to lie on either side of the critical value $\kappa_{c_{2}}(\delta=0)\approx 0.59$ for which striped order melts in the monolayer.  In Fig.\ \ref{M_striped_fix-J2}(a) we also show the corresponding LSUB$\infty$' extrapolated curve based on the restricted LSUB$n$ data set $n=\{6,8,10\}$.  The agreement with the LSUB$\infty$ curve based on the full LSUB$n$ data set $n=\{4,6,8,10\}$ is again observed to be excellent.  Similar levels of agreement are found for all other values of $\kappa$.  Once again, for the special case of the monolayer LSUB12 calculations have also been performed \cite{Richter:2015_ccm_J1J2sq_spinGap} based on the striped model state.  In this case we find $\kappa_{c_{2}}(0)\approx 0.587$ based on the extrapolation of Eq.\ (\ref{M_extrapo_frustrated}) with the LSUB$n$ data set $n=\{4,6,8,10\}$ as input, while the corresponding result based on the set $n=\{4,6,8,10,12\}$ yielded the almost identical value $\kappa_{c_{2}}(0)\approx 0.588$ \cite{Richter:2015_ccm_J1J2sq_spinGap}.

The similarity between Fig.\ \ref{M_neel_fix-J2}(a) and \ref{M_striped_fix-J2}(a), and also between Fig.\ \ref{M_neel_fix-J2}(b) and \ref{M_striped_fix-J2}(b), is self-evident.  In this case we find that the lower and upper critical values, $\delta_{c_{2}}^{<}(\kappa)$ and $\delta_{c_{2}}^{>}(\kappa)$, as seen in Fig.\ \ref{M_striped_fix-J2}(b) in the region $\delta>0$ for the value $\kappa=0.57 < \kappa_{c_{2}}(0)$ again move towards one another as $\kappa$ is now slowly decreased beyond this value, until they merge at a value $\kappa_{2}^{\mathrm{min}} \approx 0.555$, where $\delta_{c_{2}}^{<}(\kappa_{2}^{\mathrm{min}})=\delta_{c_{2}}^{{\mathrm >}}(\kappa_{2}^{\mathrm{min}}) \approx 0.3$.  For all values $\kappa < \kappa_{2}^{\mathrm{min}}$ of the intralayer frustration parameter striped order is then absent, whatever the value $\delta > 0$ of the AFM interlayer coupling.

In  Figs.\ \ref{M_contours_fix-J2}(a) and \ref{M_contours_fix-J2}(b) we now show sets of extrapolated (LSUB$\infty$) curves for the magnetic order parameter $M$ as a function of the interlayer coupling parameter $\delta$ for each layer with N\'{e}el order and striped order, respectively, for various fixed values of the intralayer frustration $\kappa$.  These correspond, respectively, to curves such as those shown in Figs.\ \ref{M_neel_fix-J2} and \ref{M_striped_fix-J2}.  From Fig.\ \ref{M_contours_fix-J2}(b) it is clear that the position of the cusp at $\delta=0$ for the striped-ordered phase rather rapidly approaches a limiting value for $M$ as $\kappa$ is increased.  This corresponds to the limit ($\kappa \to \infty$) of the model where each layer corresponds to two independent, and equivalent, interpenetrating square sublattices, each of which is N\'{e}el-ordered.  A recent CCM calculation \cite{Farnell:2018_archimedean_multiSpins} for the spin-$\frac{1}{2}$ Heisenberg antiferromagnet on the square lattice utilized an LSUB$n$ data set $n=\{4,6,8,10,12\}$ to yield the extrapolated value $M\approx 0.3093$.  By comparison, we find here the remarkably close value $M(\delta=0)\approx 0.3094$ from using Eq.\ (\ref{M_extrapo_frustrated}) with the LSUB$n$ data set $n=\{4,6,8,10\}$, and as shown in Fig.\ \ref{M_contours_fix-J2}(b), for the curve $\kappa=1.0$, which value is itself already extremely close to the limiting value obtained as $\kappa \to \infty$.

We also note from Fig.\ \ref{M_contours_fix-J2}(a) that for the case of zero intralayer frustration ($\kappa=0$) our extrapolation of Eq.\ (\ref{M_extrapo_frustrated}) leads to an upper critical value, $\delta_{c_{1}}^{>}(\kappa=0)\approx 2.84$, of the interlayer coupling parameter, beyond which N\'{e}el order melts and a phase with IDVBC order is stabilized.  This may be compared with the corresponding value of 2.5220 obtained from a large-scale QMC simulation \cite{Wang:2006_SqLatt_bilayer} of the spin-$\frac{1}{2}$ $J_{1}$--$J_{1}^{\perp}$ model discussed in Sec.\ \ref{model_sec}.  In this context it is worth noting that the location of the phase boundary is less accurately determined in our CCM calculations for the region where the order-disorder transition is essentially driven by singlet-dimerization than in the region where it is essentially due to frustration.

\begin{figure}[!t]
  \includegraphics[width=9cm]{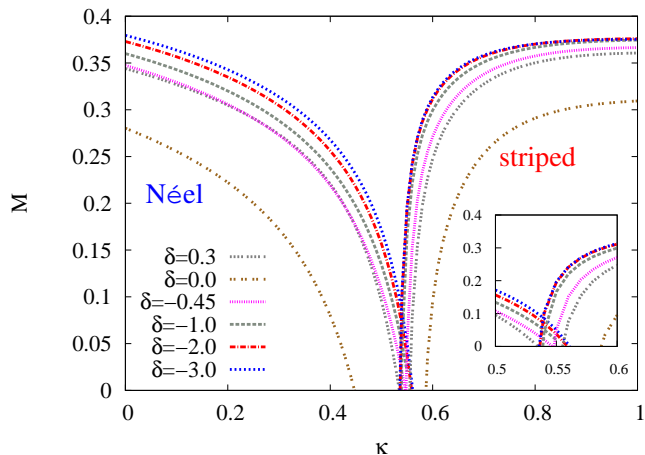}
  \caption{CCM results for the GS magnetic order parameter $M$ vs the intralayer frustration parameter, $\kappa \equiv J_{2}/J_{1}$, for the spin-$\frac{1}{2}$ $J_{1}$--$J_{2}$--$J_{1}^{\perp}$ model on the square-lattice bilayer (with $J_{1}>0$), for a variety of values of the scaled interlayer exchange coupling constant, $\delta \equiv J_{1}^{\perp}/J_{1}$, using (a) the N\'{e}el state and (b) the striped state as the CCM model state on each monolayer, and the two layers coupled so that NN spins between them are antiparallel to one another in the cases where $\delta > 0$ and parallel to one another in the cases where $\delta < 0$.  In each case we show extrapolated results, obtained from using Eq.\ (\ref{M_extrapo_frustrated}) with the respective LSUB$n$ data sets $n=\{4,6,8,10\}$.}
\label{M_contours_fix-J1perp}
\end{figure}

Particularly in the half-plane $\delta<0$, corresponding to FM interlayer coupling, it is also convenient to investigate the magnetic order parameter $M$ for the two quasiclassical phases with AFM orderings on each monolayer as functions of the intralayer frustration parameter $\kappa$, for various fixed values of the interlayer coupling parameter $\delta$.  In Fig.\ \ref{M_contours_fix-J1perp} we show a set of such curves for both AFM monolayer orderings.
The curves for $\delta=0$ again exhibit the corresponding critical points, $\kappa_{c_{1}}(\delta=0)\approx 0.447$ and $\kappa_{c_{2}}(\delta=0)\approx 0.587$, for the melting of N\'{e}el and striped order, respectively, for the spin-$\frac{1}{2}$ $J_{1}$--$J_{2}$ model on the square-lattice monolayer.  We also show the corresponding curves for $\delta=0.3$ in the region of AFM coupling between the layers, which is approximately the value of $\delta$ for which the phase boundaries of the two quasiclassical phases are closest together in this region $\delta>0$.  Thus, at $\delta=0.3$, the paramagnet state exists only in the very narrow regime $0.535 \lesssim \kappa \lesssim 0.555$ of the frustration parameter.

We also show in Fig.\ \ref{M_contours_fix-J1perp} similar curves for various values of $\delta$ in the half-plane $\delta<0$, which corresponds to FM interlayer coupling.  
One sees clearly that for all values $\delta < \delta_{{\mathrm T}}^{{\mathrm F}} \approx -0.45$ the respective curves $M=M(\kappa)$ for the two quasiclassical phases cross one another at a value $\kappa^{{\mathrm F}}(\delta)>0$, indicating a direct first-order transition between them.  The value $\kappa^{{\mathrm F}}(\delta)$ is seen to be almost independent of $\delta$ in this regime.  For example $\kappa^{{\mathrm F}}(-3) \approx 0.539$, which may be compared with the expected value ${\mathrm{lim}}_{\delta\to -\infty}\kappa^{{\mathrm F}}(\delta) \approx 0.549$, viz., the value that corresponds to the critical coupling for the direct transition between the two states in the spin-1 $J_{1}$--$J_{2}$ model on the square-lattice monolayer \cite{Haghshenas:2018_SqLatt_J1J2mod_s1}, as discussed in Sec.\ \ref{model_sec}.  From Fig.\ \ref{M_contours_fix-J1perp} we see that the QTP where the N\'{e}el, striped, and paramagnetic phases meet is situated at $(\kappa_{{\mathrm T}}^{{\mathrm F}},\delta_{{\mathrm T}}^{{\mathrm F}})\approx (0.547,-0.45)$.

Of course, in view of the observation that the direct transition between the two quasiclassical phases for $\delta \lesssim -0.45$ is of first-order type, we can also corroborate our results in this regime by using the fact that the GS energies of the two states should also cross at the phase boundary.  We use the appropriate extrapolation scheme for the LSUB$n$ approximants $e(n)$ to the GS energy per spin, $e \equiv E/N$, which is well known to be given by
\begin{equation}
e(n) = e_{0}+e_{1}n^{-2}+e_{2}n^{-4}\,.     \label{E_extrapo}
\end{equation}
We may thus make use of Eq.\ (\ref{E_extrapo}) with the same LSUB$n$ input data sets $n=\{4,6,8,10\}$ as used in Fig.\ \ref{M_contours_fix-J1perp} to corroborate the points shown in Fig.\ \ref{M_contours_fix-J1perp} by cross ($\times$) symbols, which denote the points where the respective $M=M(\kappa)$ curves for the two quasiclassical phases cross one another for a given value of $\delta$ at the value $\kappa=\kappa^{\mathrm{F}}(\delta)$.  We find, for example, at the value $\delta=-3$, the two corresponding LSUB$\infty$ extrapolated curves $e=e(\kappa)$ cross one another at the value $\kappa \approx 0.558$, which may be compared with the value $\kappa^{\mathrm{F}}(-3) \approx 0.539$ cited above, where the respective $M=M(\kappa)$ curves cross.  The agreement between these two essentially independent calculations is good.  This clearly also provides internal confirmation of the robustness of the extrapolation schemes that form the sole approximation made in our results

\begin{figure}[!t]
  \includegraphics[width=9cm]{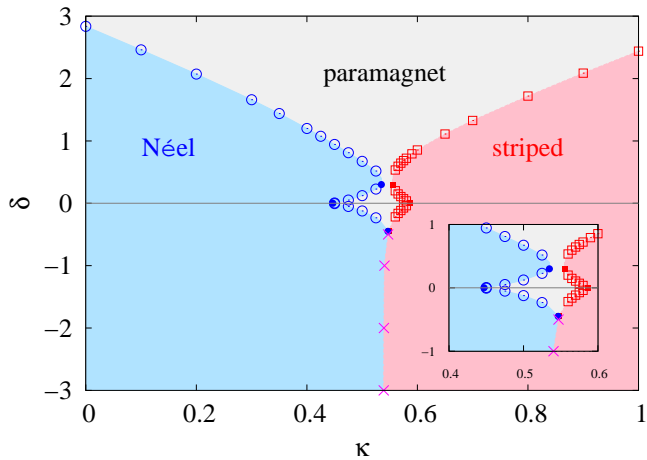}
  \caption{$T=0$ phase diagram of the spin-$\frac{1}{2}$
    $J_{1}$--$J_{2}$--$J_{1}^{\perp}$ model on the bilayer square
    lattice with $J_{1}>0$, $\delta \equiv J_{1}^{\perp}/J_{1}$, and
    $\kappa \equiv J_{2}/J_{1}$.  The blue and pink regions are the
    quasiclassical phases with AFM N\'{e}el and striped orders, respectively, while in the grey region quasiclassical collinear order is absent.  The filled and empty circle (and square)
    symbols are points at which the
    extrapolated GS magnetic order parameter $M$ for the N\'{e}el (and striped) phases
    vanishes, for specified values of $\delta$ and $\kappa$,
    respectively.  By contrast, the cross ($\times$) symbols indicate points at which the corresponding two curves $M=M(\kappa)$, for a specified value of $\delta$, cross one another.  In each
    case the N\'{e}el or striped state on each monolayer is used as CCM model state, and Eq.\
    (\ref{M_extrapo_frustrated}) is used for the extrapolations with
    the corresponding LSUB$n$ data sets $n=\{4,6,8,10\}$.}
\label{phase_diag_neel_striped_SqLatt_bilayer}
\end{figure}

By combining results from curves such as those shown in Figs.\ \ref{M_neel_fix-J2}--\ref{M_contours_fix-J1perp} to extract the points where the extrapolated (LSUB$\infty$) GS magnetic order parameter $M$ vanishes for the two quasiclassical AFM phases, we may finally construct the zero-temperature ($T=0$) quantum phase diagram of our spin-$\frac{1}{2}$ $J_{1}$--$J_{2}$--$J_{1}^{\perp}$ model on the bilayer square lattice.  It is shown in Fig.\ \ref{phase_diag_neel_striped_SqLatt_bilayer} in the $\kappa$--$\delta$ half-plane with $\kappa>0$. 
We note that different symbols are used in Fig.\ \ref{phase_diag_neel_striped_SqLatt_bilayer} to distinguish between points on the phase boundaries that have been extracted from calculations done at fixed values of $\kappa$ and those extracted from calculations done at fixed values of $\delta$.  It is clear by visual inspection that these two sets of critical points lie very accurately on a smooth boundary curve for each collinear AFM state.  This again provides good internal evidence that our extrapolations are robust and accurate.

Our results are now discussed and summarized in Sec.\ \ref{discuss_summary_sec}.

\section{DISCUSSION AND SUMMARY}
\label{discuss_summary_sec}
We see from Fig.\ \ref{phase_diag_neel_striped_SqLatt_bilayer} that the phase boundary for each of the AFM quasiclassical states is rather accurately linear in the $\kappa$--$\delta$ plane for large enough values of AFM interlayer coupling (viz., for $\delta \gtrsim 1.5$).  For the N\'{e}el phase boundary the slope of the curve at large values of $\delta$ is ${\mathrm d}\delta/{\mathrm d}\kappa \approx -3.9$, while the corresponding value for the striped phase boundary is ${\mathrm d}\delta/{\mathrm d}\kappa \approx 3.7$.  If this linear behavior would continue unchanged to smaller values of $\delta$ the two curves would cross at a point $(\kappa,\delta) \approx (0.54,0.7)$.  Instead, as we see from Fig.\ \ref{phase_diag_neel_striped_SqLatt_bilayer}, the two curves turn against each other before this point, in a typical ``avoided crossing'' manner, although they do approach one another rather closely in the region around $\delta \approx 0.3$.  This has the effect that the entire disordered paramagnetic regime in Fig.\ \ref{phase_diag_neel_striped_SqLatt_bilayer} is singly connected.

At small values of $\delta$ both phase boundaries exhibit the reentrant behavior expected from our discussion in Sec.\ \ref{model_sec}, with both displaying cusps at $\delta=0$.  However, a closer inspection of Fig.\ \ref{phase_diag_neel_striped_SqLatt_bilayer} shows that the nature of the cusps is quite different for the two phase boundaries, with that on the N\'{e}el side much ``sharper'' than its counterpart on the striped side.  More quantitatively, on the striped side the slopes of the curves on each side of the cusp at the point $\kappa = \kappa_{c_{2}}(0), \delta=0$ are clearly nonzero.  By contrast, on the N\'{e}el side the corresponding slopes of the curves of the cusp at the point $\kappa=\kappa_{c_{1}}(0),\delta=0$ appear to be zero (or very close) to zero.  This difference would certainly explain why the critical parameter $\kappa_{c_{1}}(0)$ is more difficult to calculate accurately than the corresponding parameter $\kappa_{c_{2}}(0)$ for the spin-$\frac{1}{2}$ $J_{1}$--$J_{2}$ model on the square lattice, as has been discussed previously in Sec.\ \ref{model_sec}.  The difference is surely also a direct reflection of the different natures of the two QPTs in that model.  Thus, our results are in clear accord with the consensual view that, while the transition at $\kappa_{c_{2}}(0)$ between the striped and paramagnetic phases is a first-order one, that at $\kappa_{c_{1}}(0)$ between the N\'{e}el and paramagnetic phases is continuous.

In the half-plane $\delta<0$ the phase boundaries of the two quasiclassical AFM phases end at a QTP where they meet a line of direct first-order transitions between the two phases.  Starting at the QTP, which we have calculated as being located at $(\kappa_{\mathrm{T}}^{\mathrm{F}}\approx 0.547,\delta_{\mathrm{T}}^{\mathrm{F}}\approx -0.45)$, this line of first-order transitions is very nearly a vertical straight line in the $\kappa$--$\delta$ plane.  At large negative values of $\delta$ it approaches the value $\kappa^{\mathrm{F}}(\delta \to -\infty)\approx0.539$, which is itself an accurate estimate of the QCP for the direct transition between the two quasiclassical phases of the spin-1 $J_{1}$--$J_{2}$ model on the square lattice.

It is completely beyond the scope of the present paper to investigate in detail the nature of the phases in the (grey shaded) paramagnetic regime in Fig.\ \ref{phase_diag_neel_striped_SqLatt_bilayer}, outside the respective regimes in which we have calculated that the N\'{e}el state or the striped state on each monolayer forms the stable GS phase.  Nevertheless we conclude with a few comments on this issue.

We have already discussed in Sec.\ \ref{introd_sec} the lack of overall consensus for the nature of the GS phase or phases in the region $\kappa_{c_{1}}(0) < \kappa < \kappa_{c_{2}}(0)$ for the spin-$\frac{1}{2}$ $J_{1}$--$J_{2}$ model on the square-lattice monolayer.  However, the results of two independent high-order techniques, viz., the CCM \cite{Darradi:2008_J1J2mod,Richter:2015_ccm_J1J2sq_spinGap} and the DMRG method \cite{Gong:2014_J1J2mod_sqLatt}, applied to the model, are both compatible with the existence of two phases in this paramagnetic region.  Both CCM and DMRG calculations also agree on the critical values $\kappa_{c_{1}}(0) \approx 0.45(1)$ and $\kappa_{c_{2}}(0) \approx 0.60(1)$.  Furthermore, both can be consistently interpreted with the hypothesis of a gapped plaquette-ordered VBC (PVBC) ground state in the region $0.5 \lesssim \kappa < \kappa_{c_{2}}(0)$, and a ground state in the region $\kappa_{c_{1}}(0) < \kappa \lesssim 0.5$ that could be a gapless QSL state.  In view of the single-connectedness of the entire paramagnetic regime that we have found, it is clear that, based on the above scenario being true for $\delta=0$, this paramagnetic  regime for the bilayer should include at least three phases, viz., QSL, PVBC, and IDVBC.  It will clearly be of great future interest to study the boundaries of these phases in detail.

\section*{ACKNOWLEDGMENTS}
We thank the University of Minnesota Supercomputing Institute for the
grant of supercomputing facilities, on which some of the work reported here
was performed.

% If we want the longbibliography option which include paper titles, we need to not include (or comment out) any \bibliographystyle{} command.
%\bibliographystyle{apsrev4-1}
%\bibliographystyle{plain}
%\bibliographystyle{h-physrev3}
\bibliography{bib_general}

\providecommand{\noopsort}[1]{}\providecommand{\singleletter}[1]{#1}%
\begin{thebibliography}{116}%
\makeatletter
\providecommand \@ifxundefined [1]{%
 \@ifx{#1\undefined}
}%
\providecommand \@ifnum [1]{%
 \ifnum #1\expandafter \@firstoftwo
 \else \expandafter \@secondoftwo
 \fi
}%
\providecommand \@ifx [1]{%
 \ifx #1\expandafter \@firstoftwo
 \else \expandafter \@secondoftwo
 \fi
}%
\providecommand \natexlab [1]{#1}%
\providecommand \enquote  [1]{``#1''}%
\providecommand \bibnamefont  [1]{#1}%
\providecommand \bibfnamefont [1]{#1}%
\providecommand \citenamefont [1]{#1}%
\providecommand \href@noop [0]{\@secondoftwo}%
\providecommand \href [0]{\begingroup \@sanitize@url \@href}%
\providecommand \@href[1]{\@@startlink{#1}\@@href}%
\providecommand \@@href[1]{\endgroup#1\@@endlink}%
\providecommand \@sanitize@url [0]{\catcode `\\12\catcode `\$12\catcode
  `\&12\catcode `\#12\catcode `\^12\catcode `\_12\catcode `\%12\relax}%
\providecommand \@@startlink[1]{}%
\providecommand \@@endlink[0]{}%
\providecommand \url  [0]{\begingroup\@sanitize@url \@url }%
\providecommand \@url [1]{\endgroup\@href {#1}{\urlprefix }}%
\providecommand \urlprefix  [0]{URL }%
\providecommand \Eprint [0]{\href }%
\providecommand \doibase [0]{http://dx.doi.org/}%
\providecommand \selectlanguage [0]{\@gobble}%
\providecommand \bibinfo  [0]{\@secondoftwo}%
\providecommand \bibfield  [0]{\@secondoftwo}%
\providecommand \translation [1]{[#1]}%
\providecommand \BibitemOpen [0]{}%
\providecommand \bibitemStop [0]{}%
\providecommand \bibitemNoStop [0]{.\EOS\space}%
\providecommand \EOS [0]{\spacefactor3000\relax}%
\providecommand \BibitemShut  [1]{\csname bibitem#1\endcsname}%
\let\auto@bib@innerbib\@empty
%</preamble>
\bibitem [{\citenamefont {Chandra}\ and\ \citenamefont
  {Doucot}(1988)}]{Chandra:1988}%
  \BibitemOpen
  \bibfield  {author} {\bibinfo {author} {\bibfnamefont {P.}~\bibnamefont
  {Chandra}}\ and\ \bibinfo {author} {\bibfnamefont {B.}~\bibnamefont
  {Doucot}},\ }\bibfield  {title} {\enquote {\bibinfo {title} {Possible
  spin-liquid state at large ${S}$ for the frustrated square {H}eisenberg
  lattice},}\ }\href {\doibase 10.1103/PhysRevB.38.9335} {\bibfield  {journal}
  {\bibinfo  {journal} {Phys. Rev. B}\ }\textbf {\bibinfo {volume} {38}},\
  \bibinfo {pages} {9335(R)--9338(R)} (\bibinfo {year} {1988})}\BibitemShut
  {NoStop}%
\bibitem [{\citenamefont {Dagotto}\ and\ \citenamefont
  {Moreo}(1989)}]{Dagotto:1989}%
  \BibitemOpen
  \bibfield  {author} {\bibinfo {author} {\bibfnamefont {Elbio}\ \bibnamefont
  {Dagotto}}\ and\ \bibinfo {author} {\bibfnamefont {Adriana}\ \bibnamefont
  {Moreo}},\ }\bibfield  {title} {\enquote {\bibinfo {title} {Phase diagram of
  the frustrated spin-$\frac{1}{2}$ {H}eisenberg antiferromagnet in two
  dimensions},}\ }\href {\doibase 10.1103/PhysRevLett.63.2148} {\bibfield
  {journal} {\bibinfo  {journal} {Phys. Rev. Lett.}\ }\textbf {\bibinfo
  {volume} {63}},\ \bibinfo {pages} {2148--2151} (\bibinfo {year}
  {1989})}\BibitemShut {NoStop}%
\bibitem [{\citenamefont {Gelfand}\ \emph {et~al.}(1989)\citenamefont
  {Gelfand}, \citenamefont {Singh},\ and\ \citenamefont {Huse}}]{Gelfand:1989}%
  \BibitemOpen
  \bibfield  {author} {\bibinfo {author} {\bibfnamefont {Martin~P.}\
  \bibnamefont {Gelfand}}, \bibinfo {author} {\bibfnamefont {Rajiv R.~P.}\
  \bibnamefont {Singh}}, \ and\ \bibinfo {author} {\bibfnamefont {David~A.}\
  \bibnamefont {Huse}},\ }\bibfield  {title} {\enquote {\bibinfo {title}
  {Zero-temperature ordering in two-dimensional frustrated quantum {H}eisenberg
  antiferromagnets},}\ }\href {\doibase 10.1103/PhysRevB.40.10801} {\bibfield
  {journal} {\bibinfo  {journal} {Phys. Rev. B}\ }\textbf {\bibinfo {volume}
  {40}},\ \bibinfo {pages} {10801--10809} (\bibinfo {year} {1989})}\BibitemShut
  {NoStop}%
\bibitem [{\citenamefont {Sachdev}\ and\ \citenamefont
  {Bhatt}(1990)}]{Sachdev:1990}%
  \BibitemOpen
  \bibfield  {author} {\bibinfo {author} {\bibfnamefont {Subir}\ \bibnamefont
  {Sachdev}}\ and\ \bibinfo {author} {\bibfnamefont {R.~N.}\ \bibnamefont
  {Bhatt}},\ }\bibfield  {title} {\enquote {\bibinfo {title} {Bond-operator
  representation of quantum spins: Mean-field theory of frustrated quantum
  {H}eisenberg antiferromagnets},}\ }\href {\doibase 10.1103/PhysRevB.41.9323}
  {\bibfield  {journal} {\bibinfo  {journal} {Phys. Rev. B}\ }\textbf {\bibinfo
  {volume} {41}},\ \bibinfo {pages} {9323--9329} (\bibinfo {year}
  {1990})}\BibitemShut {NoStop}%
\bibitem [{\citenamefont {Chubukov}\ and\ \citenamefont
  {Jolicoeur}(1991)}]{Chubukov:1991}%
  \BibitemOpen
  \bibfield  {author} {\bibinfo {author} {\bibfnamefont {Andrey~V.}\
  \bibnamefont {Chubukov}}\ and\ \bibinfo {author} {\bibfnamefont {{\relax
  Th}.}~\bibnamefont {Jolicoeur}},\ }\bibfield  {title} {\enquote {\bibinfo
  {title} {Dimer stability region in a frustrated quantum {H}eisenberg
  antiferromagnet},}\ }\href {\doibase 10.1103/PhysRevB.44.12050} {\bibfield
  {journal} {\bibinfo  {journal} {Phys. Rev. B}\ }\textbf {\bibinfo {volume}
  {44}},\ \bibinfo {pages} {12050(R)--12053(R)} (\bibinfo {year}
  {1991})}\BibitemShut {NoStop}%
\bibitem [{\citenamefont {Read}\ and\ \citenamefont
  {Sachdev}(1991)}]{Read:1991}%
  \BibitemOpen
  \bibfield  {author} {\bibinfo {author} {\bibfnamefont {N.}~\bibnamefont
  {Read}}\ and\ \bibinfo {author} {\bibfnamefont {Subir}\ \bibnamefont
  {Sachdev}},\ }\bibfield  {title} {\enquote {\bibinfo {title} {Large-${N}$
  expansion for frustrated quantum antiferromagnets},}\ }\href {\doibase
  10.1103/PhysRevLett.66.1773} {\bibfield  {journal} {\bibinfo  {journal}
  {Phys. Rev. Lett.}\ }\textbf {\bibinfo {volume} {66}},\ \bibinfo {pages}
  {1773--1776} (\bibinfo {year} {1991})}\BibitemShut {NoStop}%
\bibitem [{\citenamefont {Richter}(1993)}]{Richter:1993}%
  \BibitemOpen
  \bibfield  {author} {\bibinfo {author} {\bibfnamefont {Johannes}\
  \bibnamefont {Richter}},\ }\bibfield  {title} {\enquote {\bibinfo {title}
  {Zero-temperature magnetic ordering in the inhomogeneously frustrated quantum
  {H}eisenberg antiferromagnet on a square lattice},}\ }\href {\doibase
  10.1103/PhysRevB.47.5794} {\bibfield  {journal} {\bibinfo  {journal} {Phys.
  Rev. B}\ }\textbf {\bibinfo {volume} {47}},\ \bibinfo {pages} {5794--5804}
  (\bibinfo {year} {1993})}\BibitemShut {NoStop}%
\bibitem [{\citenamefont {Richter}\ \emph {et~al.}(1994)\citenamefont
  {Richter}, \citenamefont {Ivanov},\ and\ \citenamefont
  {Retzlaff}}]{Richter:1994}%
  \BibitemOpen
  \bibfield  {author} {\bibinfo {author} {\bibfnamefont {J.}~\bibnamefont
  {Richter}}, \bibinfo {author} {\bibfnamefont {N.~B.}\ \bibnamefont {Ivanov}},
  \ and\ \bibinfo {author} {\bibfnamefont {K.}~\bibnamefont {Retzlaff}},\
  }\bibfield  {title} {\enquote {\bibinfo {title} {On the violation of
  {M}arshall-{P}eierls sign rule in the frustrated ${J}_{1}$-${J}_{2}$
  {H}eisenberg antiferromagnet},}\ }\href {\doibase 10.1209/0295-5075/25/7/012}
  {\bibfield  {journal} {\bibinfo  {journal} {Europhys. Lett.}\ }\textbf
  {\bibinfo {volume} {25}},\ \bibinfo {pages} {545--550} (\bibinfo {year}
  {1994})}\BibitemShut {NoStop}%
\bibitem [{\citenamefont {Ivanov}\ and\ \citenamefont
  {Richter}(1994)}]{Ivanov:1994_J1J2mod}%
  \BibitemOpen
  \bibfield  {author} {\bibinfo {author} {\bibfnamefont {N.~B.}\ \bibnamefont
  {Ivanov}}\ and\ \bibinfo {author} {\bibfnamefont {J.}~\bibnamefont
  {Richter}},\ }\bibfield  {title} {\enquote {\bibinfo {title}
  {${J}_{1}$-${J}_{2}$ quantum {H}eisenberg antiferromagnet: improved spin-wave
  theories versus exact-diagonalization data},}\ }\href {\doibase
  10.1088/0953-8984/6/20/018} {\bibfield  {journal} {\bibinfo  {journal} {J.
  Phys.: Condens. Matter}\ }\textbf {\bibinfo {volume} {6}},\ \bibinfo {pages}
  {3785--3792} (\bibinfo {year} {1994})}\BibitemShut {NoStop}%
\bibitem [{\citenamefont {Schulz}\ \emph {et~al.}(1996)\citenamefont {Schulz},
  \citenamefont {Ziman},\ and\ \citenamefont {Poilblanc}}]{Schulz:1996}%
  \BibitemOpen
  \bibfield  {author} {\bibinfo {author} {\bibfnamefont {H.~J.}\ \bibnamefont
  {Schulz}}, \bibinfo {author} {\bibfnamefont {T.~A.~L.}\ \bibnamefont
  {Ziman}}, \ and\ \bibinfo {author} {\bibfnamefont {D.}~\bibnamefont
  {Poilblanc}},\ }\bibfield  {title} {\enquote {\bibinfo {title} {Magnetic
  order and disorder in the frustrated quantum {H}eisenberg antiferromagnet in
  two dimensions},}\ }\href {\doibase 10.1051/jp1:1996236} {\bibfield
  {journal} {\bibinfo  {journal} {J. Phys. I France}\ }\textbf {\bibinfo
  {volume} {6}},\ \bibinfo {pages} {675--703} (\bibinfo {year}
  {1996})}\BibitemShut {NoStop}%
\bibitem [{\citenamefont {Oitmaa}\ and\ \citenamefont
  {Weihong}(1996)}]{Oitmaa:1996}%
  \BibitemOpen
  \bibfield  {author} {\bibinfo {author} {\bibfnamefont {J.}~\bibnamefont
  {Oitmaa}}\ and\ \bibinfo {author} {\bibfnamefont {Zheng}\ \bibnamefont
  {Weihong}},\ }\bibfield  {title} {\enquote {\bibinfo {title} {Series
  expansion for the ${J}_{1}$-${J}_{2}$ {H}eisenberg antiferromagnet on a
  square lattice},}\ }\href {\doibase 10.1103/PhysRevB.54.3022} {\bibfield
  {journal} {\bibinfo  {journal} {Phys. Rev. B}\ }\textbf {\bibinfo {volume}
  {54}},\ \bibinfo {pages} {3022--3025} (\bibinfo {year} {1996})}\BibitemShut
  {NoStop}%
\bibitem [{\citenamefont {Zhitomirsky}\ and\ \citenamefont
  {Ueda}(1996)}]{Zhitomirsky:1996}%
  \BibitemOpen
  \bibfield  {author} {\bibinfo {author} {\bibfnamefont {M.~E.}\ \bibnamefont
  {Zhitomirsky}}\ and\ \bibinfo {author} {\bibfnamefont {Kazuo}\ \bibnamefont
  {Ueda}},\ }\bibfield  {title} {\enquote {\bibinfo {title} {Valence-bond
  crystal phase of a frustrated spin-$\frac{1}{2}$ square-lattice
  antiferromagnet},}\ }\href {\doibase 10.1103/PhysRevB.54.9007} {\bibfield
  {journal} {\bibinfo  {journal} {Phys. Rev. B}\ }\textbf {\bibinfo {volume}
  {54}},\ \bibinfo {pages} {9007--9010} (\bibinfo {year} {1996})}\BibitemShut
  {NoStop}%
\bibitem [{\citenamefont {Trumper}\ \emph {et~al.}(1997)\citenamefont
  {Trumper}, \citenamefont {Manuel}, \citenamefont {Gazza},\ and\ \citenamefont
  {Ceccatto}}]{Trumper:1997_J1J2mod}%
  \BibitemOpen
  \bibfield  {author} {\bibinfo {author} {\bibfnamefont {A.~E.}\ \bibnamefont
  {Trumper}}, \bibinfo {author} {\bibfnamefont {L.~O.}\ \bibnamefont {Manuel}},
  \bibinfo {author} {\bibfnamefont {C.~J.}\ \bibnamefont {Gazza}}, \ and\
  \bibinfo {author} {\bibfnamefont {H.~A.}\ \bibnamefont {Ceccatto}},\
  }\bibfield  {title} {\enquote {\bibinfo {title} {Schwinger-boson approach to
  quantum spin systems: Gaussian fluctuations in the ``natural'' gauge},}\
  }\href {\doibase 10.1103/PhysRevLett.78.2216} {\bibfield  {journal} {\bibinfo
   {journal} {Phys. Rev. Lett.}\ }\textbf {\bibinfo {volume} {78}},\ \bibinfo
  {pages} {2216--2219} (\bibinfo {year} {1997})}\BibitemShut {NoStop}%
\bibitem [{\citenamefont {Bishop}\ \emph {et~al.}(1998)\citenamefont {Bishop},
  \citenamefont {Farnell},\ and\ \citenamefont
  {Parkinson}}]{Bishop:1998_J1J2mod}%
  \BibitemOpen
  \bibfield  {author} {\bibinfo {author} {\bibfnamefont {R.~F.}\ \bibnamefont
  {Bishop}}, \bibinfo {author} {\bibfnamefont {D.~J.~J.}\ \bibnamefont
  {Farnell}}, \ and\ \bibinfo {author} {\bibfnamefont {J.~B.}\ \bibnamefont
  {Parkinson}},\ }\bibfield  {title} {\enquote {\bibinfo {title} {Phase
  transitions in the spin-half ${J}_{1}$--${J}_{2}$ model},}\ }\href {\doibase
  10.1103/PhysRevB.58.6394} {\bibfield  {journal} {\bibinfo  {journal} {Phys.
  Rev. B}\ }\textbf {\bibinfo {volume} {58}},\ \bibinfo {pages} {6394--6402}
  (\bibinfo {year} {1998})}\BibitemShut {NoStop}%
\bibitem [{\citenamefont {Singh}\ \emph {et~al.}(1999)\citenamefont {Singh},
  \citenamefont {Weihong}, \citenamefont {Hamer},\ and\ \citenamefont
  {Oitmaa}}]{Singh:1999}%
  \BibitemOpen
  \bibfield  {author} {\bibinfo {author} {\bibfnamefont {Rajiv R.~P.}\
  \bibnamefont {Singh}}, \bibinfo {author} {\bibfnamefont {Zheng}\ \bibnamefont
  {Weihong}}, \bibinfo {author} {\bibfnamefont {C.~J.}\ \bibnamefont {Hamer}},
  \ and\ \bibinfo {author} {\bibfnamefont {J.}~\bibnamefont {Oitmaa}},\
  }\bibfield  {title} {\enquote {\bibinfo {title} {Dimer order with striped
  correlations in the ${J}_{1}$-${J}_{2}$ {H}eisenberg model},}\ }\href
  {\doibase 10.1103/PhysRevB.60.7278} {\bibfield  {journal} {\bibinfo
  {journal} {Phys. Rev. B}\ }\textbf {\bibinfo {volume} {60}},\ \bibinfo
  {pages} {7278--7283} (\bibinfo {year} {1999})}\BibitemShut {NoStop}%
\bibitem [{\citenamefont {Kotov}\ \emph {et~al.}(1999)\citenamefont {Kotov},
  \citenamefont {Oitmaa}, \citenamefont {Sushkov},\ and\ \citenamefont
  {Weihong}}]{Kotov:1999}%
  \BibitemOpen
  \bibfield  {author} {\bibinfo {author} {\bibfnamefont {Valeri~N.}\
  \bibnamefont {Kotov}}, \bibinfo {author} {\bibfnamefont {J.}~\bibnamefont
  {Oitmaa}}, \bibinfo {author} {\bibfnamefont {Oleg~P.}\ \bibnamefont
  {Sushkov}}, \ and\ \bibinfo {author} {\bibfnamefont {Zheng}\ \bibnamefont
  {Weihong}},\ }\bibfield  {title} {\enquote {\bibinfo {title} {Low-energy
  singlet and triplet excitations in the spin-liquid phase of the
  two-dimensional ${J}_{1}$-${J}_{2}$ model},}\ }\href {\doibase
  10.1103/PhysRevB.60.14613} {\bibfield  {journal} {\bibinfo  {journal} {Phys.
  Rev. B}\ }\textbf {\bibinfo {volume} {60}},\ \bibinfo {pages} {14613--14616}
  (\bibinfo {year} {1999})}\BibitemShut {NoStop}%
\bibitem [{\citenamefont {Capriotti}\ and\ \citenamefont
  {Sorella}(2000)}]{Capriotti:2000}%
  \BibitemOpen
  \bibfield  {author} {\bibinfo {author} {\bibfnamefont {Luca}\ \bibnamefont
  {Capriotti}}\ and\ \bibinfo {author} {\bibfnamefont {Sandro}\ \bibnamefont
  {Sorella}},\ }\bibfield  {title} {\enquote {\bibinfo {title} {Spontaneous
  plaquette dimerization in the ${J}_{1}$--${J}_{2}$ {H}eisenberg model},}\
  }\href {\doibase 10.1103/PhysRevLett.84.3173} {\bibfield  {journal} {\bibinfo
   {journal} {Phys. Rev. Lett.}\ }\textbf {\bibinfo {volume} {84}},\ \bibinfo
  {pages} {3173--3176} (\bibinfo {year} {2000})}\BibitemShut {NoStop}%
\bibitem [{\citenamefont {Capriotti}\ \emph {et~al.}(2001)\citenamefont
  {Capriotti}, \citenamefont {Becca}, \citenamefont {Parola},\ and\
  \citenamefont {Sorella}}]{Capriotti:2001}%
  \BibitemOpen
  \bibfield  {author} {\bibinfo {author} {\bibfnamefont {Luca}\ \bibnamefont
  {Capriotti}}, \bibinfo {author} {\bibfnamefont {Federico}\ \bibnamefont
  {Becca}}, \bibinfo {author} {\bibfnamefont {Alberto}\ \bibnamefont {Parola}},
  \ and\ \bibinfo {author} {\bibfnamefont {Sandro}\ \bibnamefont {Sorella}},\
  }\bibfield  {title} {\enquote {\bibinfo {title} {Resonating valence bond wave
  functions for strongly frustrated spin systems},}\ }\href {\doibase
  10.1103/PhysRevLett.87.097201} {\bibfield  {journal} {\bibinfo  {journal}
  {Phys. Rev. Lett.}\ }\textbf {\bibinfo {volume} {87}},\ \bibinfo {pages}
  {097201} (\bibinfo {year} {2001})}\BibitemShut {NoStop}%
\bibitem [{\citenamefont {Takano}\ \emph {et~al.}(2003)\citenamefont {Takano},
  \citenamefont {Kito}, \citenamefont {{\={O}}no},\ and\ \citenamefont
  {Sano}}]{Takano:2003}%
  \BibitemOpen
  \bibfield  {author} {\bibinfo {author} {\bibfnamefont {Ken'ichi}\
  \bibnamefont {Takano}}, \bibinfo {author} {\bibfnamefont {Yoshiya}\
  \bibnamefont {Kito}}, \bibinfo {author} {\bibfnamefont {Yoshiaki}\
  \bibnamefont {{\={O}}no}}, \ and\ \bibinfo {author} {\bibfnamefont
  {Kazuhiro}\ \bibnamefont {Sano}},\ }\bibfield  {title} {\enquote {\bibinfo
  {title} {Nonlinear $\sigma$ model method for the ${J}_{1}$-${J}_{2}$
  {H}eisenberg model: Disordered ground state with plaquette symmetry},}\
  }\href {\doibase 10.1103/PhysRevLett.91.197202} {\bibfield  {journal}
  {\bibinfo  {journal} {Phys. Rev. Lett.}\ }\textbf {\bibinfo {volume} {91}},\
  \bibinfo {pages} {197202} (\bibinfo {year} {2003})}\BibitemShut {NoStop}%
\bibitem [{\citenamefont {Roscilde}\ \emph {et~al.}(2004)\citenamefont
  {Roscilde}, \citenamefont {Feiguin}, \citenamefont {Chernyshev},
  \citenamefont {Liu},\ and\ \citenamefont {Haas}}]{Roscilde:2004}%
  \BibitemOpen
  \bibfield  {author} {\bibinfo {author} {\bibfnamefont {Tommaso}\ \bibnamefont
  {Roscilde}}, \bibinfo {author} {\bibfnamefont {Adrian}\ \bibnamefont
  {Feiguin}}, \bibinfo {author} {\bibfnamefont {Alexande~L.}\ \bibnamefont
  {Chernyshev}}, \bibinfo {author} {\bibfnamefont {Shiu}\ \bibnamefont {Liu}},
  \ and\ \bibinfo {author} {\bibfnamefont {Stephan}\ \bibnamefont {Haas}},\
  }\bibfield  {title} {\enquote {\bibinfo {title} {Anisotropy-induced ordering
  in the quantum ${J}_{1}$-${J}_{2}$ antiferromagnet},}\ }\href {\doibase
  10.1103/PhysRevLett.93.017203} {\bibfield  {journal} {\bibinfo  {journal}
  {Phys. Rev. Lett.}\ }\textbf {\bibinfo {volume} {93}},\ \bibinfo {pages}
  {017203} (\bibinfo {year} {2004})}\BibitemShut {NoStop}%
\bibitem [{\citenamefont {Lante}\ and\ \citenamefont
  {Parola}(2006)}]{Lante:2006_J1J2mod_sqLatt}%
  \BibitemOpen
  \bibfield  {author} {\bibinfo {author} {\bibfnamefont {Valeria}\ \bibnamefont
  {Lante}}\ and\ \bibinfo {author} {\bibfnamefont {Alberto}\ \bibnamefont
  {Parola}},\ }\bibfield  {title} {\enquote {\bibinfo {title} {Ising phase in
  the ${J}_{1}$-${J}_{2}$ {H}eisenberg model},}\ }\href {\doibase
  10.1103/PhysRevB.73.094427} {\bibfield  {journal} {\bibinfo  {journal} {Phys.
  Rev. B}\ }\textbf {\bibinfo {volume} {73}},\ \bibinfo {pages} {094427}
  (\bibinfo {year} {2006})}\BibitemShut {NoStop}%
\bibitem [{\citenamefont {Sirker}\ \emph {et~al.}(2006)\citenamefont {Sirker},
  \citenamefont {Weihong}, \citenamefont {Sushkov},\ and\ \citenamefont
  {Oitmaa}}]{Sirker:2006}%
  \BibitemOpen
  \bibfield  {author} {\bibinfo {author} {\bibfnamefont {J.}~\bibnamefont
  {Sirker}}, \bibinfo {author} {\bibfnamefont {Zheng}\ \bibnamefont {Weihong}},
  \bibinfo {author} {\bibfnamefont {O.~P.}\ \bibnamefont {Sushkov}}, \ and\
  \bibinfo {author} {\bibfnamefont {J.}~\bibnamefont {Oitmaa}},\ }\bibfield
  {title} {\enquote {\bibinfo {title} {${J}1$--${J}2$ model: First-order phase
  transition versus deconfinement of spinons},}\ }\href {\doibase
  10.1103/PhysRevB.73.184420} {\bibfield  {journal} {\bibinfo  {journal} {Phys.
  Rev. B}\ }\textbf {\bibinfo {volume} {73}},\ \bibinfo {pages} {184420}
  (\bibinfo {year} {2006})}\BibitemShut {NoStop}%
\bibitem [{\citenamefont {Schmalfu{\ss}}\ \emph {et~al.}(2006)\citenamefont
  {Schmalfu{\ss}}, \citenamefont {Darradi}, \citenamefont {Richter},
  \citenamefont {Schulenburg},\ and\ \citenamefont
  {Ihle}}]{Schm:2006_stackSqLatt}%
  \BibitemOpen
  \bibfield  {author} {\bibinfo {author} {\bibfnamefont {D.}~\bibnamefont
  {Schmalfu{\ss}}}, \bibinfo {author} {\bibfnamefont {R.}~\bibnamefont
  {Darradi}}, \bibinfo {author} {\bibfnamefont {J.}~\bibnamefont {Richter}},
  \bibinfo {author} {\bibfnamefont {J.}~\bibnamefont {Schulenburg}}, \ and\
  \bibinfo {author} {\bibfnamefont {D.}~\bibnamefont {Ihle}},\ }\bibfield
  {title} {\enquote {\bibinfo {title} {Quantum ${J}_{1}$--${J}_{2}$
  antiferromagnet on a stacked square lattice: Influence of the interlayer
  coupling on the ground-state magnetic ordering},}\ }\href {\doibase
  10.1103/PhysRevLett.97.157201} {\bibfield  {journal} {\bibinfo  {journal}
  {Phys. Rev. Lett.}\ }\textbf {\bibinfo {volume} {97}},\ \bibinfo {pages}
  {157201} (\bibinfo {year} {2006})}\BibitemShut {NoStop}%
\bibitem [{\citenamefont {Mambrini}\ \emph {et~al.}(2006)\citenamefont
  {Mambrini}, \citenamefont {L{\"{a}}uchli}, \citenamefont {Poilblanc},\ and\
  \citenamefont {Mila}}]{Mambrini:2006}%
  \BibitemOpen
  \bibfield  {author} {\bibinfo {author} {\bibfnamefont {Matthieu}\
  \bibnamefont {Mambrini}}, \bibinfo {author} {\bibfnamefont {Andreas}\
  \bibnamefont {L{\"{a}}uchli}}, \bibinfo {author} {\bibfnamefont {Didier}\
  \bibnamefont {Poilblanc}}, \ and\ \bibinfo {author} {\bibfnamefont
  {Fr{\'{e}}d{\'{e}}ric}\ \bibnamefont {Mila}},\ }\bibfield  {title} {\enquote
  {\bibinfo {title} {Plaquette valence-bond crystal in the frustrated
  {H}eisenberg quantum antiferromagnet on the square lattice},}\ }\href
  {\doibase 10.1103/PhysRevB.74.144422} {\bibfield  {journal} {\bibinfo
  {journal} {Phys. Rev. B}\ }\textbf {\bibinfo {volume} {74}},\ \bibinfo
  {pages} {144422} (\bibinfo {year} {2006})}\BibitemShut {NoStop}%
\bibitem [{\citenamefont {Bishop}\ \emph
  {et~al.}(2008{\natexlab{a}})\citenamefont {Bishop}, \citenamefont {Li},
  \citenamefont {Darradi},\ and\ \citenamefont
  {Richter}}]{Bi:2008_JPCM_J1J1primeJ2}%
  \BibitemOpen
  \bibfield  {author} {\bibinfo {author} {\bibfnamefont {R.~F.}\ \bibnamefont
  {Bishop}}, \bibinfo {author} {\bibfnamefont {P.~H.~Y.}\ \bibnamefont {Li}},
  \bibinfo {author} {\bibfnamefont {R.}~\bibnamefont {Darradi}}, \ and\
  \bibinfo {author} {\bibfnamefont {J.}~\bibnamefont {Richter}},\ }\bibfield
  {title} {\enquote {\bibinfo {title} {The quantum
  ${J}_{1}$--${J}_{1}'$--${J}_{2}$ spin-$1/2$ {H}eisenberg model: influence of
  the interchain coupling on the ground-state magnetic ordering in two
  dimensions},}\ }\href {\doibase 10.1088/0953-8984/20/25/255251} {\bibfield
  {journal} {\bibinfo  {journal} {J. Phys.: Condens. Matter}\ }\textbf
  {\bibinfo {volume} {20}},\ \bibinfo {pages} {255251} (\bibinfo {year}
  {2008}{\natexlab{a}})}\BibitemShut {NoStop}%
\bibitem [{\citenamefont {Bishop}\ \emph
  {et~al.}(2008{\natexlab{b}})\citenamefont {Bishop}, \citenamefont {Li},
  \citenamefont {Darradi}, \citenamefont {Schulenburg},\ and\ \citenamefont
  {Richter}}]{Bi:2008_PRB_J1xxzJ2xxz}%
  \BibitemOpen
  \bibfield  {author} {\bibinfo {author} {\bibfnamefont {R.~F.}\ \bibnamefont
  {Bishop}}, \bibinfo {author} {\bibfnamefont {P.~H.~Y.}\ \bibnamefont {Li}},
  \bibinfo {author} {\bibfnamefont {R.}~\bibnamefont {Darradi}}, \bibinfo
  {author} {\bibfnamefont {J.}~\bibnamefont {Schulenburg}}, \ and\ \bibinfo
  {author} {\bibfnamefont {J.}~\bibnamefont {Richter}},\ }\bibfield  {title}
  {\enquote {\bibinfo {title} {Effect of anisotropy on the ground-state
  magnetic ordering of the spin-half quantum ${J}_{1}^{XXZ}$--${J}_{2}^{XXZ}$
  model on the square lattice},}\ }\href {\doibase 10.1103/PhysRevB.78.054412}
  {\bibfield  {journal} {\bibinfo  {journal} {Phys. Rev. B}\ }\textbf {\bibinfo
  {volume} {78}},\ \bibinfo {pages} {054412} (\bibinfo {year}
  {2008}{\natexlab{b}})}\BibitemShut {NoStop}%
\bibitem [{\citenamefont {Darradi}\ \emph {et~al.}(2008)\citenamefont
  {Darradi}, \citenamefont {Derzhko}, \citenamefont {Zinke}, \citenamefont
  {Schulenburg}, \citenamefont {Kr{\"{u}}ger},\ and\ \citenamefont
  {Richter}}]{Darradi:2008_J1J2mod}%
  \BibitemOpen
  \bibfield  {author} {\bibinfo {author} {\bibfnamefont {R.}~\bibnamefont
  {Darradi}}, \bibinfo {author} {\bibfnamefont {O.}~\bibnamefont {Derzhko}},
  \bibinfo {author} {\bibfnamefont {R.}~\bibnamefont {Zinke}}, \bibinfo
  {author} {\bibfnamefont {J.}~\bibnamefont {Schulenburg}}, \bibinfo {author}
  {\bibfnamefont {S.~E.}\ \bibnamefont {Kr{\"{u}}ger}}, \ and\ \bibinfo
  {author} {\bibfnamefont {J.}~\bibnamefont {Richter}},\ }\bibfield  {title}
  {\enquote {\bibinfo {title} {Ground state phases of the spin-1/2
  ${J}_{1}$--${J}_{2}$ {H}eisenberg antiferromagnet on the square lattice: A
  high-order coupled cluster treatment},}\ }\href {\doibase
  10.1103/PhysRevB.78.214415} {\bibfield  {journal} {\bibinfo  {journal} {Phys.
  Rev. B}\ }\textbf {\bibinfo {volume} {78}},\ \bibinfo {pages} {214415}
  (\bibinfo {year} {2008})}\BibitemShut {NoStop}%
\bibitem [{\citenamefont {Isaev}\ \emph {et~al.}(2009)\citenamefont {Isaev},
  \citenamefont {Ortiz},\ and\ \citenamefont {Dukelsky}}]{Isaev:2009_J1J2mod}%
  \BibitemOpen
  \bibfield  {author} {\bibinfo {author} {\bibfnamefont {L.}~\bibnamefont
  {Isaev}}, \bibinfo {author} {\bibfnamefont {G.}~\bibnamefont {Ortiz}}, \ and\
  \bibinfo {author} {\bibfnamefont {J.}~\bibnamefont {Dukelsky}},\ }\bibfield
  {title} {\enquote {\bibinfo {title} {Hierarchical mean-field approach to the
  ${J}_{1}$-${J}_{2}$ {H}eisenberg model on a square lattice},}\ }\href
  {\doibase 10.1103/PhysRevB.79.024409} {\bibfield  {journal} {\bibinfo
  {journal} {Phys. Rev. B}\ }\textbf {\bibinfo {volume} {79}},\ \bibinfo
  {pages} {024409} (\bibinfo {year} {2009})}\BibitemShut {NoStop}%
\bibitem [{\citenamefont {Murg}\ \emph {et~al.}(2009)\citenamefont {Murg},
  \citenamefont {Verstraete},\ and\ \citenamefont {Cirac}}]{Murg:2009_peps}%
  \BibitemOpen
  \bibfield  {author} {\bibinfo {author} {\bibfnamefont {V.}~\bibnamefont
  {Murg}}, \bibinfo {author} {\bibfnamefont {F.}~\bibnamefont {Verstraete}}, \
  and\ \bibinfo {author} {\bibfnamefont {J.~I.}\ \bibnamefont {Cirac}},\
  }\bibfield  {title} {\enquote {\bibinfo {title} {Exploring frustrated spin
  systems using projected entangled pair states},}\ }\href {\doibase
  10.1103/PhysRevB.79.195119} {\bibfield  {journal} {\bibinfo  {journal} {Phys.
  Rev. B}\ }\textbf {\bibinfo {volume} {79}},\ \bibinfo {pages} {195119}
  (\bibinfo {year} {2009})}\BibitemShut {NoStop}%
\bibitem [{\citenamefont {Ralko}\ \emph {et~al.}(2009)\citenamefont {Ralko},
  \citenamefont {Mambrini},\ and\ \citenamefont {Poilblanc}}]{Ralko:2009}%
  \BibitemOpen
  \bibfield  {author} {\bibinfo {author} {\bibfnamefont {A.}~\bibnamefont
  {Ralko}}, \bibinfo {author} {\bibfnamefont {M.}~\bibnamefont {Mambrini}}, \
  and\ \bibinfo {author} {\bibfnamefont {D.}~\bibnamefont {Poilblanc}},\
  }\bibfield  {title} {\enquote {\bibinfo {title} {Generalized quantum dimer
  model applied to the frustrated {H}eisenberg model on the square lattice:
  Emergence of a mixed columnar-plaquette phase},}\ }\href {\doibase
  10.1103/PhysRevB.80.184427} {\bibfield  {journal} {\bibinfo  {journal} {Phys.
  Rev. B}\ }\textbf {\bibinfo {volume} {80}},\ \bibinfo {pages} {184427}
  (\bibinfo {year} {2009})}\BibitemShut {NoStop}%
\bibitem [{\citenamefont {Richter}\ and\ \citenamefont
  {Schulenburg}(2010)}]{Richter:2010_ED40}%
  \BibitemOpen
  \bibfield  {author} {\bibinfo {author} {\bibfnamefont {J.}~\bibnamefont
  {Richter}}\ and\ \bibinfo {author} {\bibfnamefont {J.}~\bibnamefont
  {Schulenburg}},\ }\bibfield  {title} {\enquote {\bibinfo {title} {The
  spin-1/2 ${J}_{1}$--${J}_{2}$ {H}eisenberg antiferromagnet on the square
  lattice: Exact diagonalization for ${N} = 40$ spins},}\ }\href {\doibase
  10.1140/epjb/e2009-00400-4} {\bibfield  {journal} {\bibinfo  {journal} {Eur.
  Phys. J. B}\ }\textbf {\bibinfo {volume} {73}},\ \bibinfo {pages} {117--124}
  (\bibinfo {year} {2010})}\BibitemShut {NoStop}%
\bibitem [{\citenamefont {Reuther}\ and\ \citenamefont
  {W{\"{o}}lfle}(2010)}]{Reuther:2010_J1J2mod}%
  \BibitemOpen
  \bibfield  {author} {\bibinfo {author} {\bibfnamefont {Johannes}\
  \bibnamefont {Reuther}}\ and\ \bibinfo {author} {\bibfnamefont {Peter}\
  \bibnamefont {W{\"{o}}lfle}},\ }\bibfield  {title} {\enquote {\bibinfo
  {title} {${J}_{1}$-${J}_{2}$ frustrated two-dimensional {H}eisenberg model:
  Random phase approximation and functional renormalization group},}\ }\href
  {\doibase 10.1103/PhysRevB.81.144410} {\bibfield  {journal} {\bibinfo
  {journal} {Phys. Rev. B}\ }\textbf {\bibinfo {volume} {81}},\ \bibinfo
  {pages} {144410} (\bibinfo {year} {2010})}\BibitemShut {NoStop}%
\bibitem [{\citenamefont {Reuther}\ \emph {et~al.}(2011)\citenamefont
  {Reuther}, \citenamefont {W{{\"o}}lfle}, \citenamefont {Darradi},
  \citenamefont {Brenig}, \citenamefont {Arlego},\ and\ \citenamefont
  {Richter}}]{Reuther:2011_J1J2J3mod}%
  \BibitemOpen
  \bibfield  {author} {\bibinfo {author} {\bibfnamefont {Johannes}\
  \bibnamefont {Reuther}}, \bibinfo {author} {\bibfnamefont {Peter}\
  \bibnamefont {W{{\"o}}lfle}}, \bibinfo {author} {\bibfnamefont {Rachid}\
  \bibnamefont {Darradi}}, \bibinfo {author} {\bibfnamefont {Wolfram}\
  \bibnamefont {Brenig}}, \bibinfo {author} {\bibfnamefont {Marcelo}\
  \bibnamefont {Arlego}}, \ and\ \bibinfo {author} {\bibfnamefont {Johannes}\
  \bibnamefont {Richter}},\ }\bibfield  {title} {\enquote {\bibinfo {title}
  {Quantum phases of the planar antiferromagnetic
  ${J}_{1}$--${J}_{2}$--${J}_{3}$ {H}eisenberg model},}\ }\href {\doibase
  10.1103/PhysRevB.83.064416} {\bibfield  {journal} {\bibinfo  {journal} {Phys.
  Rev. B}\ }\textbf {\bibinfo {volume} {83}},\ \bibinfo {pages} {064416}
  (\bibinfo {year} {2011})}\BibitemShut {NoStop}%
\bibitem [{\citenamefont {Yu}\ and\ \citenamefont {Kao}(2012)}]{Yu:2012}%
  \BibitemOpen
  \bibfield  {author} {\bibinfo {author} {\bibfnamefont {Ji-Feng}\ \bibnamefont
  {Yu}}\ and\ \bibinfo {author} {\bibfnamefont {Ying-Jer}\ \bibnamefont
  {Kao}},\ }\bibfield  {title} {\enquote {\bibinfo {title} {Spin-$\frac{1}{2}$
  ${J}_{1}$-${J}_{2}$ {H}eisenberg antiferromagnet on a square lattice: A
  plaquette renormalized tensor network study},}\ }\href {\doibase
  10.1103/PhysRevB.85.094407} {\bibfield  {journal} {\bibinfo  {journal} {Phys.
  Rev. B}\ }\textbf {\bibinfo {volume} {85}},\ \bibinfo {pages} {094407}
  (\bibinfo {year} {2012})}\BibitemShut {NoStop}%
\bibitem [{\citenamefont {G{{\"o}}tze}\ \emph {et~al.}(2012)\citenamefont
  {G{{\"o}}tze}, \citenamefont {Kr{\"{u}}ger}, \citenamefont {Fleck},
  \citenamefont {Schulenburg},\ and\ \citenamefont {Richter}}]{Gotze:2012}%
  \BibitemOpen
  \bibfield  {author} {\bibinfo {author} {\bibfnamefont {O.}~\bibnamefont
  {G{{\"o}}tze}}, \bibinfo {author} {\bibfnamefont {S.~E.}\ \bibnamefont
  {Kr{\"{u}}ger}}, \bibinfo {author} {\bibfnamefont {F.}~\bibnamefont {Fleck}},
  \bibinfo {author} {\bibfnamefont {J.}~\bibnamefont {Schulenburg}}, \ and\
  \bibinfo {author} {\bibfnamefont {J.}~\bibnamefont {Richter}},\ }\bibfield
  {title} {\enquote {\bibinfo {title} {Ground-state phase diagram of the
  spin-$\frac{1}{2}$ square-lattice ${J}_{1}$-${J}_{2}$ model with plaquette
  structure},}\ }\href {\doibase 10.1103/PhysRevB.85.224424} {\bibfield
  {journal} {\bibinfo  {journal} {Phys. Rev. B}\ }\textbf {\bibinfo {volume}
  {85}},\ \bibinfo {pages} {224424} (\bibinfo {year} {2012})}\BibitemShut
  {NoStop}%
\bibitem [{\citenamefont {Jiang}\ \emph {et~al.}(2012)\citenamefont {Jiang},
  \citenamefont {Yao},\ and\ \citenamefont {Balents}}]{Jiang:2012}%
  \BibitemOpen
  \bibfield  {author} {\bibinfo {author} {\bibfnamefont {Hong-Chen}\
  \bibnamefont {Jiang}}, \bibinfo {author} {\bibfnamefont {Hong}\ \bibnamefont
  {Yao}}, \ and\ \bibinfo {author} {\bibfnamefont {Leon}\ \bibnamefont
  {Balents}},\ }\bibfield  {title} {\enquote {\bibinfo {title} {Spin liquid
  ground state of the spin-$\frac{1}{2}$ square ${J}_{1}$-${J}_{2}$
  {H}eisenberg model},}\ }\href {\doibase 10.1103/PhysRevB.86.024424}
  {\bibfield  {journal} {\bibinfo  {journal} {Phys. Rev. B}\ }\textbf {\bibinfo
  {volume} {86}},\ \bibinfo {pages} {024424} (\bibinfo {year}
  {2012})}\BibitemShut {NoStop}%
\bibitem [{\citenamefont {Mezzacapo}(2012)}]{Mezzacapo:2012}%
  \BibitemOpen
  \bibfield  {author} {\bibinfo {author} {\bibfnamefont {Fabio}\ \bibnamefont
  {Mezzacapo}},\ }\bibfield  {title} {\enquote {\bibinfo {title} {Ground-state
  phase diagram of the quantum ${J}_{1}$-${J}_{2}$ model on the square
  lattice},}\ }\href {\doibase 10.1103/PhysRevB.86.045115} {\bibfield
  {journal} {\bibinfo  {journal} {Phys. Rev. B}\ }\textbf {\bibinfo {volume}
  {86}},\ \bibinfo {pages} {045115} (\bibinfo {year} {2012})}\BibitemShut
  {NoStop}%
\bibitem [{\citenamefont {Li}\ \emph {et~al.}(2012)\citenamefont {Li},
  \citenamefont {Becca}, \citenamefont {Hu},\ and\ \citenamefont
  {Sorella}}]{LiT:2012}%
  \BibitemOpen
  \bibfield  {author} {\bibinfo {author} {\bibfnamefont {Tao}\ \bibnamefont
  {Li}}, \bibinfo {author} {\bibfnamefont {Federico}\ \bibnamefont {Becca}},
  \bibinfo {author} {\bibfnamefont {Wenjun}\ \bibnamefont {Hu}}, \ and\
  \bibinfo {author} {\bibfnamefont {Sandro}\ \bibnamefont {Sorella}},\
  }\bibfield  {title} {\enquote {\bibinfo {title} {Gapped spin-liquid phase in
  the ${J}_{1}$-${J}_{2}$ {H}eisenberg model by a bosonic resonating
  valence-bond ansatz},}\ }\href {\doibase 10.1103/PhysRevB.86.075111}
  {\bibfield  {journal} {\bibinfo  {journal} {Phys. Rev. B}\ }\textbf {\bibinfo
  {volume} {86}},\ \bibinfo {pages} {075111} (\bibinfo {year}
  {2012})}\BibitemShut {NoStop}%
\bibitem [{\citenamefont {Wang}\ \emph {et~al.}(2013)\citenamefont {Wang},
  \citenamefont {Poilblanc}, \citenamefont {Gu}, \citenamefont {Wen},\ and\
  \citenamefont {Verstraete}}]{Wang:2013}%
  \BibitemOpen
  \bibfield  {author} {\bibinfo {author} {\bibfnamefont {Ling}\ \bibnamefont
  {Wang}}, \bibinfo {author} {\bibfnamefont {Didier}\ \bibnamefont
  {Poilblanc}}, \bibinfo {author} {\bibfnamefont {Zheng-Cheng}\ \bibnamefont
  {Gu}}, \bibinfo {author} {\bibfnamefont {Xiao-Gang}\ \bibnamefont {Wen}}, \
  and\ \bibinfo {author} {\bibfnamefont {Frank}\ \bibnamefont {Verstraete}},\
  }\bibfield  {title} {\enquote {\bibinfo {title} {Constructing a gapless
  spin-liquid state for the spin-$1/2$ ${J}_{1}$-${J}_{2}$ {H}eisenberg model
  on a square lattice},}\ }\href {\doibase 10.1103/PhysRevLett.111.037202}
  {\bibfield  {journal} {\bibinfo  {journal} {Phys. Rev. Lett.}\ }\textbf
  {\bibinfo {volume} {111}},\ \bibinfo {pages} {037202} (\bibinfo {year}
  {2013})}\BibitemShut {NoStop}%
\bibitem [{\citenamefont {Zhang}\ and\ \citenamefont
  {Beach}(2013)}]{Zhang:2013:J1J2SqLatt}%
  \BibitemOpen
  \bibfield  {author} {\bibinfo {author} {\bibfnamefont {Xiaoming}\
  \bibnamefont {Zhang}}\ and\ \bibinfo {author} {\bibfnamefont {K.~S.~D.}\
  \bibnamefont {Beach}},\ }\bibfield  {title} {\enquote {\bibinfo {title}
  {Resonating valence bond trial wave functions with both static and
  dynamically determined {M}arshall sign structure},}\ }\href {\doibase
  10.1103/PhysRevB.87.094420} {\bibfield  {journal} {\bibinfo  {journal} {Phys.
  Rev. B}\ }\textbf {\bibinfo {volume} {87}},\ \bibinfo {pages} {094420}
  (\bibinfo {year} {2013})}\BibitemShut {NoStop}%
\bibitem [{\citenamefont {Hu}\ \emph {et~al.}(2013)\citenamefont {Hu},
  \citenamefont {Becca}, \citenamefont {Parola},\ and\ \citenamefont
  {Sorella}}]{Hu:2013}%
  \BibitemOpen
  \bibfield  {author} {\bibinfo {author} {\bibfnamefont {Wen-Jun}\ \bibnamefont
  {Hu}}, \bibinfo {author} {\bibfnamefont {Federico}\ \bibnamefont {Becca}},
  \bibinfo {author} {\bibfnamefont {Alberto}\ \bibnamefont {Parola}}, \ and\
  \bibinfo {author} {\bibfnamefont {Sandro}\ \bibnamefont {Sorella}},\
  }\bibfield  {title} {\enquote {\bibinfo {title} {Direct evidence for a
  gapless ${Z}_{2}$ spin liquid by frustrating {N}{\'{e}}el
  antiferromagnetism},}\ }\href {\doibase 10.1103/PhysRevB.88.060402}
  {\bibfield  {journal} {\bibinfo  {journal} {Phys. Rev. B}\ }\textbf {\bibinfo
  {volume} {88}},\ \bibinfo {pages} {060402(R)} (\bibinfo {year}
  {2013})}\BibitemShut {NoStop}%
\bibitem [{\citenamefont {Gong}\ \emph {et~al.}(2014)\citenamefont {Gong},
  \citenamefont {Zhu}, \citenamefont {Sheng}, \citenamefont {Motrunich},\ and\
  \citenamefont {Fisher}}]{Gong:2014_J1J2mod_sqLatt}%
  \BibitemOpen
  \bibfield  {author} {\bibinfo {author} {\bibfnamefont {Shou-Shu}\
  \bibnamefont {Gong}}, \bibinfo {author} {\bibfnamefont {Wei}\ \bibnamefont
  {Zhu}}, \bibinfo {author} {\bibfnamefont {D.~N.}\ \bibnamefont {Sheng}},
  \bibinfo {author} {\bibfnamefont {Olexei~I.}\ \bibnamefont {Motrunich}}, \
  and\ \bibinfo {author} {\bibfnamefont {Matthew P.~A.}\ \bibnamefont
  {Fisher}},\ }\bibfield  {title} {\enquote {\bibinfo {title} {Plaquette
  ordered phase and quantum phase diagram in the spin-$\frac{1}{2}$
  ${J}_{1}$-${J}_{2}$ square {H}eisenberg model},}\ }\href {\doibase
  10.1103/PhysRevLett.113.027201} {\bibfield  {journal} {\bibinfo  {journal}
  {Phys. Rev. Lett.}\ }\textbf {\bibinfo {volume} {113}},\ \bibinfo {pages}
  {027201} (\bibinfo {year} {2014})}\BibitemShut {NoStop}%
\bibitem [{\citenamefont {Doretto}(2014)}]{Doretto:2014_J1J2mod_sqLatt}%
  \BibitemOpen
  \bibfield  {author} {\bibinfo {author} {\bibfnamefont {R.~L.}\ \bibnamefont
  {Doretto}},\ }\bibfield  {title} {\enquote {\bibinfo {title} {Plaquette
  valence-bond solid in the square-lattice ${J}_{1}$-${J}_{2}$ antiferromagnet
  {H}eisenberg model: A bond operator approach},}\ }\href {\doibase
  10.1103/PhysRevB.89.104415} {\bibfield  {journal} {\bibinfo  {journal} {Phys.
  Rev. B}\ }\textbf {\bibinfo {volume} {89}},\ \bibinfo {pages} {104415}
  (\bibinfo {year} {2014})}\BibitemShut {NoStop}%
\bibitem [{\citenamefont {Qi}\ and\ \citenamefont
  {Gu}(2014)}]{Qi:2014_J1J2SqLatt}%
  \BibitemOpen
  \bibfield  {author} {\bibinfo {author} {\bibfnamefont {Yang}\ \bibnamefont
  {Qi}}\ and\ \bibinfo {author} {\bibfnamefont {Zheng-Cheng}\ \bibnamefont
  {Gu}},\ }\bibfield  {title} {\enquote {\bibinfo {title} {Continuous phase
  transition from {N}\'{e}el state to ${Z}_{2}$ spin-liquid state on a square
  lattice},}\ }\href {\doibase 10.1103/PhysRevB.89.235122} {\bibfield
  {journal} {\bibinfo  {journal} {Phys. Rev. B}\ }\textbf {\bibinfo {volume}
  {89}},\ \bibinfo {pages} {235122} (\bibinfo {year} {2014})}\BibitemShut
  {NoStop}%
\bibitem [{\citenamefont {Metavitsiadis}\ \emph {et~al.}(2014)\citenamefont
  {Metavitsiadis}, \citenamefont {Sellmann},\ and\ \citenamefont
  {Eggert}}]{Metavitsiadis_Eggert:2014_J1J2mod_sqLatt}%
  \BibitemOpen
  \bibfield  {author} {\bibinfo {author} {\bibfnamefont {Alexandros}\
  \bibnamefont {Metavitsiadis}}, \bibinfo {author} {\bibfnamefont {Daniel}\
  \bibnamefont {Sellmann}}, \ and\ \bibinfo {author} {\bibfnamefont
  {Sebastian}\ \bibnamefont {Eggert}},\ }\bibfield  {title} {\enquote {\bibinfo
  {title} {Spin-liquid versus dimer phases in an anisotropic
  ${J}_{1}$-${J}_{2}$ frustrated square antiferromagnet},}\ }\href {\doibase
  10.1103/PhysRevB.89.241104} {\bibfield  {journal} {\bibinfo  {journal} {Phys.
  Rev. B}\ }\textbf {\bibinfo {volume} {89}},\ \bibinfo {pages} {241104(R)}
  (\bibinfo {year} {2014})}\BibitemShut {NoStop}%
\bibitem [{\citenamefont {Ren}\ \emph {et~al.}(2014)\citenamefont {Ren},
  \citenamefont {Tong},\ and\ \citenamefont {Xie}}]{Ren:2014_J1J2SqLatt}%
  \BibitemOpen
  \bibfield  {author} {\bibinfo {author} {\bibfnamefont {Yong-Zhi}\
  \bibnamefont {Ren}}, \bibinfo {author} {\bibfnamefont {Ning-Hua}\
  \bibnamefont {Tong}}, \ and\ \bibinfo {author} {\bibfnamefont {Xin-Cheng}\
  \bibnamefont {Xie}},\ }\bibfield  {title} {\enquote {\bibinfo {title}
  {Cluster mean-field theory study of ${J}_{1}$--${J}_{2}$ {H}eisenberg model
  on a square lattice},}\ }\href {\doibase 10.1088/0953-8984/26/11/115601}
  {\bibfield  {journal} {\bibinfo  {journal} {J. Phys.: Condens. Matter}\
  }\textbf {\bibinfo {volume} {26}},\ \bibinfo {pages} {115601} (\bibinfo
  {year} {2014})}\BibitemShut {NoStop}%
\bibitem [{\citenamefont {{Wang}}(2014)}]{Wang:2014_J1J2SqLatt}%
  \BibitemOpen
  \bibfield  {author} {\bibinfo {author} {\bibfnamefont {Ling}\ \bibnamefont
  {{Wang}}},\ }\href@noop {} {\enquote {\bibinfo {title} {{Correlated valence
  bond state and its study of the spin-1/2 ${J}_{1}$--${J}_{2}$
  antiferromagnetic Heisenberg model on a square lattice}},}\ } (\bibinfo
  {year} {2014}),\ \Eprint {http://arxiv.org/abs/1402.3564} {arXiv:1402.3564
  [cond-mat.str-el]} \BibitemShut {NoStop}%
\bibitem [{\citenamefont {Chou}\ and\ \citenamefont
  {Chen}(2014)}]{Chou:2014_J1J2SqLatt}%
  \BibitemOpen
  \bibfield  {author} {\bibinfo {author} {\bibfnamefont {Chung-Pin}\
  \bibnamefont {Chou}}\ and\ \bibinfo {author} {\bibfnamefont {Hong-Yi}\
  \bibnamefont {Chen}},\ }\bibfield  {title} {\enquote {\bibinfo {title}
  {Simulating a two-dimensional frustrated spin system with fermionic
  resonating-valence-bond states},}\ }\href {\doibase
  10.1103/PhysRevB.90.041106} {\bibfield  {journal} {\bibinfo  {journal} {Phys.
  Rev. B}\ }\textbf {\bibinfo {volume} {90}},\ \bibinfo {pages} {041106(R)}
  (\bibinfo {year} {2014})}\BibitemShut {NoStop}%
\bibitem [{\citenamefont {Morita}\ \emph {et~al.}(2015)\citenamefont {Morita},
  \citenamefont {Kaneko},\ and\ \citenamefont
  {Imada}}]{Morita:2015_J1J2SqLatt}%
  \BibitemOpen
  \bibfield  {author} {\bibinfo {author} {\bibfnamefont {Satoshi}\ \bibnamefont
  {Morita}}, \bibinfo {author} {\bibfnamefont {Ryui}\ \bibnamefont {Kaneko}}, \
  and\ \bibinfo {author} {\bibfnamefont {Masatoshi}\ \bibnamefont {Imada}},\
  }\bibfield  {title} {\enquote {\bibinfo {title} {Quantum spin liquid in
  spin-1/2 ${J}_{1}$-${J}_{2}$ {H}eisenberg model on square lattice:
  {M}any-variable variational {M}onte {C}arlo study combined with
  quantum-number projections},}\ }\href {\doibase 10.7566/JPSJ.84.024720}
  {\bibfield  {journal} {\bibinfo  {journal} {J. Phys. Soc. Jpn.}\ }\textbf
  {\bibinfo {volume} {84}},\ \bibinfo {pages} {024720} (\bibinfo {year}
  {2015})}\BibitemShut {NoStop}%
\bibitem [{\citenamefont {Richter}\ \emph {et~al.}(2015)\citenamefont
  {Richter}, \citenamefont {Zinke},\ and\ \citenamefont
  {Farnell}}]{Richter:2015_ccm_J1J2sq_spinGap}%
  \BibitemOpen
  \bibfield  {author} {\bibinfo {author} {\bibfnamefont {Johannes}\
  \bibnamefont {Richter}}, \bibinfo {author} {\bibfnamefont {Ronald}\
  \bibnamefont {Zinke}}, \ and\ \bibinfo {author} {\bibfnamefont {Damian
  J.~J.}\ \bibnamefont {Farnell}},\ }\bibfield  {title} {\enquote {\bibinfo
  {title} {The spin-1/2 square-lattice ${J}_{1}$--${J}_{2}$ model: the spin-gap
  issue},}\ }\href {\doibase 10.1140/epjb/e2014-50589-x} {\bibfield  {journal}
  {\bibinfo  {journal} {Eur. Phys. J. B}\ }\textbf {\bibinfo {volume} {88}},\
  \bibinfo {pages} {2} (\bibinfo {year} {2015})}\BibitemShut {NoStop}%
\bibitem [{\citenamefont {Wang}\ \emph {et~al.}(2016)\citenamefont {Wang},
  \citenamefont {Gu}, \citenamefont {Verstraete},\ and\ \citenamefont
  {Wen}}]{Wang:2016_J1J2mod}%
  \BibitemOpen
  \bibfield  {author} {\bibinfo {author} {\bibfnamefont {Ling}\ \bibnamefont
  {Wang}}, \bibinfo {author} {\bibfnamefont {Zheng-Cheng}\ \bibnamefont {Gu}},
  \bibinfo {author} {\bibfnamefont {Frank}\ \bibnamefont {Verstraete}}, \ and\
  \bibinfo {author} {\bibfnamefont {Xiao-Gang}\ \bibnamefont {Wen}},\
  }\bibfield  {title} {\enquote {\bibinfo {title} {Tensor-product state
  approach to spin-$\frac{1}{2}$ square ${J}_{1}\text{\ensuremath{-}}{J}_{2}$
  antiferromagnetic {H}eisenberg model: Evidence for deconfined quantum
  criticality},}\ }\href {\doibase 10.1103/PhysRevB.94.075143} {\bibfield
  {journal} {\bibinfo  {journal} {Phys. Rev. B}\ }\textbf {\bibinfo {volume}
  {94}},\ \bibinfo {pages} {075143} (\bibinfo {year} {2016})}\BibitemShut
  {NoStop}%
\bibitem [{\citenamefont {Poilblanc}\ and\ \citenamefont
  {Mambrini}(2017)}]{Poilblanc:2017_J1J2mod}%
  \BibitemOpen
  \bibfield  {author} {\bibinfo {author} {\bibfnamefont {Didier}\ \bibnamefont
  {Poilblanc}}\ and\ \bibinfo {author} {\bibfnamefont {Matthieu}\ \bibnamefont
  {Mambrini}},\ }\bibfield  {title} {\enquote {\bibinfo {title} {Quantum
  critical phase with infinite projected entangled paired states},}\ }\href
  {\doibase 10.1103/PhysRevB.96.014414} {\bibfield  {journal} {\bibinfo
  {journal} {Phys. Rev. B}\ }\textbf {\bibinfo {volume} {96}},\ \bibinfo
  {pages} {014414} (\bibinfo {year} {2017})}\BibitemShut {NoStop}%
\bibitem [{\citenamefont {Haghshenas}\ and\ \citenamefont
  {Sheng}(2018)}]{Haghshenas:2018_J1J2mod}%
  \BibitemOpen
  \bibfield  {author} {\bibinfo {author} {\bibfnamefont {R.}~\bibnamefont
  {Haghshenas}}\ and\ \bibinfo {author} {\bibfnamefont {D.~N.}\ \bibnamefont
  {Sheng}},\ }\bibfield  {title} {\enquote {\bibinfo {title}
  {${U}(1)$-symmetric infinite projected entangled-pair states study of the
  spin-1/2 square ${J}_{1}\text{\ensuremath{-}}{J}_{2}$ {H}eisenberg model},}\
  }\href {\doibase 10.1103/PhysRevB.97.174408} {\bibfield  {journal} {\bibinfo
  {journal} {Phys. Rev. B}\ }\textbf {\bibinfo {volume} {97}},\ \bibinfo
  {pages} {174408} (\bibinfo {year} {2018})}\BibitemShut {NoStop}%
\bibitem [{\citenamefont {Yu}\ \emph {et~al.}(2018)\citenamefont {Yu},
  \citenamefont {Wang}, \citenamefont {Dong}, \citenamefont {Yao},\ and\
  \citenamefont {Li}}]{Yu:2018_J1J2mod}%
  \BibitemOpen
  \bibfield  {author} {\bibinfo {author} {\bibfnamefont {Shun-Li}\ \bibnamefont
  {Yu}}, \bibinfo {author} {\bibfnamefont {Wei}\ \bibnamefont {Wang}}, \bibinfo
  {author} {\bibfnamefont {Zhao-Yang}\ \bibnamefont {Dong}}, \bibinfo {author}
  {\bibfnamefont {Zi-Jian}\ \bibnamefont {Yao}}, \ and\ \bibinfo {author}
  {\bibfnamefont {Jian-Xin}\ \bibnamefont {Li}},\ }\bibfield  {title} {\enquote
  {\bibinfo {title} {Deconfinement of spinons in frustrated spin systems:
  Spectral perspective},}\ }\href {\doibase 10.1103/PhysRevB.98.134410}
  {\bibfield  {journal} {\bibinfo  {journal} {Phys. Rev. B}\ }\textbf {\bibinfo
  {volume} {98}},\ \bibinfo {pages} {134410} (\bibinfo {year}
  {2018})}\BibitemShut {NoStop}%
\bibitem [{\citenamefont {Wang}\ and\ \citenamefont
  {Sandvik}(2018)}]{Wang:2018_J1J2mod}%
  \BibitemOpen
  \bibfield  {author} {\bibinfo {author} {\bibfnamefont {Ling}\ \bibnamefont
  {Wang}}\ and\ \bibinfo {author} {\bibfnamefont {Anders~W.}\ \bibnamefont
  {Sandvik}},\ }\bibfield  {title} {\enquote {\bibinfo {title} {Critical level
  crossings and gapless spin liquid in the square-lattice spin-$1/2$
  ${J}_{1}$--${J}_{2}$ {H}eisenberg antiferromagnet},}\ }\href {\doibase
  10.1103/PhysRevLett.121.107202} {\bibfield  {journal} {\bibinfo  {journal}
  {Phys. Rev. Lett.}\ }\textbf {\bibinfo {volume} {121}},\ \bibinfo {pages}
  {107202} (\bibinfo {year} {2018})}\BibitemShut {NoStop}%
\bibitem [{\citenamefont {Liu}\ \emph {et~al.}(2018)\citenamefont {Liu},
  \citenamefont {Dong}, \citenamefont {Wang}, \citenamefont {Han},
  \citenamefont {An}, \citenamefont {Guo},\ and\ \citenamefont
  {He}}]{Liu:2018_J1J2mod}%
  \BibitemOpen
  \bibfield  {author} {\bibinfo {author} {\bibfnamefont {Wen-Yuan}\
  \bibnamefont {Liu}}, \bibinfo {author} {\bibfnamefont {Shaojun}\ \bibnamefont
  {Dong}}, \bibinfo {author} {\bibfnamefont {Chao}\ \bibnamefont {Wang}},
  \bibinfo {author} {\bibfnamefont {Yongjian}\ \bibnamefont {Han}}, \bibinfo
  {author} {\bibfnamefont {Hong}\ \bibnamefont {An}}, \bibinfo {author}
  {\bibfnamefont {Guang-Can}\ \bibnamefont {Guo}}, \ and\ \bibinfo {author}
  {\bibfnamefont {Lixin}\ \bibnamefont {He}},\ }\bibfield  {title} {\enquote
  {\bibinfo {title} {Gapless spin liquid ground state of the spin-$\frac{1}{2}$
  ${J}_{1}$-${J}_{2}$ {H}eisenberg model on square lattices},}\ }\href
  {\doibase 10.1103/PhysRevB.98.241109} {\bibfield  {journal} {\bibinfo
  {journal} {Phys. Rev. B}\ }\textbf {\bibinfo {volume} {98}},\ \bibinfo
  {pages} {241109(R)} (\bibinfo {year} {2018})}\BibitemShut {NoStop}%
\bibitem [{\citenamefont {Anderson}(1987)}]{Anderson:1987_QSL}%
  \BibitemOpen
  \bibfield  {author} {\bibinfo {author} {\bibfnamefont {P.~W.}\ \bibnamefont
  {Anderson}},\ }\bibfield  {title} {\enquote {\bibinfo {title} {The resonating
  valence bond state in {L}a$_{2}${C}u{O}$_{4}$ and superconductivity},}\
  }\href {\doibase 10.1126/science.235.4793.1196} {\bibfield  {journal}
  {\bibinfo  {journal} {Science}\ }\textbf {\bibinfo {volume} {235}},\ \bibinfo
  {pages} {1196--1198} (\bibinfo {year} {1987})}\BibitemShut {NoStop}%
\bibitem [{\citenamefont {Lee}\ \emph {et~al.}(2006)\citenamefont {Lee},
  \citenamefont {Nagaosa},\ and\ \citenamefont {Wen}}]{Lee:2006_QSL}%
  \BibitemOpen
  \bibfield  {author} {\bibinfo {author} {\bibfnamefont {Patrick~A.}\
  \bibnamefont {Lee}}, \bibinfo {author} {\bibfnamefont {Naoto}\ \bibnamefont
  {Nagaosa}}, \ and\ \bibinfo {author} {\bibfnamefont {Xiao-Gang}\ \bibnamefont
  {Wen}},\ }\bibfield  {title} {\enquote {\bibinfo {title} {Doping a {M}ott
  insulator: Physics of high-temperature superconductivity},}\ }\href {\doibase
  10.1103/RevModPhys.78.17} {\bibfield  {journal} {\bibinfo  {journal} {Rev.
  Mod. Phys.}\ }\textbf {\bibinfo {volume} {78}},\ \bibinfo {pages} {17--85}
  (\bibinfo {year} {2006})}\BibitemShut {NoStop}%
\bibitem [{\citenamefont {Melzi}\ \emph {et~al.}(2000)\citenamefont {Melzi},
  \citenamefont {Carretta}, \citenamefont {Lascialfari}, \citenamefont
  {Mambrini}, \citenamefont {Troyer}, \citenamefont {Millet},\ and\
  \citenamefont {Mila}}]{Melzi:2000_sqLatt_J1J2mod_merge}%
  \BibitemOpen
  \bibfield  {author} {\bibinfo {author} {\bibfnamefont {R.}~\bibnamefont
  {Melzi}}, \bibinfo {author} {\bibfnamefont {P.}~\bibnamefont {Carretta}},
  \bibinfo {author} {\bibfnamefont {A.}~\bibnamefont {Lascialfari}}, \bibinfo
  {author} {\bibfnamefont {M.}~\bibnamefont {Mambrini}}, \bibinfo {author}
  {\bibfnamefont {M.}~\bibnamefont {Troyer}}, \bibinfo {author} {\bibfnamefont
  {P.}~\bibnamefont {Millet}}, \ and\ \bibinfo {author} {\bibfnamefont
  {F.}~\bibnamefont {Mila}},\ }\bibfield  {title} {\enquote {\bibinfo {title}
  {{L}i$_{2}${VO}({S}i,{G}e){O}$_{4}$, a prototype of a two-dimensional
  frustrated quantum {H}eisenberg antiferromagnet},}\ }\href {\doibase
  10.1103/PhysRevLett.85.1318} {\bibfield  {journal} {\bibinfo  {journal}
  {Phys. Rev. Lett.}\ }\textbf {\bibinfo {volume} {85}},\ \bibinfo {pages}
  {1318--1321} (\bibinfo {year} {2000})}\BibitemShut {NoStop}%
\bibitem [{\citenamefont {Melzi}\ \emph {et~al.}(2001)\citenamefont {Melzi},
  \citenamefont {Aldrovandi}, \citenamefont {Tedoldi}, \citenamefont
  {Carretta}, \citenamefont {Millet},\ and\ \citenamefont
  {Mila}}]{Melzi:2001_sqLatt_J1J2mod_merge}%
  \BibitemOpen
  \bibfield  {author} {\bibinfo {author} {\bibfnamefont {R.}~\bibnamefont
  {Melzi}}, \bibinfo {author} {\bibfnamefont {S.}~\bibnamefont {Aldrovandi}},
  \bibinfo {author} {\bibfnamefont {F.}~\bibnamefont {Tedoldi}}, \bibinfo
  {author} {\bibfnamefont {P.}~\bibnamefont {Carretta}}, \bibinfo {author}
  {\bibfnamefont {P.}~\bibnamefont {Millet}}, \ and\ \bibinfo {author}
  {\bibfnamefont {F.}~\bibnamefont {Mila}},\ }\bibfield  {title} {\enquote
  {\bibinfo {title} {Magnetic and thermodynamic properties of
  {L}i$_{2}${VO}{S}i{O}$_{4}$: A two-dimensional ${S}=1/2$ frustrated
  antiferromagnet on a square lattice},}\ }\href {\doibase
  10.1103/PhysRevB.64.024409} {\bibfield  {journal} {\bibinfo  {journal} {Phys.
  Rev. B}\ }\textbf {\bibinfo {volume} {64}},\ \bibinfo {pages} {024409}
  (\bibinfo {year} {2001})}\BibitemShut {NoStop}%
\bibitem [{\citenamefont {Rosner}\ \emph {et~al.}(2003)\citenamefont {Rosner},
  \citenamefont {Singh}, \citenamefont {Zheng}, \citenamefont {Oitmaa},\ and\
  \citenamefont {Pickett}}]{Rosner:2003_sqLatt_J1J2mod_merge}%
  \BibitemOpen
  \bibfield  {author} {\bibinfo {author} {\bibfnamefont {H.}~\bibnamefont
  {Rosner}}, \bibinfo {author} {\bibfnamefont {R.~R.~P.}\ \bibnamefont
  {Singh}}, \bibinfo {author} {\bibfnamefont {W.~H.}\ \bibnamefont {Zheng}},
  \bibinfo {author} {\bibfnamefont {J.}~\bibnamefont {Oitmaa}}, \ and\ \bibinfo
  {author} {\bibfnamefont {W.~E.}\ \bibnamefont {Pickett}},\ }\bibfield
  {title} {\enquote {\bibinfo {title} {High-temperature expansions for the
  ${J}_{1}$-${J}_{2}$ {H}eisenberg models: Applications to {{\it ab initio}}
  calculated models for {L}i$_{2}${VO}{S}i{O}$_{4}$ and
  {L}i$_{2}${VO}{G}e{O}$_{4}$},}\ }\href {\doibase 10.1103/PhysRevB.67.014416}
  {\bibfield  {journal} {\bibinfo  {journal} {Phys. Rev. B}\ }\textbf {\bibinfo
  {volume} {67}},\ \bibinfo {pages} {014416} (\bibinfo {year}
  {2003})}\BibitemShut {NoStop}%
\bibitem [{\citenamefont {Todate}\ \emph {et~al.}(2007)\citenamefont {Todate},
  \citenamefont {Higemoto}, \citenamefont {Nishiyama},\ and\ \citenamefont
  {Hirota}}]{Todate:2007_sqLatt_J1J2mod}%
  \BibitemOpen
  \bibfield  {author} {\bibinfo {author} {\bibfnamefont {Yoshiei}\ \bibnamefont
  {Todate}}, \bibinfo {author} {\bibfnamefont {Wataru}\ \bibnamefont
  {Higemoto}}, \bibinfo {author} {\bibfnamefont {Kusuo}\ \bibnamefont
  {Nishiyama}}, \ and\ \bibinfo {author} {\bibfnamefont {Kazuma}\ \bibnamefont
  {Hirota}},\ }\bibfield  {title} {\enquote {\bibinfo {title} {Magnetic
  ordering in ordered complex {C}u perovskite probed by $\mu${SR} and neutron
  diffraction},}\ }\href {\doibase 10.1016/j.jpcs.2007.08.027} {\bibfield
  {journal} {\bibinfo  {journal} {J. Phys. Chem. Solids}\ }\textbf {\bibinfo
  {volume} {68}},\ \bibinfo {pages} {2107--2110} (\bibinfo {year}
  {2007})}\BibitemShut {NoStop}%
\bibitem [{\citenamefont {Vasala}\ \emph
  {et~al.}(2014{\natexlab{a}})\citenamefont {Vasala}, \citenamefont {Saadaoui},
  \citenamefont {Morenzoni}, \citenamefont {Chmaissem}, \citenamefont {Chan},
  \citenamefont {Chen}, \citenamefont {Hsu}, \citenamefont {Yamauchi},\ and\
  \citenamefont {Karppinen}}]{Vasala:2014_sqLatt_J1J2mod}%
  \BibitemOpen
  \bibfield  {author} {\bibinfo {author} {\bibfnamefont {Sami}\ \bibnamefont
  {Vasala}}, \bibinfo {author} {\bibfnamefont {Hassan}\ \bibnamefont
  {Saadaoui}}, \bibinfo {author} {\bibfnamefont {Elvezio}\ \bibnamefont
  {Morenzoni}}, \bibinfo {author} {\bibfnamefont {Omar}\ \bibnamefont
  {Chmaissem}}, \bibinfo {author} {\bibfnamefont {Ting-Shan}\ \bibnamefont
  {Chan}}, \bibinfo {author} {\bibfnamefont {Jin-Ming}\ \bibnamefont {Chen}},
  \bibinfo {author} {\bibfnamefont {Ying-Ya}\ \bibnamefont {Hsu}}, \bibinfo
  {author} {\bibfnamefont {Hisao}\ \bibnamefont {Yamauchi}}, \ and\ \bibinfo
  {author} {\bibfnamefont {Maarit}\ \bibnamefont {Karppinen}},\ }\bibfield
  {title} {\enquote {\bibinfo {title} {Characterization of magnetic properties
  of {S}r$_{2}${C}u{WO}$_{6}$ and {S}r$_{2}${C}u{M}o{O}$_{6}$},}\ }\href
  {\doibase 10.1103/PhysRevB.89.134419} {\bibfield  {journal} {\bibinfo
  {journal} {Phys. Rev. B}\ }\textbf {\bibinfo {volume} {89}},\ \bibinfo
  {pages} {134419} (\bibinfo {year} {2014}{\natexlab{a}})}\BibitemShut
  {NoStop}%
\bibitem [{\citenamefont {Vasala}\ \emph
  {et~al.}(2014{\natexlab{b}})\citenamefont {Vasala}, \citenamefont {Avdeev},
  \citenamefont {Danilkin}, \citenamefont {Chmaissem},\ and\ \citenamefont
  {Karppinen}}]{Vasala:2014_sqLatt_J1J2mod_JPCM26}%
  \BibitemOpen
  \bibfield  {author} {\bibinfo {author} {\bibfnamefont {S.}~\bibnamefont
  {Vasala}}, \bibinfo {author} {\bibfnamefont {M.}~\bibnamefont {Avdeev}},
  \bibinfo {author} {\bibfnamefont {S.}~\bibnamefont {Danilkin}}, \bibinfo
  {author} {\bibfnamefont {O.}~\bibnamefont {Chmaissem}}, \ and\ \bibinfo
  {author} {\bibfnamefont {M.}~\bibnamefont {Karppinen}},\ }\bibfield  {title}
  {\enquote {\bibinfo {title} {Magnetic structure of
  {S}r$_{2}${C}u{WO}$_{6}$},}\ }\href {\doibase 10.1088/0953-8984/26/49/496001}
  {\bibfield  {journal} {\bibinfo  {journal} {J. Phys.: Condens. Matter}\
  }\textbf {\bibinfo {volume} {26}},\ \bibinfo {pages} {496001} (\bibinfo
  {year} {2014}{\natexlab{b}})}\BibitemShut {NoStop}%
\bibitem [{\citenamefont {Koga}\ \emph {et~al.}(2016)\citenamefont {Koga},
  \citenamefont {Kurita}, \citenamefont {Avdeev}, \citenamefont {Danilkin},
  \citenamefont {Sato},\ and\ \citenamefont
  {Tanaka}}]{Koga:2016_sqLatt_J1J2mod}%
  \BibitemOpen
  \bibfield  {author} {\bibinfo {author} {\bibfnamefont {Tomoyuki}\
  \bibnamefont {Koga}}, \bibinfo {author} {\bibfnamefont {Nobuyuki}\
  \bibnamefont {Kurita}}, \bibinfo {author} {\bibfnamefont {Maxim}\
  \bibnamefont {Avdeev}}, \bibinfo {author} {\bibfnamefont {Sergey}\
  \bibnamefont {Danilkin}}, \bibinfo {author} {\bibfnamefont {Taku~J.}\
  \bibnamefont {Sato}}, \ and\ \bibinfo {author} {\bibfnamefont {Hidekazu}\
  \bibnamefont {Tanaka}},\ }\bibfield  {title} {\enquote {\bibinfo {title}
  {Magnetic structure of the ${S}=\frac{1}{2}$ quasi-two-dimensional
  square-lattice {H}eisenberg antiferromagnet {S}r$_{2}${C}u{T}e{O}$_{6}$},}\
  }\href {\doibase 10.1103/PhysRevB.93.054426} {\bibfield  {journal} {\bibinfo
  {journal} {Phys. Rev. B}\ }\textbf {\bibinfo {volume} {93}},\ \bibinfo
  {pages} {054426} (\bibinfo {year} {2016})}\BibitemShut {NoStop}%
\bibitem [{\citenamefont {Chandra}\ \emph {et~al.}(1990)\citenamefont
  {Chandra}, \citenamefont {Coleman},\ and\ \citenamefont
  {Larkin}}]{Chandra:1990_ordByDisord_experimental}%
  \BibitemOpen
  \bibfield  {author} {\bibinfo {author} {\bibfnamefont {P.}~\bibnamefont
  {Chandra}}, \bibinfo {author} {\bibfnamefont {P.}~\bibnamefont {Coleman}}, \
  and\ \bibinfo {author} {\bibfnamefont {A.~I.}\ \bibnamefont {Larkin}},\
  }\bibfield  {title} {\enquote {\bibinfo {title} {Ising transition in
  frustrated {H}eisenberg models},}\ }\href {\doibase
  10.1103/PhysRevLett.64.88} {\bibfield  {journal} {\bibinfo  {journal} {Phys.
  Rev. Lett.}\ }\textbf {\bibinfo {volume} {64}},\ \bibinfo {pages} {88--91}
  (\bibinfo {year} {1990})}\BibitemShut {NoStop}%
\bibitem [{\citenamefont {Villain}(1977)}]{Villain:1977_ordByDisord_merge}%
  \BibitemOpen
  \bibfield  {author} {\bibinfo {author} {\bibfnamefont {J.}~\bibnamefont
  {Villain}},\ }\bibfield  {title} {\enquote {\bibinfo {title} {A magnetic
  analogue of stereoisomerism: application to helimagnetism in two
  dimensions},}\ }\href {\doibase 10.1051/jphys:01977003804038500} {\bibfield
  {journal} {\bibinfo  {journal} {J. Phys. (France)}\ }\textbf {\bibinfo
  {volume} {38}},\ \bibinfo {pages} {385--391} (\bibinfo {year}
  {1977})}\BibitemShut {NoStop}%
\bibitem [{\citenamefont {Villain}\ \emph {et~al.}(1980)\citenamefont
  {Villain}, \citenamefont {Bidaux}, \citenamefont {Carton},\ and\
  \citenamefont {Conte}}]{Villain:1980_ordByDisord_merge}%
  \BibitemOpen
  \bibfield  {author} {\bibinfo {author} {\bibfnamefont {J.}~\bibnamefont
  {Villain}}, \bibinfo {author} {\bibfnamefont {R.}~\bibnamefont {Bidaux}},
  \bibinfo {author} {\bibfnamefont {J.-P.}\ \bibnamefont {Carton}}, \ and\
  \bibinfo {author} {\bibfnamefont {R.}~\bibnamefont {Conte}},\ }\bibfield
  {title} {\enquote {\bibinfo {title} {Order as an effect of disorder},}\
  }\href {\doibase 10.1051/jphys:0198000410110126300} {\bibfield  {journal}
  {\bibinfo  {journal} {J. Phys. (France)}\ }\textbf {\bibinfo {volume} {41}},\
  \bibinfo {pages} {1263--1272} (\bibinfo {year} {1980})}\BibitemShut {NoStop}%
\bibitem [{\citenamefont {Shender}(1982)}]{Shender:1982_ordByDisord}%
  \BibitemOpen
  \bibfield  {author} {\bibinfo {author} {\bibfnamefont {E.~F.}\ \bibnamefont
  {Shender}},\ }\bibfield  {title} {\enquote {\bibinfo {title}
  {Antiferromagnetic garnets with fluctuationally interacting sublattices},}\
  }\href {http://www.jetp.ac.ru/cgi-bin/e/index/r/83/1/p326?a=list} {\bibfield
  {journal} {\bibinfo  {journal} {Zh. Eksp. Teor. Fiz.}\ }\textbf {\bibinfo
  {volume} {83}},\ \bibinfo {pages} {326--337} (\bibinfo {year} {1982})},\
  \translation{Sov. Phys. JETP \textbf{56}, 178 (1982)}\BibitemShut {NoStop}%
\bibitem [{\citenamefont {Sandvik}(2012)}]{Sandvik:2012_SqLatt_J-Q_model}%
  \BibitemOpen
  \bibfield  {author} {\bibinfo {author} {\bibfnamefont {Anders~W.}\
  \bibnamefont {Sandvik}},\ }\bibfield  {title} {\enquote {\bibinfo {title}
  {Finite-size scaling and boundary effects in two-dimensional valence-bond
  solids},}\ }\href {\doibase 10.1103/PhysRevB.85.134407} {\bibfield  {journal}
  {\bibinfo  {journal} {Phys. Rev. B}\ }\textbf {\bibinfo {volume} {85}},\
  \bibinfo {pages} {134407} (\bibinfo {year} {2012})}\BibitemShut {NoStop}%
\bibitem [{\citenamefont {Coester}(1958)}]{Coester:1958_ccm}%
  \BibitemOpen
  \bibfield  {author} {\bibinfo {author} {\bibfnamefont {F.}~\bibnamefont
  {Coester}},\ }\bibfield  {title} {\enquote {\bibinfo {title} {Bound states of
  a many-particle system},}\ }\href {\doibase 10.1016/0029-5582(58)90280-3}
  {\bibfield  {journal} {\bibinfo  {journal} {Nucl. Phys.}\ }\textbf {\bibinfo
  {volume} {7}},\ \bibinfo {pages} {421--424} (\bibinfo {year}
  {1958})}\BibitemShut {NoStop}%
\bibitem [{\citenamefont {Coester}\ and\ \citenamefont
  {K{\"{u}}mmel}(1960)}]{Coester:1960_ccm}%
  \BibitemOpen
  \bibfield  {author} {\bibinfo {author} {\bibfnamefont {F.}~\bibnamefont
  {Coester}}\ and\ \bibinfo {author} {\bibfnamefont {H.}~\bibnamefont
  {K{\"{u}}mmel}},\ }\bibfield  {title} {\enquote {\bibinfo {title}
  {Short-range correlations in nuclear wave functions},}\ }\href {\doibase
  10.1016/0029-5582(60)90140-1} {\bibfield  {journal} {\bibinfo  {journal}
  {Nucl. Phys.}\ }\textbf {\bibinfo {volume} {17}},\ \bibinfo {pages}
  {477--485} (\bibinfo {year} {1960})}\BibitemShut {NoStop}%
\bibitem [{\citenamefont {{\v{C}}i{\v{z}}ek}(1966)}]{Cizek:1966_ccm}%
  \BibitemOpen
  \bibfield  {author} {\bibinfo {author} {\bibfnamefont {Ji{\v{r}}i}\
  \bibnamefont {{\v{C}}i{\v{z}}ek}},\ }\bibfield  {title} {\enquote {\bibinfo
  {title} {On the correlation problem in atomic and molecular systems.
  {C}alculation of wavefunction components in {U}rsell-type expansion using
  quantum-field theoretical methods},}\ }\href {\doibase 10.1063/1.1727484}
  {\bibfield  {journal} {\bibinfo  {journal} {J. Chem. Phys.}\ }\textbf
  {\bibinfo {volume} {45}},\ \bibinfo {pages} {4256--4266} (\bibinfo {year}
  {1966})}\BibitemShut {NoStop}%
\bibitem [{\citenamefont {K{\"{u}}mmel}\ \emph {et~al.}(1978)\citenamefont
  {K{\"{u}}mmel}, \citenamefont {L{\"{u}}hrmann},\ and\ \citenamefont
  {Zabolitzky}}]{Kummel:1978_ccm}%
  \BibitemOpen
  \bibfield  {author} {\bibinfo {author} {\bibfnamefont {H.}~\bibnamefont
  {K{\"{u}}mmel}}, \bibinfo {author} {\bibfnamefont {K.~H.}\ \bibnamefont
  {L{\"{u}}hrmann}}, \ and\ \bibinfo {author} {\bibfnamefont {J.~G.}\
  \bibnamefont {Zabolitzky}},\ }\bibfield  {title} {\enquote {\bibinfo {title}
  {Many-fermion theory in exp${S}$ (or coupled cluster) form},}\ }\href
  {\doibase 10.1016/0370-1573(78)90081-9} {\bibfield  {journal} {\bibinfo
  {journal} {Phys Rep.}\ }\textbf {\bibinfo {volume} {36}},\ \bibinfo {pages}
  {1--63} (\bibinfo {year} {1978})}\BibitemShut {NoStop}%
\bibitem [{\citenamefont {Bishop}\ and\ \citenamefont
  {L{\"{u}}hrmann}(1978)}]{Bishop:1978_ccm}%
  \BibitemOpen
  \bibfield  {author} {\bibinfo {author} {\bibfnamefont {R.~F.}\ \bibnamefont
  {Bishop}}\ and\ \bibinfo {author} {\bibfnamefont {K.~H.}\ \bibnamefont
  {L{\"{u}}hrmann}},\ }\bibfield  {title} {\enquote {\bibinfo {title} {Electron
  correlations: I. {G}round-state results in the high-density regime},}\ }\href
  {\doibase 10.1103/PhysRevB.17.3757} {\bibfield  {journal} {\bibinfo
  {journal} {Phys. Rev. B}\ }\textbf {\bibinfo {volume} {17}},\ \bibinfo
  {pages} {3757--3780} (\bibinfo {year} {1978})}\BibitemShut {NoStop}%
\bibitem [{\citenamefont {Bishop}\ and\ \citenamefont
  {L{\"{u}}hrmann}(1982)}]{Bishop:1982_ccm}%
  \BibitemOpen
  \bibfield  {author} {\bibinfo {author} {\bibfnamefont {R.~F.}\ \bibnamefont
  {Bishop}}\ and\ \bibinfo {author} {\bibfnamefont {K.~H.}\ \bibnamefont
  {L{\"{u}}hrmann}},\ }\bibfield  {title} {\enquote {\bibinfo {title} {Electron
  correlations. {II}. {G}round-state results at low and metallic densities},}\
  }\href {\doibase 10.1103/PhysRevB.26.5523} {\bibfield  {journal} {\bibinfo
  {journal} {Phys. Rev. B}\ }\textbf {\bibinfo {volume} {26}},\ \bibinfo
  {pages} {5523--5557} (\bibinfo {year} {1982})}\BibitemShut {NoStop}%
\bibitem [{\citenamefont {Arponen}(1983)}]{Arponen:1983_ccm}%
  \BibitemOpen
  \bibfield  {author} {\bibinfo {author} {\bibfnamefont {Jouko}\ \bibnamefont
  {Arponen}},\ }\bibfield  {title} {\enquote {\bibinfo {title} {Variational
  principles and linked-cluster exp ${S}$ expansions for static and dynamic
  many-body problems},}\ }\href {\doibase 10.1016/0003-4916(83)90284-1}
  {\bibfield  {journal} {\bibinfo  {journal} {Ann. Phys. (N.Y.)}\ }\textbf
  {\bibinfo {volume} {151}},\ \bibinfo {pages} {311--382} (\bibinfo {year}
  {1983})}\BibitemShut {NoStop}%
\bibitem [{\citenamefont {Bishop}\ and\ \citenamefont
  {K{\"{u}}mmel}(1987)}]{Bishop:1987_ccm}%
  \BibitemOpen
  \bibfield  {author} {\bibinfo {author} {\bibfnamefont {R.~F.}\ \bibnamefont
  {Bishop}}\ and\ \bibinfo {author} {\bibfnamefont {H.~G.}\ \bibnamefont
  {K{\"{u}}mmel}},\ }\bibfield  {title} {\enquote {\bibinfo {title} {The
  coupled-cluster method},}\ }\href {\doibase 10.1063/1.881103} {\bibfield
  {journal} {\bibinfo  {journal} {Phys. Today}\ }\textbf {\bibinfo {volume}
  {40(3)}},\ \bibinfo {pages} {52--60} (\bibinfo {year} {1987})}\BibitemShut
  {NoStop}%
\bibitem [{\citenamefont {Arponen}\ \emph
  {et~al.}(1987{\natexlab{a}})\citenamefont {Arponen}, \citenamefont {Bishop},\
  and\ \citenamefont {Pajanne}}]{Arponen:1987_ccm}%
  \BibitemOpen
  \bibfield  {author} {\bibinfo {author} {\bibfnamefont {J.~S.}\ \bibnamefont
  {Arponen}}, \bibinfo {author} {\bibfnamefont {R.~F.}\ \bibnamefont {Bishop}},
  \ and\ \bibinfo {author} {\bibfnamefont {E.}~\bibnamefont {Pajanne}},\
  }\bibfield  {title} {\enquote {\bibinfo {title} {Extended coupled-cluster
  method. {I}. {G}eneralized coherent bosonization as a mapping of quantum
  theory into classical {H}amiltonian mechanics},}\ }\href {\doibase
  10.1103/PhysRevA.36.2519} {\bibfield  {journal} {\bibinfo  {journal} {Phys.
  Rev. A}\ }\textbf {\bibinfo {volume} {36}},\ \bibinfo {pages} {2519--2538}
  (\bibinfo {year} {1987}{\natexlab{a}})}\BibitemShut {NoStop}%
\bibitem [{\citenamefont {Arponen}\ \emph
  {et~al.}(1987{\natexlab{b}})\citenamefont {Arponen}, \citenamefont {Bishop},\
  and\ \citenamefont {Pajanne}}]{Arponen:1987_ccm_2}%
  \BibitemOpen
  \bibfield  {author} {\bibinfo {author} {\bibfnamefont {J.~S.}\ \bibnamefont
  {Arponen}}, \bibinfo {author} {\bibfnamefont {R.~F.}\ \bibnamefont {Bishop}},
  \ and\ \bibinfo {author} {\bibfnamefont {E.}~\bibnamefont {Pajanne}},\
  }\bibfield  {title} {\enquote {\bibinfo {title} {Extended coupled-cluster
  method. {II}. {E}xcited states and generalized random-phase approximation},}\
  }\href {\doibase 10.1103/PhysRevA.36.2539} {\bibfield  {journal} {\bibinfo
  {journal} {Phys. Rev. A}\ }\textbf {\bibinfo {volume} {36}},\ \bibinfo
  {pages} {2539--2549} (\bibinfo {year} {1987}{\natexlab{b}})}\BibitemShut
  {NoStop}%
\bibitem [{\citenamefont {Bartlett}(1989)}]{Bartlett:1989_ccm}%
  \BibitemOpen
  \bibfield  {author} {\bibinfo {author} {\bibfnamefont {R.~J.}\ \bibnamefont
  {Bartlett}},\ }\bibfield  {title} {\enquote {\bibinfo {title}
  {Coupled-cluster approach to molecular structure and spectra: A step toward
  predictive quantum chemistry},}\ }\href {\doibase 10.1021/j100342a008}
  {\bibfield  {journal} {\bibinfo  {journal} {J. Phys. Chem.}\ }\textbf
  {\bibinfo {volume} {93}},\ \bibinfo {pages} {1697--1708} (\bibinfo {year}
  {1989})}\BibitemShut {NoStop}%
\bibitem [{\citenamefont {Arponen}\ and\ \citenamefont
  {Bishop}(1991)}]{Arponen:1991_ccm}%
  \BibitemOpen
  \bibfield  {author} {\bibinfo {author} {\bibfnamefont {J.~S.}\ \bibnamefont
  {Arponen}}\ and\ \bibinfo {author} {\bibfnamefont {R.~F.}\ \bibnamefont
  {Bishop}},\ }\bibfield  {title} {\enquote {\bibinfo {title}
  {Independent-cluster parametrizations of wave functions in model field
  theories. {I}. {I}ntroduction to their holomorphic representations},}\ }\href
  {\doibase 10.1016/0003-4916(91)90183-9} {\bibfield  {journal} {\bibinfo
  {journal} {Ann. Phys. (N.Y.)}\ }\textbf {\bibinfo {volume} {207}},\ \bibinfo
  {pages} {171--217} (\bibinfo {year} {1991})}\BibitemShut {NoStop}%
\bibitem [{\citenamefont {Bishop}(1991)}]{Bishop:1991_TheorChimActa_QMBT}%
  \BibitemOpen
  \bibfield  {author} {\bibinfo {author} {\bibfnamefont {R.~F.}\ \bibnamefont
  {Bishop}},\ }\bibfield  {title} {\enquote {\bibinfo {title} {An overview of
  coupled cluster theory and its applications in physics},}\ }\href {\doibase
  10.1007/BF01119617} {\bibfield  {journal} {\bibinfo  {journal} {Theor. Chim.
  Acta}\ }\textbf {\bibinfo {volume} {80}},\ \bibinfo {pages} {95--148}
  (\bibinfo {year} {1991})}\BibitemShut {NoStop}%
\bibitem [{\citenamefont {Bishop}(1998)}]{Bishop:1998_QMBT_coll}%
  \BibitemOpen
  \bibfield  {author} {\bibinfo {author} {\bibfnamefont {R.~F.}\ \bibnamefont
  {Bishop}},\ }\bibfield  {title} {\enquote {\bibinfo {title} {The coupled
  cluster method},}\ }in\ \href {\doibase 10.1007/BFb0104523} {\emph {\bibinfo
  {booktitle} {Microscopic Quantum Many-Body Theories and Their
  Applications}}},\ \bibinfo {series and number} {Lecture Notes in Physics Vol.
  510},\ \bibinfo {editor} {edited by\ \bibinfo {editor} {\bibfnamefont
  {J.}~\bibnamefont {Navarro}}\ and\ \bibinfo {editor} {\bibfnamefont
  {A.}~\bibnamefont {Polls}}}\ (\bibinfo  {publisher} {Springer-Verlag},\
  \bibinfo {address} {Berlin},\ \bibinfo {year} {1998})\ pp.\ \bibinfo {pages}
  {1--70}\BibitemShut {NoStop}%
\bibitem [{\citenamefont {Zeng}\ \emph {et~al.}(1998)\citenamefont {Zeng},
  \citenamefont {Farnell},\ and\ \citenamefont
  {Bishop}}]{Zeng:1998_SqLatt_TrianLatt}%
  \BibitemOpen
  \bibfield  {author} {\bibinfo {author} {\bibfnamefont {C.}~\bibnamefont
  {Zeng}}, \bibinfo {author} {\bibfnamefont {D.~J.~J.}\ \bibnamefont
  {Farnell}}, \ and\ \bibinfo {author} {\bibfnamefont {R.~F.}\ \bibnamefont
  {Bishop}},\ }\bibfield  {title} {\enquote {\bibinfo {title} {An efficient
  implementation of high-order coupled-cluster techniques applied to quantum
  magnets},}\ }\href {\doibase 10.1023/A:1023220222019} {\bibfield  {journal}
  {\bibinfo  {journal} {J. Stat. Phys.}\ }\textbf {\bibinfo {volume} {90}},\
  \bibinfo {pages} {327--361} (\bibinfo {year} {1998})}\BibitemShut {NoStop}%
\bibitem [{\citenamefont {Farnell}\ and\ \citenamefont
  {Bishop}(2004)}]{Fa:2004_QM-coll}%
  \BibitemOpen
  \bibfield  {author} {\bibinfo {author} {\bibfnamefont {D.~J.~J.}\
  \bibnamefont {Farnell}}\ and\ \bibinfo {author} {\bibfnamefont {R.~F.}\
  \bibnamefont {Bishop}},\ }\bibfield  {title} {\enquote {\bibinfo {title} {The
  coupled cluster method applied to quantum magnetism},}\ }in\ \href {\doibase
  10.1007/BFb0119597} {\emph {\bibinfo {booktitle} {Quantum Magnetism}}},\
  \bibinfo {series and number} {Lecture Notes in Physics Vol. 645},\ \bibinfo
  {editor} {edited by\ \bibinfo {editor} {\bibfnamefont {Ulrich}\ \bibnamefont
  {Schollw{\"{o}}ck}}, \bibinfo {editor} {\bibfnamefont {Johannes}\
  \bibnamefont {Richter}}, \bibinfo {editor} {\bibfnamefont {Damian J.~J.}\
  \bibnamefont {Farnell}}, \ and\ \bibinfo {editor} {\bibfnamefont
  {Raymond~F.}\ \bibnamefont {Bishop}}}\ (\bibinfo  {publisher}
  {Springer-Verlag},\ \bibinfo {address} {Berlin},\ \bibinfo {year} {2004})\
  pp.\ \bibinfo {pages} {307--348}\BibitemShut {NoStop}%
\bibitem [{\citenamefont {Bartlett}\ and\ \citenamefont
  {Musia{\l{}}}(2007)}]{Bartlett:2007_ccm}%
  \BibitemOpen
  \bibfield  {author} {\bibinfo {author} {\bibfnamefont {Rodney~J.}\
  \bibnamefont {Bartlett}}\ and\ \bibinfo {author} {\bibfnamefont {Monika}\
  \bibnamefont {Musia{\l{}}}},\ }\bibfield  {title} {\enquote {\bibinfo {title}
  {Coupled-cluster theory in quantum chemistry},}\ }\href {\doibase
  10.1103/RevModPhys.79.291} {\bibfield  {journal} {\bibinfo  {journal} {Rev.
  Mod. Phys.}\ }\textbf {\bibinfo {volume} {79}},\ \bibinfo {pages} {291--352}
  (\bibinfo {year} {2007})}\BibitemShut {NoStop}%
\bibitem [{\citenamefont {Bishop}\ \emph {et~al.}(2014)\citenamefont {Bishop},
  \citenamefont {Li},\ and\ \citenamefont
  {Campbell}}]{Bishop:2014_honey_XXZ_nmp14}%
  \BibitemOpen
  \bibfield  {author} {\bibinfo {author} {\bibfnamefont {R.~F.}\ \bibnamefont
  {Bishop}}, \bibinfo {author} {\bibfnamefont {P.~H.~Y.}\ \bibnamefont {Li}}, \
  and\ \bibinfo {author} {\bibfnamefont {C.~E.}\ \bibnamefont {Campbell}},\
  }\bibfield  {title} {\enquote {\bibinfo {title} {Highly frustrated
  spin-lattice models of magnetism and their quantum phase transitions: A
  microscopic treatment via the coupled cluster method},}\ }\href {\doibase
  10.1063/1.4899216} {\bibfield  {journal} {\bibinfo  {journal} {AIP Conf.
  Proc.}\ }\textbf {\bibinfo {volume} {1619}},\ \bibinfo {pages} {40--50}
  (\bibinfo {year} {2014})}\BibitemShut {NoStop}%
\bibitem [{\citenamefont {Hida}(1990)}]{Kazuo:1990_SqLatt_bilayer}%
  \BibitemOpen
  \bibfield  {author} {\bibinfo {author} {\bibfnamefont {Kazuo}\ \bibnamefont
  {Hida}},\ }\bibfield  {title} {\enquote {\bibinfo {title} {Low temperature
  properties of the double layer quantum {H}eisenberg antiferromagnet:
  {M}odified spin wave method},}\ }\href {\doibase 10.1143/JPSJ.59.2230}
  {\bibfield  {journal} {\bibinfo  {journal} {J. Phys. Soc. Jpn.}\ }\textbf
  {\bibinfo {volume} {59}},\ \bibinfo {pages} {2230--2236} (\bibinfo {year}
  {1990})}\BibitemShut {NoStop}%
\bibitem [{\citenamefont {Hida}(1992)}]{Kazuo:1992_SqLatt_bilayer}%
  \BibitemOpen
  \bibfield  {author} {\bibinfo {author} {\bibfnamefont {Kazuo}\ \bibnamefont
  {Hida}},\ }\bibfield  {title} {\enquote {\bibinfo {title} {Quantum disordered
  state without frustration in the double layer {H}eisenberg antiferromagnet:
  {D}imer expansion and projector {M}onte {C}arlo study},}\ }\href {\doibase
  10.1143/JPSJ.61.1013} {\bibfield  {journal} {\bibinfo  {journal} {J. Phys.
  Soc. Jpn.}\ }\textbf {\bibinfo {volume} {61}},\ \bibinfo {pages} {1013--1018}
  (\bibinfo {year} {1992})}\BibitemShut {NoStop}%
\bibitem [{\citenamefont {Millis}\ and\ \citenamefont
  {Monien}(1993)}]{Millis:1993_SqLatt_bilayer}%
  \BibitemOpen
  \bibfield  {author} {\bibinfo {author} {\bibfnamefont {A.~J.}\ \bibnamefont
  {Millis}}\ and\ \bibinfo {author} {\bibfnamefont {H.}~\bibnamefont
  {Monien}},\ }\bibfield  {title} {\enquote {\bibinfo {title} {Spin gaps and
  spin dynamics in {L}a$_{2-x}${S}r$_{x}${C}u{O}$_{4}$ and
  {YB}a$_{2}${C}u$_{3}${O}$_{7-\delta}$},}\ }\href {\doibase
  10.1103/PhysRevLett.70.2810} {\bibfield  {journal} {\bibinfo  {journal}
  {Phys. Rev. Lett.}\ }\textbf {\bibinfo {volume} {70}},\ \bibinfo {pages}
  {2810--2813} (\bibinfo {year} {1993})}\BibitemShut {NoStop}%
\bibitem [{\citenamefont {Millis}\ and\ \citenamefont
  {Monien}(1994)}]{Millis:1994_SqLatt_bilayer}%
  \BibitemOpen
  \bibfield  {author} {\bibinfo {author} {\bibfnamefont {A.~J.}\ \bibnamefont
  {Millis}}\ and\ \bibinfo {author} {\bibfnamefont {H.}~\bibnamefont
  {Monien}},\ }\bibfield  {title} {\enquote {\bibinfo {title} {Spin gaps and
  bilayer coupling in {YB}a$_{2}${C}u$_{3}${O}$_{7-\delta}$ and
  {YB}a$_{2}${C}u$_{4}${O}$_{\delta}$},}\ }\href {\doibase
  10.1103/PhysRevB.50.16606} {\bibfield  {journal} {\bibinfo  {journal} {Phys.
  Rev. B}\ }\textbf {\bibinfo {volume} {50}},\ \bibinfo {pages} {16606--16622}
  (\bibinfo {year} {1994})}\BibitemShut {NoStop}%
\bibitem [{\citenamefont {Sandvik}\ and\ \citenamefont
  {Scalapino}(1994)}]{Sandvik:1994_SqLatt_bilayer}%
  \BibitemOpen
  \bibfield  {author} {\bibinfo {author} {\bibfnamefont {A.~W.}\ \bibnamefont
  {Sandvik}}\ and\ \bibinfo {author} {\bibfnamefont {D.~J.}\ \bibnamefont
  {Scalapino}},\ }\bibfield  {title} {\enquote {\bibinfo {title}
  {Order-disorder transition in a two-layer quantum antiferromagnet},}\ }\href
  {\doibase 10.1103/PhysRevLett.72.2777} {\bibfield  {journal} {\bibinfo
  {journal} {Phys. Rev. Lett.}\ }\textbf {\bibinfo {volume} {72}},\ \bibinfo
  {pages} {2777--2780} (\bibinfo {year} {1994})}\BibitemShut {NoStop}%
\bibitem [{\citenamefont {Sandvik}\ \emph {et~al.}(1995)\citenamefont
  {Sandvik}, \citenamefont {Chubukov},\ and\ \citenamefont
  {Sachdev}}]{Sandvik:1995_SqLatt_bilayer}%
  \BibitemOpen
  \bibfield  {author} {\bibinfo {author} {\bibfnamefont {Anders~W.}\
  \bibnamefont {Sandvik}}, \bibinfo {author} {\bibfnamefont {Andrey~V.}\
  \bibnamefont {Chubukov}}, \ and\ \bibinfo {author} {\bibfnamefont {Subir}\
  \bibnamefont {Sachdev}},\ }\bibfield  {title} {\enquote {\bibinfo {title}
  {Quantum critical behavior in a two-layer antiferromagnet},}\ }\href
  {\doibase 10.1103/PhysRevB.51.16483} {\bibfield  {journal} {\bibinfo
  {journal} {Phys. Rev. B}\ }\textbf {\bibinfo {volume} {51}},\ \bibinfo
  {pages} {16483(R)--16486(R)} (\bibinfo {year} {1995})}\BibitemShut {NoStop}%
\bibitem [{\citenamefont {Chubukov}\ and\ \citenamefont
  {Morr}(1995)}]{Chubukov:1995_SqLatt_bilayer}%
  \BibitemOpen
  \bibfield  {author} {\bibinfo {author} {\bibfnamefont {Andrey~V.}\
  \bibnamefont {Chubukov}}\ and\ \bibinfo {author} {\bibfnamefont {Dirk~K.}\
  \bibnamefont {Morr}},\ }\bibfield  {title} {\enquote {\bibinfo {title} {Phase
  transition, longitudinal spin fluctuations, and scaling in a two layer
  antiferromagnet},}\ }\href {\doibase 10.1103/PhysRevB.52.3521} {\bibfield
  {journal} {\bibinfo  {journal} {Phys. Rev. B}\ }\textbf {\bibinfo {volume}
  {52}},\ \bibinfo {pages} {3521--3532} (\bibinfo {year} {1995})}\BibitemShut
  {NoStop}%
\bibitem [{\citenamefont {Weihong}(1997)}]{Zheng:1997_SqLatt_bilayer}%
  \BibitemOpen
  \bibfield  {author} {\bibinfo {author} {\bibfnamefont {Zheng}\ \bibnamefont
  {Weihong}},\ }\bibfield  {title} {\enquote {\bibinfo {title} {Various series
  expansions for the bilayer ${S}=\frac{1}{2}$ {H}eisenberg antiferromagnet},}\
  }\href {\doibase 10.1103/PhysRevB.55.12267} {\bibfield  {journal} {\bibinfo
  {journal} {Phys. Rev. B}\ }\textbf {\bibinfo {volume} {55}},\ \bibinfo
  {pages} {12267--12275} (\bibinfo {year} {1997})}\BibitemShut {NoStop}%
\bibitem [{\citenamefont {Shevchenko}\ and\ \citenamefont
  {Sushkov}(1999)}]{Shevchenko:1999_SqLatt_bilayer}%
  \BibitemOpen
  \bibfield  {author} {\bibinfo {author} {\bibfnamefont {P.~V.}\ \bibnamefont
  {Shevchenko}}\ and\ \bibinfo {author} {\bibfnamefont {O.~P.}\ \bibnamefont
  {Sushkov}},\ }\bibfield  {title} {\enquote {\bibinfo {title} {Brueckner
  approach to the spin-wave gap critical index for the two-layer {H}eisenberg
  antiferromagnet},}\ }\href {\doibase 10.1103/PhysRevB.59.8383} {\bibfield
  {journal} {\bibinfo  {journal} {Phys. Rev. B}\ }\textbf {\bibinfo {volume}
  {59}},\ \bibinfo {pages} {8383--8386} (\bibinfo {year} {1999})}\BibitemShut
  {NoStop}%
\bibitem [{\citenamefont {Shevchenko}\ \emph {et~al.}(2000)\citenamefont
  {Shevchenko}, \citenamefont {Sandvik},\ and\ \citenamefont
  {Sushkov}}]{Shevchenko:2000_SqLatt_bilayer}%
  \BibitemOpen
  \bibfield  {author} {\bibinfo {author} {\bibfnamefont {P.~V.}\ \bibnamefont
  {Shevchenko}}, \bibinfo {author} {\bibfnamefont {A.~W.}\ \bibnamefont
  {Sandvik}}, \ and\ \bibinfo {author} {\bibfnamefont {O.~P.}\ \bibnamefont
  {Sushkov}},\ }\bibfield  {title} {\enquote {\bibinfo {title} {Double-layer
  {H}eisenberg antiferromagnet at finite temperature: Brueckner theory and
  quantum {M}onte {C}arlo simulations},}\ }\href {\doibase
  10.1103/PhysRevB.61.3475} {\bibfield  {journal} {\bibinfo  {journal} {Phys.
  Rev. B}\ }\textbf {\bibinfo {volume} {61}},\ \bibinfo {pages} {3475--3487}
  (\bibinfo {year} {2000})}\BibitemShut {NoStop}%
\bibitem [{\citenamefont {Wang}\ \emph {et~al.}(2006)\citenamefont {Wang},
  \citenamefont {Beach},\ and\ \citenamefont
  {Sandvik}}]{Wang:2006_SqLatt_bilayer}%
  \BibitemOpen
  \bibfield  {author} {\bibinfo {author} {\bibfnamefont {Ling}\ \bibnamefont
  {Wang}}, \bibinfo {author} {\bibfnamefont {K.~S.~D.}\ \bibnamefont {Beach}},
  \ and\ \bibinfo {author} {\bibfnamefont {Anders~W.}\ \bibnamefont
  {Sandvik}},\ }\bibfield  {title} {\enquote {\bibinfo {title} {High-precision
  finite-size scaling analysis of the quantum-critical point of ${S}=1/2$
  {H}eisenberg antiferromagnetic bilayers},}\ }\href {\doibase
  10.1103/PhysRevB.73.014431} {\bibfield  {journal} {\bibinfo  {journal} {Phys.
  Rev. B}\ }\textbf {\bibinfo {volume} {73}},\ \bibinfo {pages} {014431}
  (\bibinfo {year} {2006})}\BibitemShut {NoStop}%
\bibitem [{\citenamefont {Collins}\ and\ \citenamefont
  {Hamer}(2008)}]{Collins:2008_SqLatt_bilayer}%
  \BibitemOpen
  \bibfield  {author} {\bibinfo {author} {\bibfnamefont {A.}~\bibnamefont
  {Collins}}\ and\ \bibinfo {author} {\bibfnamefont {C.~J.}\ \bibnamefont
  {Hamer}},\ }\bibfield  {title} {\enquote {\bibinfo {title} {Two-particle
  bound states and one-particle structure factor in a {H}eisenberg bilayer
  system},}\ }\href {\doibase 10.1103/PhysRevB.78.054419} {\bibfield  {journal}
  {\bibinfo  {journal} {Phys. Rev. B}\ }\textbf {\bibinfo {volume} {78}},\
  \bibinfo {pages} {054419} (\bibinfo {year} {2008})}\BibitemShut {NoStop}%
\bibitem [{\citenamefont {Fritz}\ \emph {et~al.}(2011)\citenamefont {Fritz},
  \citenamefont {Doretto}, \citenamefont {Wessel}, \citenamefont {Wenzel},
  \citenamefont {Burdin},\ and\ \citenamefont
  {Vojta}}]{Fritz:2011_dimerized_AFM}%
  \BibitemOpen
  \bibfield  {author} {\bibinfo {author} {\bibfnamefont {L.}~\bibnamefont
  {Fritz}}, \bibinfo {author} {\bibfnamefont {R.~L.}\ \bibnamefont {Doretto}},
  \bibinfo {author} {\bibfnamefont {S.}~\bibnamefont {Wessel}}, \bibinfo
  {author} {\bibfnamefont {S.}~\bibnamefont {Wenzel}}, \bibinfo {author}
  {\bibfnamefont {S.}~\bibnamefont {Burdin}}, \ and\ \bibinfo {author}
  {\bibfnamefont {M.}~\bibnamefont {Vojta}},\ }\bibfield  {title} {\enquote
  {\bibinfo {title} {Cubic interactions and quantum criticality in dimerized
  antiferromagnets},}\ }\href {\doibase 10.1103/PhysRevB.83.174416} {\bibfield
  {journal} {\bibinfo  {journal} {Phys. Rev. B}\ }\textbf {\bibinfo {volume}
  {83}},\ \bibinfo {pages} {174416} (\bibinfo {year} {2011})}\BibitemShut
  {NoStop}%
\bibitem [{\citenamefont {Ganesh}\ \emph {et~al.}(2011)\citenamefont {Ganesh},
  \citenamefont {Isakov},\ and\ \citenamefont
  {Paramekanti}}]{Ganesh:2011_honey_bilayer_PRB84}%
  \BibitemOpen
  \bibfield  {author} {\bibinfo {author} {\bibfnamefont {R.}~\bibnamefont
  {Ganesh}}, \bibinfo {author} {\bibfnamefont {Sergei~V.}\ \bibnamefont
  {Isakov}}, \ and\ \bibinfo {author} {\bibfnamefont {Arun}\ \bibnamefont
  {Paramekanti}},\ }\bibfield  {title} {\enquote {\bibinfo {title} {N{\'{e}}el
  to dimer transition in spin-${S}$ antiferromagnets: Comparing bond operator
  theory with quantum {M}onte {C}arlo simulations for bilayer {H}eisenberg
  models},}\ }\href {\doibase 10.1103/PhysRevB.84.214412} {\bibfield  {journal}
  {\bibinfo  {journal} {Phys. Rev. B}\ }\textbf {\bibinfo {volume} {84}},\
  \bibinfo {pages} {214412} (\bibinfo {year} {2011})}\BibitemShut {NoStop}%
\bibitem [{\citenamefont {Helmes}\ and\ \citenamefont
  {Wessel}(2014)}]{Helmes:2014_honey_bilayer}%
  \BibitemOpen
  \bibfield  {author} {\bibinfo {author} {\bibfnamefont {Johannes}\
  \bibnamefont {Helmes}}\ and\ \bibinfo {author} {\bibfnamefont {Stefan}\
  \bibnamefont {Wessel}},\ }\bibfield  {title} {\enquote {\bibinfo {title}
  {Entanglement entropy scaling in the bilayer {H}eisenberg spin system},}\
  }\href {\doibase 10.1103/PhysRevB.89.245120} {\bibfield  {journal} {\bibinfo
  {journal} {Phys. Rev. B}\ }\textbf {\bibinfo {volume} {89}},\ \bibinfo
  {pages} {245120} (\bibinfo {year} {2014})}\BibitemShut {NoStop}%
\bibitem [{\citenamefont {Devakul}\ and\ \citenamefont
  {Singh}(2014)}]{Devakul:2014_honey_bilayer}%
  \BibitemOpen
  \bibfield  {author} {\bibinfo {author} {\bibfnamefont {Trithep}\ \bibnamefont
  {Devakul}}\ and\ \bibinfo {author} {\bibfnamefont {Rajiv R.~P.}\ \bibnamefont
  {Singh}},\ }\bibfield  {title} {\enquote {\bibinfo {title} {Quantum critical
  universality and singular corner entanglement entropy of bilayer
  {H}eisenberg-{I}sing model},}\ }\href {\doibase 10.1103/PhysRevB.90.064424}
  {\bibfield  {journal} {\bibinfo  {journal} {Phys. Rev. B}\ }\textbf {\bibinfo
  {volume} {90}},\ \bibinfo {pages} {064424} (\bibinfo {year}
  {2014})}\BibitemShut {NoStop}%
\bibitem [{\citenamefont {Loh{\"{o}}fer}\ \emph {et~al.}(2015)\citenamefont
  {Loh{\"{o}}fer}, \citenamefont {Coletta}, \citenamefont {Joshi},
  \citenamefont {Assaad}, \citenamefont {Vojta}, \citenamefont {Wessel},\ and\
  \citenamefont {Mila}}]{Lohofer:2015_honey_bilayer}%
  \BibitemOpen
  \bibfield  {author} {\bibinfo {author} {\bibfnamefont {M.}~\bibnamefont
  {Loh{\"{o}}fer}}, \bibinfo {author} {\bibfnamefont {T.}~\bibnamefont
  {Coletta}}, \bibinfo {author} {\bibfnamefont {D.~G.}\ \bibnamefont {Joshi}},
  \bibinfo {author} {\bibfnamefont {F.~F.}\ \bibnamefont {Assaad}}, \bibinfo
  {author} {\bibfnamefont {M.}~\bibnamefont {Vojta}}, \bibinfo {author}
  {\bibfnamefont {S.}~\bibnamefont {Wessel}}, \ and\ \bibinfo {author}
  {\bibfnamefont {F.}~\bibnamefont {Mila}},\ }\bibfield  {title} {\enquote
  {\bibinfo {title} {Dynamical structure factors and excitation modes of the
  bilayer {H}eisenberg model},}\ }\href {\doibase 10.1103/PhysRevB.92.245137}
  {\bibfield  {journal} {\bibinfo  {journal} {Phys. Rev. B}\ }\textbf {\bibinfo
  {volume} {92}},\ \bibinfo {pages} {245137} (\bibinfo {year}
  {2015})}\BibitemShut {NoStop}%
\bibitem [{\citenamefont {Hida}(1996)}]{Kazuo:1996_SqLatt_bilayer_J1J2J1perp}%
  \BibitemOpen
  \bibfield  {author} {\bibinfo {author} {\bibfnamefont {Kazuo}\ \bibnamefont
  {Hida}},\ }\bibfield  {title} {\enquote {\bibinfo {title} {Modified spin wave
  theory of the bilayer square lattice frustrated quantum {H}eisenberg
  antiferromagnet},}\ }\href {\doibase 10.1143/JPSJ.65.594} {\bibfield
  {journal} {\bibinfo  {journal} {J. Phys. Soc. Jpn.}\ }\textbf {\bibinfo
  {volume} {65}},\ \bibinfo {pages} {594--600} (\bibinfo {year}
  {1996})}\BibitemShut {NoStop}%
\bibitem [{\citenamefont {Hida}(1998)}]{Kazuo:1998_SqLatt_bilayer_J1J2J1perp}%
  \BibitemOpen
  \bibfield  {author} {\bibinfo {author} {\bibfnamefont {Kazuo}\ \bibnamefont
  {Hida}},\ }\bibfield  {title} {\enquote {\bibinfo {title} {Dimer expansion
  study of the bilayer square lattice frustrated quantum {H}eisenberg
  antiferromagnet},}\ }\href {\doibase 10.1143/JPSJ.67.1540} {\bibfield
  {journal} {\bibinfo  {journal} {J. Phys. Soc. Jpn.}\ }\textbf {\bibinfo
  {volume} {67}},\ \bibinfo {pages} {1540--1543} (\bibinfo {year}
  {1998})}\BibitemShut {NoStop}%
\bibitem [{\citenamefont {Stapmanns}\ \emph {et~al.}(2018)\citenamefont
  {Stapmanns}, \citenamefont {Corboz}, \citenamefont {Mila}, \citenamefont
  {Honecker}, \citenamefont {Normand},\ and\ \citenamefont
  {Wessel}}]{Stapmanns:2018_SqLatt_bilayer_J1J1perpJ2perp}%
  \BibitemOpen
  \bibfield  {author} {\bibinfo {author} {\bibfnamefont {J.}~\bibnamefont
  {Stapmanns}}, \bibinfo {author} {\bibfnamefont {P.}~\bibnamefont {Corboz}},
  \bibinfo {author} {\bibfnamefont {F.}~\bibnamefont {Mila}}, \bibinfo {author}
  {\bibfnamefont {A.}~\bibnamefont {Honecker}}, \bibinfo {author}
  {\bibfnamefont {B.}~\bibnamefont {Normand}}, \ and\ \bibinfo {author}
  {\bibfnamefont {S.}~\bibnamefont {Wessel}},\ }\bibfield  {title} {\enquote
  {\bibinfo {title} {Thermal critical points and quantum critical end point in
  the frustrated bilayer {H}eisenberg antiferromagnet},}\ }\href {\doibase
  10.1103/PhysRevLett.121.127201} {\bibfield  {journal} {\bibinfo  {journal}
  {Phys. Rev. Lett.}\ }\textbf {\bibinfo {volume} {121}},\ \bibinfo {pages}
  {127201} (\bibinfo {year} {2018})}\BibitemShut {NoStop}%
\bibitem [{\citenamefont {Alet}\ \emph {et~al.}(2016)\citenamefont {Alet},
  \citenamefont {Damle},\ and\ \citenamefont
  {Pujari}}]{Alet:2016_SqLatt_bilayer}%
  \BibitemOpen
  \bibfield  {author} {\bibinfo {author} {\bibfnamefont {Fabien}\ \bibnamefont
  {Alet}}, \bibinfo {author} {\bibfnamefont {Kedar}\ \bibnamefont {Damle}}, \
  and\ \bibinfo {author} {\bibfnamefont {Sumiran}\ \bibnamefont {Pujari}},\
  }\bibfield  {title} {\enquote {\bibinfo {title} {Sign-problem-free {M}onte
  {C}arlo simulation of certain frustrated quantum magnets},}\ }\href {\doibase
  10.1103/PhysRevLett.117.197203} {\bibfield  {journal} {\bibinfo  {journal}
  {Phys. Rev. Lett.}\ }\textbf {\bibinfo {volume} {117}},\ \bibinfo {pages}
  {197203} (\bibinfo {year} {2016})}\BibitemShut {NoStop}%
\bibitem [{\citenamefont {Bishop}\ \emph
  {et~al.}(2008{\natexlab{c}})\citenamefont {Bishop}, \citenamefont {Li},
  \citenamefont {Darradi},\ and\ \citenamefont
  {Richter}}]{Bi:2008_EPL_J1J1primeJ2_s1}%
  \BibitemOpen
  \bibfield  {author} {\bibinfo {author} {\bibfnamefont {R.~F.}\ \bibnamefont
  {Bishop}}, \bibinfo {author} {\bibfnamefont {P.~H.~Y.}\ \bibnamefont {Li}},
  \bibinfo {author} {\bibfnamefont {R.}~\bibnamefont {Darradi}}, \ and\
  \bibinfo {author} {\bibfnamefont {J.}~\bibnamefont {Richter}},\ }\bibfield
  {title} {\enquote {\bibinfo {title} {The quantum
  ${J}_{1}$--${J}_{1}'$--${J}_{2}$ spin-$1$ {H}eisenberg model: Influence of
  the interchain coupling on the ground-state magnetic ordering in 2{D}},}\
  }\href {\doibase 10.1209/0295-5075/83/47004} {\bibfield  {journal} {\bibinfo
  {journal} {EPL}\ }\textbf {\bibinfo {volume} {83}},\ \bibinfo {pages} {47004}
  (\bibinfo {year} {2008}{\natexlab{c}})}\BibitemShut {NoStop}%
\bibitem [{\citenamefont {Bishop}\ \emph
  {et~al.}(2008{\natexlab{d}})\citenamefont {Bishop}, \citenamefont {Li},
  \citenamefont {Darradi}, \citenamefont {Richter},\ and\ \citenamefont
  {Campbell}}]{Bi:2008_JPCM_J1xxzJ2xxz_s1}%
  \BibitemOpen
  \bibfield  {author} {\bibinfo {author} {\bibfnamefont {R.~F.}\ \bibnamefont
  {Bishop}}, \bibinfo {author} {\bibfnamefont {P.~H.~Y.}\ \bibnamefont {Li}},
  \bibinfo {author} {\bibfnamefont {R.}~\bibnamefont {Darradi}}, \bibinfo
  {author} {\bibfnamefont {J.}~\bibnamefont {Richter}}, \ and\ \bibinfo
  {author} {\bibfnamefont {C.~E.}\ \bibnamefont {Campbell}},\ }\bibfield
  {title} {\enquote {\bibinfo {title} {The effect of anisotropy on the
  ground-state magnetic ordering of the spin-1 quantum
  ${J}_{1}^{XXZ}$--${J}_{2}^{XXZ}$ model on the square lattice},}\ }\href
  {\doibase 10.1088/0953-8984/20/41/415213} {\bibfield  {journal} {\bibinfo
  {journal} {J. Phys.: Condens. Matter}\ }\textbf {\bibinfo {volume} {20}},\
  \bibinfo {pages} {415213} (\bibinfo {year} {2008}{\natexlab{d}})}\BibitemShut
  {NoStop}%
\bibitem [{\citenamefont {Haghshenas}\ \emph {et~al.}(2018)\citenamefont
  {Haghshenas}, \citenamefont {Lan}, \citenamefont {Gong},\ and\ \citenamefont
  {Sheng}}]{Haghshenas:2018_SqLatt_J1J2mod_s1}%
  \BibitemOpen
  \bibfield  {author} {\bibinfo {author} {\bibfnamefont {R.}~\bibnamefont
  {Haghshenas}}, \bibinfo {author} {\bibfnamefont {Wang-Wei}\ \bibnamefont
  {Lan}}, \bibinfo {author} {\bibfnamefont {Shou-Shu}\ \bibnamefont {Gong}}, \
  and\ \bibinfo {author} {\bibfnamefont {D.~N.}\ \bibnamefont {Sheng}},\
  }\bibfield  {title} {\enquote {\bibinfo {title} {Quantum phase diagram of
  spin-1 ${J}_{1}$-${J}_{2}$ {H}eisenberg model on the square lattice: An
  infinite projected entangled-pair state and density matrix renormalization
  group study},}\ }\href {\doibase 10.1103/PhysRevB.97.184436} {\bibfield
  {journal} {\bibinfo  {journal} {Phys. Rev. B}\ }\textbf {\bibinfo {volume}
  {97}},\ \bibinfo {pages} {184436} (\bibinfo {year} {2018})}\BibitemShut
  {NoStop}%
\bibitem [{\citenamefont {Jiang}\ \emph {et~al.}(2009)\citenamefont {Jiang},
  \citenamefont {Kr{\"{u}}ger}, \citenamefont {Moore}, \citenamefont {Sheng},
  \citenamefont {Zaanen},\ and\ \citenamefont
  {Weng}}]{Jiang:2009_SqLatt_J1J2mod_s1}%
  \BibitemOpen
  \bibfield  {author} {\bibinfo {author} {\bibfnamefont {H.~C.}\ \bibnamefont
  {Jiang}}, \bibinfo {author} {\bibfnamefont {F.}~\bibnamefont {Kr{\"{u}}ger}},
  \bibinfo {author} {\bibfnamefont {J.~E.}\ \bibnamefont {Moore}}, \bibinfo
  {author} {\bibfnamefont {D.~N.}\ \bibnamefont {Sheng}}, \bibinfo {author}
  {\bibfnamefont {J.}~\bibnamefont {Zaanen}}, \ and\ \bibinfo {author}
  {\bibfnamefont {Z.~Y.}\ \bibnamefont {Weng}},\ }\bibfield  {title} {\enquote
  {\bibinfo {title} {Phase diagram of the frustrated spatially-anisotropic
  ${S}=1$ antiferromagnet on a square lattice},}\ }\href {\doibase
  10.1103/PhysRevB.79.174409} {\bibfield  {journal} {\bibinfo  {journal} {Phys.
  Rev. B}\ }\textbf {\bibinfo {volume} {79}},\ \bibinfo {pages} {174409}
  (\bibinfo {year} {2009})}\BibitemShut {NoStop}%
\bibitem [{\citenamefont {Li}\ and\ \citenamefont
  {Bishop}(2019)}]{Li:2019_honeycomb_bilayer_J1J2J1perp_neel-II}%
  \BibitemOpen
  \bibfield  {author} {\bibinfo {author} {\bibfnamefont {P.~H.~Y.}\
  \bibnamefont {Li}}\ and\ \bibinfo {author} {\bibfnamefont {R.~F.}\
  \bibnamefont {Bishop}},\ }\bibfield  {title} {\enquote {\bibinfo {title}
  {Collinear antiferromagnetic phases of a frustrated spin-$\frac{1}{2}$
  ${J}_{1}$--${J}_{2}$--${J}_{1}^{\perp}$ {H}eisenberg model on an
  ${AA}$-stacked bilayer honeycomb lattice},}\ }\href {\doibase
  10.1016/j.jmmm.2019.03.033} {\bibfield  {journal} {\bibinfo  {journal} {J.
  Magn. Magn. Mater.}\ }\textbf {\bibinfo {volume} {482}},\ \bibinfo {pages}
  {262--273} (\bibinfo {year} {2019})}\BibitemShut {NoStop}%
\bibitem [{ccm()}]{ccm_code}%
  \BibitemOpen
  \href {http://www-e.uni-magdeburg.de/jschulen/ccm/index.html} {}\bibinfo
  {note} {We use the program package CCCM of D.~J.~J. Farnell and
  J.~Schulenburg, see
  http://www-e.uni-magdeburg.de/jschulen/ccm/index.html}\BibitemShut {NoStop}%
\bibitem [{\citenamefont {Farnell}\ \emph {et~al.}(2018)\citenamefont
  {Farnell}, \citenamefont {G{\"{o}}tze}, \citenamefont {Schulenburg},
  \citenamefont {Zinke}, \citenamefont {Bishop},\ and\ \citenamefont
  {Li}}]{Farnell:2018_archimedean_multiSpins}%
  \BibitemOpen
  \bibfield  {author} {\bibinfo {author} {\bibfnamefont {D.~J.~J.}\
  \bibnamefont {Farnell}}, \bibinfo {author} {\bibfnamefont {O.}~\bibnamefont
  {G{\"{o}}tze}}, \bibinfo {author} {\bibfnamefont {J.}~\bibnamefont
  {Schulenburg}}, \bibinfo {author} {\bibfnamefont {R.}~\bibnamefont {Zinke}},
  \bibinfo {author} {\bibfnamefont {R.~F.}\ \bibnamefont {Bishop}}, \ and\
  \bibinfo {author} {\bibfnamefont {P.~H.~Y.}\ \bibnamefont {Li}},\ }\bibfield
  {title} {\enquote {\bibinfo {title} {Interplay between lattice topology,
  frustration, and spin quantum number in quantum antiferromagnets on
  {A}rchimedean lattices},}\ }\href {\doibase 10.1103/PhysRevB.98.224402}
  {\bibfield  {journal} {\bibinfo  {journal} {Phys. Rev. B}\ }\textbf {\bibinfo
  {volume} {98}},\ \bibinfo {pages} {224402} (\bibinfo {year}
  {2018})}\BibitemShut {NoStop}%
\end{thebibliography}%

\end{document}